\documentclass{jfm}
\usepackage{graphicx}
\usepackage{epstopdf, epsfig}
\usepackage{multirow}
\usepackage[dvipsnames]{xcolor}
\usepackage{amsmath}
\usepackage{amssymb}
\usepackage{lineno}

\definecolor{col1}{rgb}{0         0    1.0000}
\definecolor{col2}{rgb}{0.2500         0    0.7500}
\definecolor{col3}{rgb}{0.5000         0    0.5000}
\definecolor{col4}{rgb}{0.7500         0    0.2500}
\definecolor{col5}{rgb}{1.0000         0         0}
\definecolor{col6}{rgb}{     0    0.4470    0.7410}
\definecolor{col7}{rgb}{0.8500    0.3250    0.0980}
\definecolor{col8}{rgb}{0.9290    0.6940    0.1250}
\definecolor{col9}{rgb}{0.4940    0.1840    0.5560}
\definecolor{col10}{rgb}{0.4660    0.6740    0.1880}
\definecolor{col11}{rgb}{0.3010    0.7450    0.9330}

\definecolor{colblue}{rgb}{0   0   1}
\definecolor{colmagenta}{rgb}{1   0   1}
\definecolor{colgrey}{rgb}{0.75 0.75 0.75}

\definecolor{colyellow}{rgb}{1.0000    0.8390    0.0390}
\definecolor{colred}{rgb}{0.6350    0.0780    0.1840}

\definecolor{col_m}{rgb}{1 0 1}

\definecolor{col_c}{rgb}{0.9290 0.6940 0.1250}

\def\aaa{{\textit{a}}}
\def\bbb{{\textit{b}}}
\def\ccc{{\textit{c}}}
\def\ddd{{\textit{d}}}
\def\eee{{\textit{e}}}
\def\fff{{\textit{f}}}
\def\ggg{{\textit{g}}}
\def\hhh{{\textit{h}}}
\def\iii{{\textit{i}}}

\newcommand{\rgm}[1]{#1}
\newcommand{\zsc}[1]{#1}
\newcommand{\oldrev}[1]{#1}
\newcommand{\rtwo}[1]{#1}

\usepackage{tikz}
\newcommand{\mdot}{%
  \tikz[baseline=-0.5ex]\draw[line width=0.7pt, black, fill=colmagenta] (0,0) circle (0.5ex);%
}
\newcommand{\rdot}{%
  \tikz[baseline=-0.5ex]\draw[line width=0.7pt, black, fill=colred] (0,0) circle (0.5ex);%
}
\newcommand{\ydot}{%
  \tikz[baseline=-0.5ex]\draw[line width=0.7pt, black, fill=colyellow] (0,0) circle (0.5ex);%
}

\shorttitle{Characterisation of canopy density}
\shortauthor{Z. Chen and R. García-Mayoral}

\title{A systematic characterisation of canopy density based on turbulent-structure penetration}

\author{Zishen Chen\aff{1}
 \and Ricardo García-Mayoral\aff{1}
 \corresp{\email{r.gmayoral@eng.cam.ac.uk}}}

\affiliation{\aff{1}Department of Engineering, University of Cambridge, Trumpington Street, Cambridge CB2 1PZ, UK}

\begin{document}

\maketitle

\begin{abstract}
Turbulent flows over canopies of rigid elements with different geometries, spacings and Reynolds numbers are investigated to identify and characterise different canopy density regimes. In the sparse regime, turbulence penetrates relatively unhindered within the canopy, whereas in the dense regime, this penetration is limited. A common measure of canopy density is the ratio of frontal to bed area, the frontal density $\lambda_f$.
While effective for conventional vegetation canopies with no preferential orientation, we observe that $\lambda_f$ does not accurately predict the density regime for some less conventional canopy topologies, suggesting that it does not necessarily encapsulate the physics governing canopy density. \oldrev{To address this, we adopt a direct approach that quantifies the degree of penetration of the overlying turbulence into the canopy. We propose density metrics based on the position and extent of individual flow eddies, in particular those of intense Reynolds shear stress $u'v'$}. We analyse a series of direct simulations for both isotropic- and anisotropic-layout canopies across a range of frontal densities $\lambda_f\approx0.01$-$2.04$, heights $h^+\approx44$-$266$, \oldrev{element width-to-pitch ratios $w/s\approx0.06$-$0.7$,} and Reynolds numbers $Re_\tau\approx180$-$2000$.
Canopies with elements closely packed in the streamwise direction but large spanwise gaps allow for significant turbulence penetration, and thus appear sparser compared to isotropic or spanwise-packed canopies with the same $\lambda_f$.
For the same spanwise gap, turbulence penetration remains similar and largely independent of the streamwise \oldrev{pitch and gap between} elements.
As the spanwise gap increases, eddies penetrate deeper and more vigorously into the canopy.
Turbulence penetration is also \oldrev{Reynolds-number-dependent;}
the same canopy topography can behave as dense at low $Re_\tau$, but \oldrev{have increasing turbulence penetration} as $Re_\tau$ increases.
\rtwo{Our results suggest that the penetration of the overlying turbulence depends
essentially on the ability of turbulent eddies as they travel downstream to fit in between canopy elements, and that this can be characterised by how the typical eddy size compares to an effective spanwise gap.}
\oldrev{Turbulence penetrates easily into the canopy when the spanwise gap is larger than the eddy size, and is essentially precluded from penetrating in the opposite case.
A penetration length can then be defined that is of the order of the \rtwo{effective} spanwise gap or the eddy size, whichever is smaller.
If the penetration length is small compared to the canopy height, the canopy behaves as dense; if it is comparable, the canopy has an intermediate behaviour; and if it is approximately equal or larger than the canopy height, the eddies penetrate all the way to the canopy bed and the canopy behaves as sparse.}
\end{abstract}


\section{Introduction and background}\label{sec:intro}
Canopies of protruding elements are ubiquitous in practical flows, holding significant environmental and engineering importance. Vegetation canopies are of great ecological importance to both terrestrial and aquatic ecosystems, mediating the concentration of nutrients, oxygen, and carbon \citep{malhi2002carbon, baldocchi2003assessing}, as reviewed by \cite{finnigan2000turbulence}, \cite{belcher2012wind}, \cite{nepf2012flow, nepf2012hydrodynamics} and \cite{brunet2020turbulent}. Canopies of pin-fin arrays are widely used in industrial applications for heat transfer enhancement \citep{ligrani2003comparison, peles2005forced, arshad2017thermal}, as summarised in \cite{mousa2021review}. The study of canopy turbulence is also relevant to a wide range of applications, including cooling urban heat islands \citep{akbari2016three, rahman2020traits}, mitigating agricultural loss due to windthrow \citep{berry2004understanding, de2008effects}, monitoring the transport of airborne pollutants \citep{belcher2005mixing, escobedo2011urban, janhall2015review}, and forecasting weather \citep{britter2003flow, blocken2015computational}. Consequently, understanding the interaction between canopies and turbulence is crucial for effectively managing such practical flows and improving the related environmental and industrial practice.

\begin{figure}
    \centering
    \includegraphics[width=0.9\textwidth]{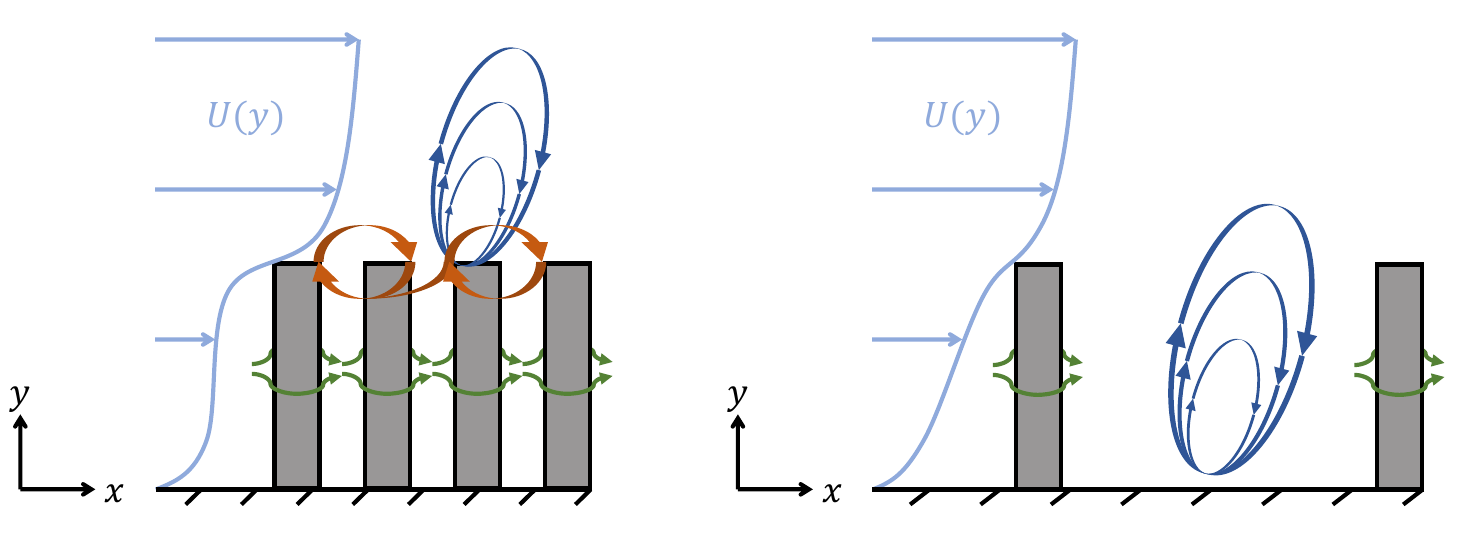}
    \put (-125mm,40mm) {(\aaa)}
    \put (-65mm,40mm) {(\bbb)}
    \caption{Graphical illustration of the flow regimes over (\aaa) dense and (\bbb) sparse canopies. The dark blue, orange and green arrows represent the background-turbulence eddies, mixing-layer-like eddies and \rgm{element-coherent} eddies, respectively. Adapted from \cite{poggi2004effect} and \cite{sharma2020turbulent}.}
    \label{fig:poggi}
    \centering
    \includegraphics[trim={0 1mm 0 11mm},clip,width=1.00\textwidth]{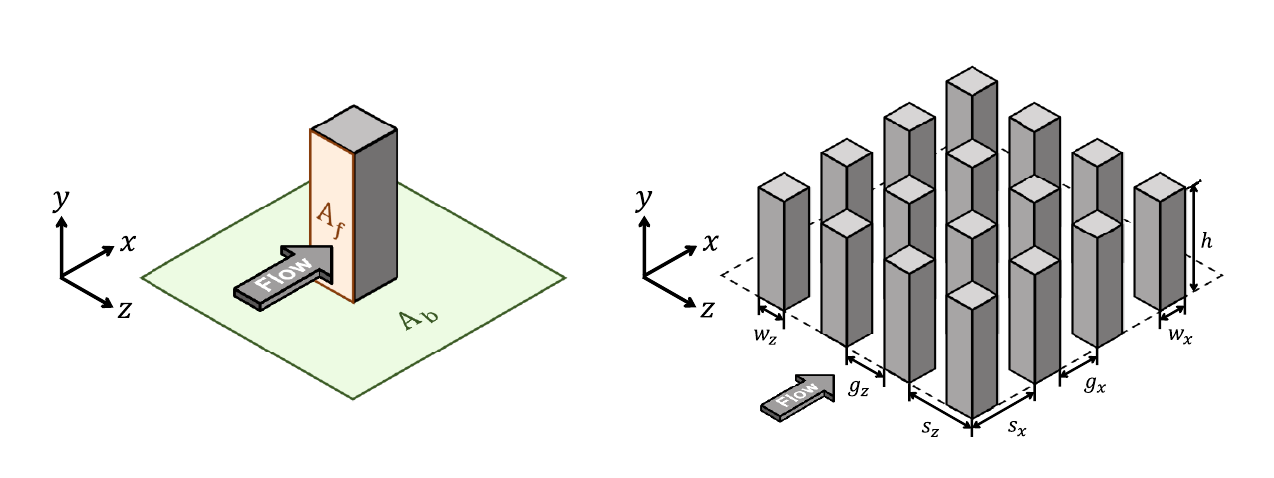}
    \put (-133mm,40mm) {(\aaa)}
    \put (-70mm,40mm) {(\bbb)}
\vspace*{-3mm}
    \caption{Schematic representation of (\aaa) a canopy element and (\bbb) a layout of canopy elements. In (\aaa), $A_f$ is the frontal area and \zsc{$A_b$ is the bed area occupied by each element.} In (\bbb), $h$ is the element height; $w_x$, $g_x$ and $s_x$ are the element \rgm{width}, gap and spacing (pitch) in the streamwise direction, respectively, and $w_z$, $g_z$ and $s_z$ are those in the spanwise direction.}
    \label{fig:lambda_parameter}
\end{figure}

Canopy density has a direct impact on the turbulent flow within and immediately above the canopy \citep{nepf2012flow, brunet2020turbulent, sharma2020scaling_a, sharma2020turbulent_b}. \cite{poggi2004effect} conducted experiments on flows over various canopy densities and identified the characteristics of different density regimes, suggesting that the flow over canopies can be broadly divided into three components. A graphical illustration of these components and their characteristic eddies for dense and sparse canopies is depicted in figure \ref{fig:poggi}.
The first component is the background turbulence characteristic of boundary layers -- this is the component that would be present even in the absence of the canopy.
The second component is the element-induced or element-coherent flow, which mainly occurs within the canopy and results from the obstruction caused by individual elements \citep{cava2008spectral, dupont2011long}. The third component is the mixing-layer-like flow that forms due to the mean-flow inflection in the vicinity of the canopy-tip plane of a dense canopy \zsc{\citep{raupach1996coherent, finnigan2000turbulence, finnigan2009turbulence}}. \cite{poggi2004effect} emphasised that the flow behaviour near the canopy-tip plane is largely dependent on the canopy density regime, and that mixing-layer-like eddies only appear when the canopy density is high enough to induce a strong and relatively homogeneous shear layer \citep{raupach1996coherent, sharma2020turbulent_b}.
\oldrev{If we consider the eddies from the background turbulence as they penetrate into the canopy,
their relative intensity and penetration depth vary depending on the canopy density \citep{nepf2012flow, brunet2020turbulent}}. As illustrated in figure \ref{fig:poggi}(\aaa), dense canopies with closely packed elements effectively limit the penetration of the background turbulence, and the flow within is essentially sheltered from that above, except for a weak footprint of the mixing-layer eddies \citep{sharma2020turbulent_b}. In contrast, as depicted in figure \ref{fig:poggi}(\bbb), sparse canopies with large element spacing allow the background turbulence to penetrate relatively unhindered into the canopy.
Based on this, turbulence over sparse canopies has typically been considered a simple superposition of the wall-bounded flow and the \rgm{element-coherent} flow, where the mean-velocity profile resembles that over a smooth wall \citep{huang2009effects, pietri2009turbulence, nepf2012flow}. 
However, this argument is valid only when the drag induced by the canopy elements is negligible compared to the total drag.
Recent direct numerical simulations (DNSs) have shown that even canopies with a frontal density as low as $\lambda_f\approx0.01$ and spacing-to-height ratios as large as $s/h\approx4$ can
result in substantial element-induced drag, accounting for more than 60\% of the total drag, and that such sparse canopies mainly modulate the background turbulence through their drag on the mean velocity
profile, as the mean shear sets the scale of turbulence locally at each height \citep{sharma2020scaling_a}.

\begin{figure}
\vspace*{1mm}
    \centering
    \includegraphics[width=0.75\textwidth]{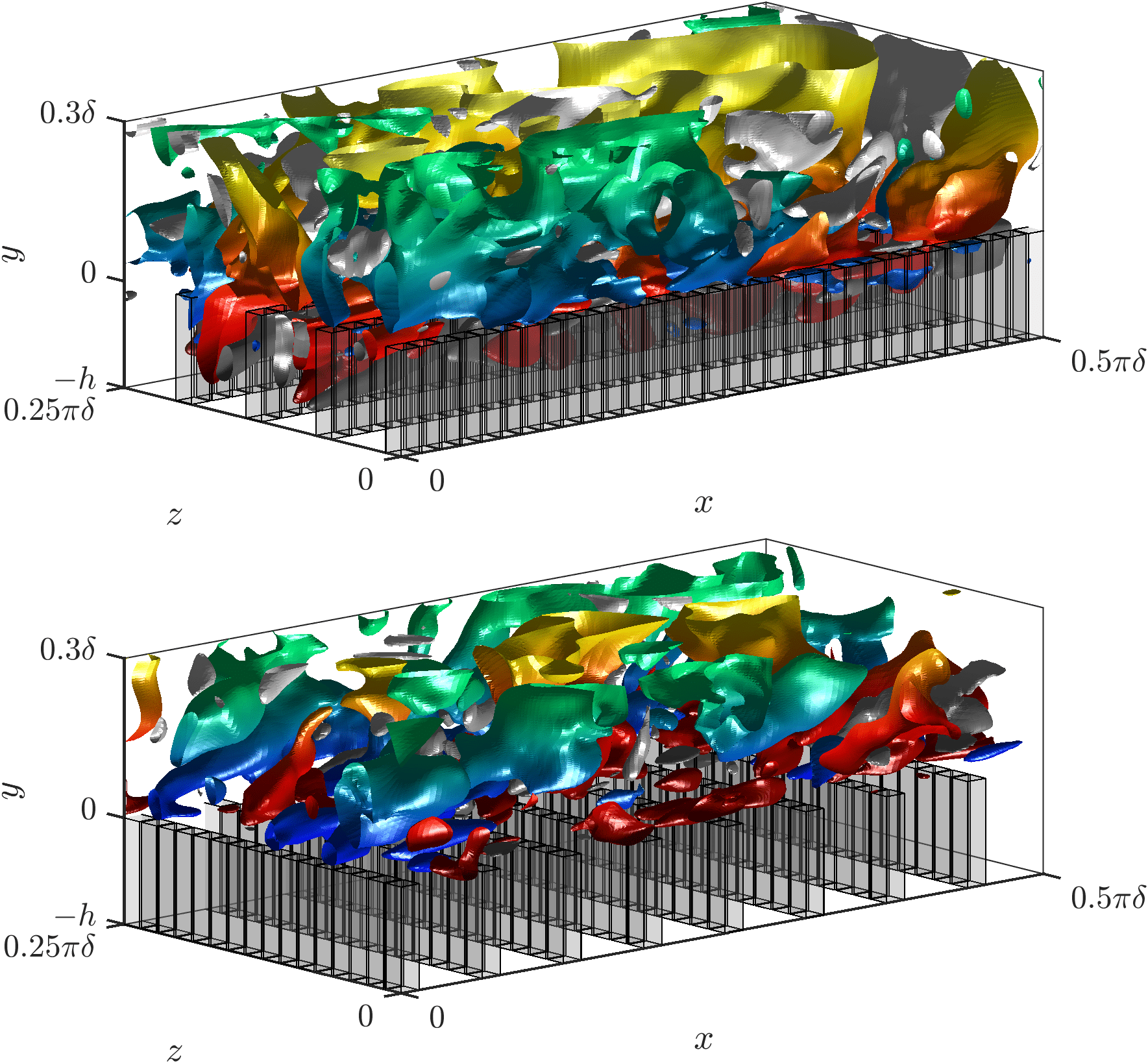}
    \caption{Instantaneous realisations of the $u'v'$ structures over and within canopies with the same number of elements per area and frontal density $\lambda_f\approx0.91$: (\aaa) streamwise-packed canopy $\mathrm{C_{S27\times108}}$ with $s_x^+\approx27$, $s_z^+\approx108$ and $h^+\approx110$, and (\bbb) spanwise-packed canopy $\mathrm{F_{S108\times27}}$ with $s_x^+\approx108$, $s_z^+\approx27$ and $h^+\approx110$. The structures are coloured by distance to the floor, and consist of ejections ($u'<0, v'>0$ blue to green), sweeps ($u'>0, v'<0$ red to yellow), and outward and inward interactions ($u'v'>0$ grey to white).}
    \label{fig:uvster_can_fen}
\end{figure}

A popular quantity to characterise canopy density is the frontal density \citep{nepf2012flow}, also known as the leaf area index in vegetated canopy literature \citep{wooding1973drag, kaimal1994atmospheric},
\begin{equation}
 \lambda_f = \frac{A_f}{A_b} = \zsc{\frac{1}{A_b}\int_{0}^{h} w_z(y)dy,}
\label{eq:lambda_f}
\end{equation}
\zsc{where $A_f/A_b$ indicates the frontal area per unit bed area, $h$ is the element height, and $w_z(y)$ is the element frontal width at each height $y$, as depicted in figure \ref{fig:lambda_parameter}(\aaa). For the present prismatic canopies, the frontal density becomes $\lambda_f=w_z h /(s_x s_z)$, 
as shown in figure \ref{fig:lambda_parameter}(\bbb), where $w$} \rgm{indicates the element width, $g$ the gap between elements, and $s$ their pitch or spacing, with the subscripts referring to} the streamwise ($x$) and spanwise ($z$) directions. \cite{nepf2012flow} noted that canopies are dense when $\lambda_f\gg0.1$, sparse when $\lambda_f\ll0.1$, and intermediate between the two limits. Permeability and porosity ($\phi$) are also popular measures of density for denser canopies and other permeable substrates, embodying the `accessibility' of the fluid region \citep{lightbody2006prediction, breugem2006influence, luhar2008interaction}.
However, these geometric parameters only provide a notional measure of canopy density, and occasionally fail to capture how different element arrangements can result in different flow regimes.
The studies by \cite{pietri2009turbulence} and \cite{bailey2013turbulence} highlighted that identical density values of $\lambda_f$ and $\phi$ do not necessarily translate to similar flow penetration behaviours and may not accurately reflect the actual canopy density regime.
This limitation is illustrated in figure \ref{fig:uvster_can_fen}, \oldrev{depicting structures of intense $u'v'$} over canopies with the same number of elements and identical $\lambda_f$ but different element layout, \oldrev{which are discussed in full detail in \S \ref{sec:fencecanyon}.}
As shown in figure \ref{fig:uvster_can_fen}(\aaa), these structures penetrate more effectively into the streamwise-packed canopy compared to the spanwise-packed canopy in figure \ref{fig:uvster_can_fen}(\bbb), showing that element layout, a factor not captured by $\lambda_f$, can significantly influence the canopy density regime.

\rgm{
The frontal density $\lambda_f$ has been proposed as a predictor for canopy density because it is a convenient surrogate for physics-based quantities, such as the drag length $\ell_d$, which characterise the impedance that the canopy exerts on the flow  \citep{nepf2012flow}.
The drag length $\ell_d$ gives a measure of how quickly the mean flow $U(y)$ decays below the canopy tips, and can be defined as \rtwo{$\ell_d=U^2/F_D$} using the mean velocity and surface drag \rtwo{$F_D$} at the plane of the tips.
It would then be natural to expect that the differences between streamwise- and spanwise-packed layouts mentioned above could be captured by parameters such as $\ell_d$ since, even for the same $\lambda_f$, the higher drag sheltering for streamwise-packed configurations results in a larger $\ell_d$.
Other lengthscales that can quantify the penetration of the mean flow are the shear length $\ell_s=U/(dU/dy)$ measured at the tip plane, which quantifies how inflexional the $U(y)$ profile is, and thus the intensity of the drag acting on it \citep{Endrikat2021};
or the  depth below the tips of the zero-plane-displacement, $d_0$, the depth perceived as the origin by the overlying turbulence \citep{chen2023examination}. 
All these lengthscales, however, characterise the effect of the canopy on the penetration of the mean velocity profile, rather than of the set of fluctuating eddies that conform the overlying turbulence.
This paper will propose a lengthscale to characterise canopy density based on the latter. 
For conventional canopy layouts, \rtwo{such as $x$-$z$-isotropic vegetation canopies and their regular surrogates often studied in the literature,} we would expect reasonable correlation between both approaches, but possibly not for counter-examples such as those of figure \ref{fig:uvster_can_fen}. 
An assessment of the above lengthscales as predictors for density can be found in appendix \ref{app:other_densities}. They provide a generally correct trend but with significant scatter, comparable to that produced using the purely-geometric $\lambda_f$.}

\cite{sharma2020scaling_a} argued that a canopy would be sparse if the spanwise spacing between elements is $s_z^+\gtrsim100$, which is large enough to fit in the near-wall streaks of characteristic width $\lambda_z^+\approx100$ \citep{kline1967structure, kim1971production}.
This is consistent with the observations of \cite{poggi2004effect} and \cite{huang2009effects}, who noted a progressive departure from smooth-wall behaviour for sparse canopies as the spacing between elements decreases.
However, \cite{sharma2020scaling_a} investigated only sparse canopies of \oldrev{isotropically laid} filaments, and did not consider the effects of a wider variety of geometric parameters;
\rtwo{a spanwise spacing can be easily defined for the collocated elements of \cite{sharma2020scaling_a}, but doing so for irregular or even staggered arrangements may be less straightforward.}
\oldrev{Canopies that are clearly sparse or clearly dense were extensively investigated in \cite{sharma2020scaling_a} and \cite{sharma2020turbulent_b}, but the transition from one limit to the other was not directly considered}. This highlights the need for a more systematic approach to measure the extent of turbulence penetration, and to characterise canopy density.
Recent studies have proposed to characterise canopy density as determined by the interchange between the relative vertical positions of the zero-plane-displacement height and the
inflection in the mean velocity profile that occurs within the canopy \citep[see e.g.][and follow-up
papers]{Monti2020}. The latter inflection occurs due to the near-bed boundary layer that forms in order for the mean flow to satisfy no-slip at the bed. The authors argued that this lower inflection played a key role in the flow dynamics, as it gave rise to an instability, reminiscent of Kelvin-Helmhotz, because the mean velocity profile satisfied Fj{\o}rtoft’s criterion for instability.
The opposite is unfortunately the case: Fj{\o}rtoft’s criterion predicts instability for a mixing-layer type, concave-to-convex inflection, like the one at the canopy tips, and stability for convex-to-concave inflections like the one near the bed. Physically, the flow is the most unstable where the mean shear is maximum, and most stable where the shear is minimum.
Furthermore, \rgm{for flows in homogeneous canopies, the lower inflection can only occur when the flow is} driven by a mean pressure gradient or a body force, \rgm{when} a Darcy-like region of uniform velocity tends to form in the canopy core, requiring a Brinkman-like boundary layer at its bottom to meet no-slip at the bed. \rgm{Beyond pressure gradients, inner inflections can also appear when there is a local maximum in the mean velocity profile within the canopy, which can be caused by inhomogeneities such as edge effects and differences between trunk and crown drag \citep[see e.g.][and references therein]{dupont2011long}. Otherwise,} 
when the intra-canopy flow is driven mainly by the shear exerted by the overlying flow, as in zero-pressure-gradient boundary layers, the Darcy-flow contribution is non-existent, and there is therefore no inner inflection. A different approach is therefore necessary \rgm{to characterise canopy density in a common framework across the broad flow typology.}

In the present study, we conduct DNSs of canopies with different layouts and element geometries. We aim to establish a metric for canopy density based on the flow dynamics to enable the quantitative diagnosis, characterisation and prediction of canopy density. For this, we quantify the extent and depth of the penetrating three-dimensional turbulence structures of intense Reynolds shear stress $u'v'$ \citep{del2006self, lozano2012three}. We focus on these structures because they are responsible for momentum transport and turbulence diffusion between the overlying and the canopy regions, and are ultimately responsible for the wall-normal distribution of the Reynolds stress and thus the shape of the mean-velocity profile.
Based on this measure, we explore the influence of the different geometric characteristics of the canopy on whether turbulence penetrates or not, and identify which are the key ones to determine canopy density.
The paper is organised as follows. \S \ref{sec:methods} presents the numerical method and the different canopies simulated. \S \ref{sec:penetration} introduces and discusses the metrics used to measure turbulence penetration.
\S \ref{sec:results} presents the DNS results, focusing on how different canopy parameters affect turbulence penetration. Finally, \S \ref{sec:conclusions} summarises the conclusions of our investigation.

\section{Methodology}\label{sec:methods}
In this section, we briefly discuss the DNS method in \S \ref{sec:dns}, and present the relevant parameters for the canopies simulated, with different element geometry and layout, in \S \ref{sec:params}.

\begin{figure}
\vspace*{1mm}
    \centering
    \includegraphics[width=.75\textwidth]{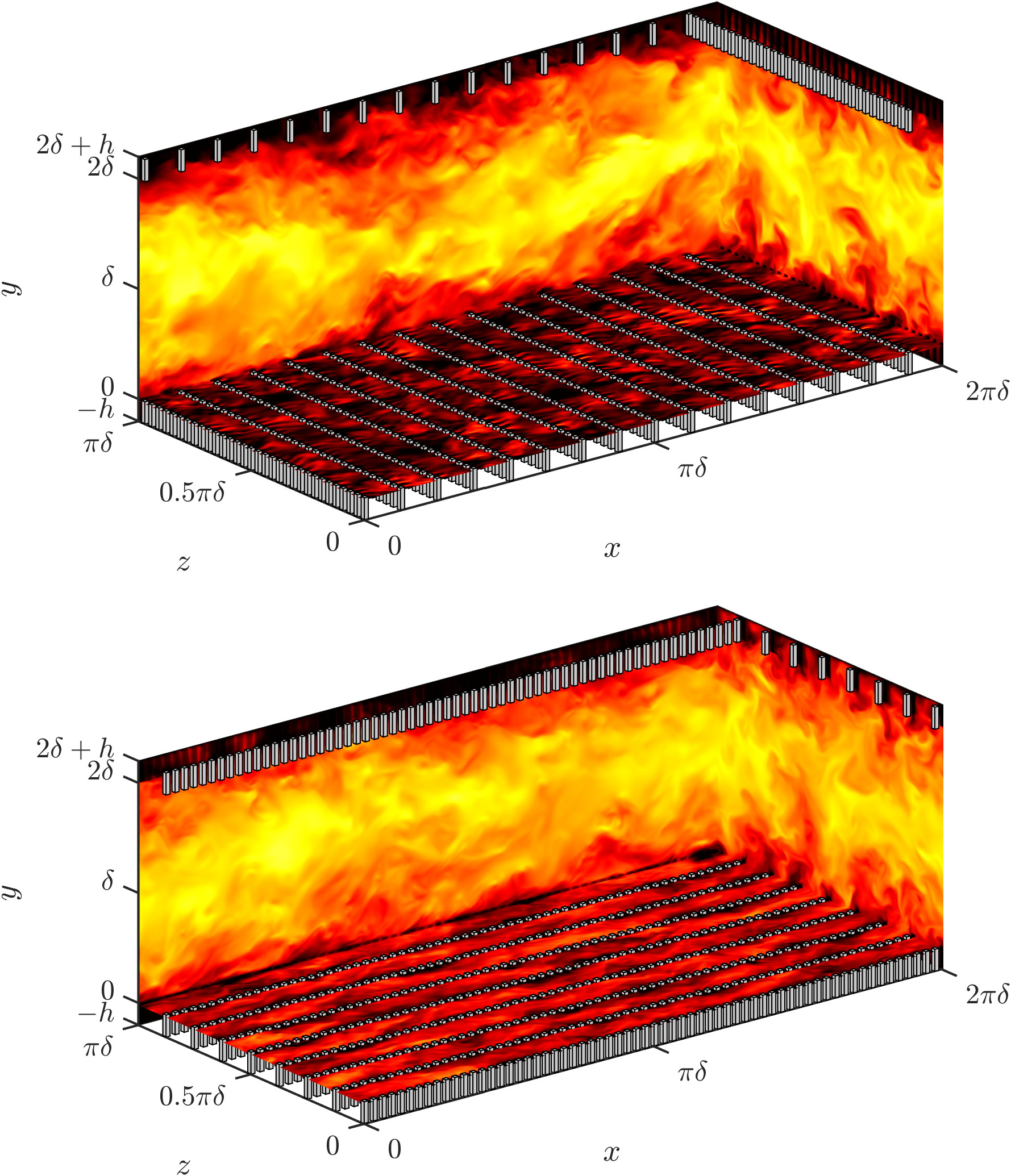}
    \put (-100mm,110mm) {(\aaa)}
    \put (-100mm,50mm) {(\bbb)}
    \caption{Schematics of the numerical channel for (\aaa) spanwise-packed canopy $\mathrm{F_{S216\times54}}$, with \rgm{streamwise spacing} $s_x^+\approx216$, \rgm{spanwise spacing} $s_z^+\approx54$ and \rgm{canopy height} $h^+\approx110$, \rgm{all in viscous units,} and (\bbb) streamwise-packed canopy $\mathrm{C_{S54\times216}}$, with $s_x^+\approx54$, $s_z^+\approx216$ and $h^+\approx110$. \rgm{The channel half-height is $\delta$.} Details of the canopy geometries are presented in table \ref{tab:canopy_param}. An instantaneous realisation of the streamwise velocity is shown in axis-orthogonal planes\rgm{, from dark to clear $u^+=0$ to $u^+=15$ in (\aaa) and $u^+=0$ to $u^+=19$ in (\bbb).}}
    \label{fig:ucontour}
\end{figure}

\subsection{Direct numerical simulations}\label{sec:dns}

We carry out DNS of \rgm{doubly periodic} closed symmetric channels with rigid filament canopies protruding from both walls. \rgm{The coordinates $x$, $y$ and $z$ represent the streamwise, wall-normal and spanwise directions, respectively.} The filament canopies and the numerical domain studied are portrayed in figure \ref{fig:ucontour}, where $\delta=1$ is the channel half-height measured from the channel centre to the canopy-tip planes. The canopy elements extend below $y=0$ on the bottom wall and above $y=2\delta$ on the top wall. \rtwo{For all simulations the channel  streamwise and spanwise dimensions are at least $L_x\times L_z=2\pi\delta\times\pi\delta$, and up to $2.8\pi\delta\times1.4\pi\delta$.} \rgm{Compared to open channels, closed channels entail roughly twice as much computational cost per time step, but they also produce twice as many statistical samples. The two setups are essentially equivalent for the near-wall flow. They can exhibit different properties in the wake region above the logarithmic layer, where turbulence activity is significantly inhibited by the no-transpiration boundary condition in open channels, while the centre of a closed channel allows eddies to flow freely, with a closer resemblance to boundary layers. For canopy flows we have reported these wake differences in \cite{chen2023examination}, where they were confined to the top $\sim$20\% of the flow thickness, and they are examined in depth across different flows, including open and closed channels, Couette flow and boundary layers, in \cite{Luchini2018,Luchini2024}.}

The DNS code in this study is from \cite{sharma2020scaling_a, sharma2020turbulent_b} and has been validated in \cite{sharma2020turbulent} and \cite{chen2023examination}. It is summarised here for reference. The numerical method resolves the three-dimensional incompressible Navier-Stokes equations,

\begin{align}
  \frac{\partial\mathbf{u}}{\partial{t}} + \mathbf{u}\cdot\nabla\mathbf{u} &= -\nabla{p} + \frac{1}{Re}\nabla^2\mathbf{u},\label{eq:ns}  \\ 
  \nabla\cdot\mathbf{u} &= 0,
\end{align}
where $\mathbf{u}=\langle u,w,v \rangle$ is the velocity vector with components in the streamwise, spanwise and wall-normal directions, respectively, $p$ is the
kinematic pressure, and $Re$ denotes the bulk Reynolds number $Re=U_b\delta/\nu$ based on the bulk velocity $U_b$, the channel half-height $\delta$, and the kinematic
viscosity $\nu$. The simulation is operated at a constant mean pressure gradient
\rgm{$-\partial_x P$}, with the flow rate adjusted to achieve the targeted friction Reynolds number \zsc{$Re_\tau=u_\tau\delta/\nu\approx180-2000$, where
$u_\tau=\sqrt{\delta \,\partial_x P }$ is the reference friction velocity at the canopy-tip plane.} 
No-slip and impermeability boundary conditions are applied at both channel walls. The rigid elements are resolved using a direct-forcing, immersed-boundary method \citep{mittal2005immersed, garcia2011hydrodynamic}. In this work, variables scaled with $u_\tau$ and $\nu$ are termed as in inner units, denoted by superscript $(\cdot)^+$, and those scaled with $U_b$ and $\delta$ are referred to as in outer units.

In the wall-normal direction, a second-order central difference scheme is applied on a staggered grid. At the channel centre, where the mean shear is the weakest, the stretched wall-normal grid has a resolution $\Delta y_{max}^+\approx3.0$ when $Re_\tau\approx180$. At higher Reynolds numbers, where dissipation occurs at larger scales \citep{jimenez2012cascades}, $\Delta y_{max}^+$ extends from 3.6 to 7.0 as $Re_\tau$ increases from 360 to 2000. The finest resolution $\Delta y_{min}^+$ is at the floor or canopy-tip plane, wherever the mean shear is the greatest. For all cases, $\Delta y_{min}^+\lesssim0.3$ is satisfied when scaled with the local, dynamically relevant friction velocity, as proposed in \cite{sharma2018turbulent, sharma2020scaling_a}.

\rgm{The discretisation in the periodic, wall-parallel directions is spectral.}
To balance computational cost \oldrev{and resolution,} the numerical domain is divided into blocks with different wall-parallel resolutions \citep{garcia2011hydrodynamic}. The wall blocks, containing the roughness sublayer, have a finer resolution than the block encompassing the channel centre. In \oldrev{the finer} blocks, the grid resolution resolves not only the turbulent scales but also the canopy geometry and \oldrev{the} \rgm{element-coherent} eddies, which are typically of the order of or smaller than the element \rgm{width} \citep{poggi2004effect}.
Consequently, the wall-parallel grid \oldrev{spacings in $x$ and $z$} are smaller than $w_x^+$ and $w_z^+$. \cite{sharma2020turbulent_b} examined wall-parallel resolutions for dense canopies, and \oldrev{concluded that a marginal resolution of 9 points per spacing and 3 points per element width} for a filament canopy with $s^+\approx24$, $w^+\approx5$ and $h^+\approx100$ could underestimate the turbulent fluctuation within the canopy, with a maximum deviation of up to 20\% observed in the wall-normal fluctuations, \oldrev{although} this discrepancy \oldrev{reduced} to 4\% just above the tips.
\oldrev{In our case, the most resolution-demanding simulations are typically those with relatively thin elements or narrow gaps, where small grid spacings are required to resolve the canopy geometry or the fluid in-between elements.
In table \ref{tab:canopy_param}, the  canopy with thinnest filaments is $\mathrm{G41_{S47\times47}}$, with $w_x^+=w_z^+\approx6$ and $s_x^+=s_z^+\approx47$, which is simulated with a resolution $\Delta x^+=\Delta z^+\approx2$. For canopy $\mathrm{IS2000_{G30\times30}}$ at $Re_\tau\approx2000$, with $w_x^+=w_z^+\approx24$ and $s_x^+=s_z^+\approx54$,} \oldrev{the resolution is also marginal at $\Delta x^+=6$ and $\Delta z^+\approx3$.}
Such marginal resolutions are applied only for the densest canopy cases, for which fluctuations within the canopy are in any event weak compared to above, and thus higher resolutions do not change the behaviour observed. For all other canopies, a minimum of 16 points per streamwise or spanwise \oldrev{pitch} are used. The height of the fine blocks is chosen such that the small and rapid \rgm{element-coherent} eddies \oldrev{decay} naturally before approaching the coarse block at the channel centre, which has a standard smooth-wall resolution of $\Delta x^+\approx8$ and $\Delta z^+\approx4$ \citep{jimenez1991minimal}. \oldrev{The height above the canopy tips of the fine-resolution blocks is set to at least twice the element width for sparse canopies and twice the gap size for dense canopies, correlating with the height of the roughness sublayer \citep{chen2023examination}.} This is validated \emph{a posteriori} by examining the spectral densities of turbulent fluctuations to ensure that any small-wavelength signal has already vanished \oldrev{well below the inter-block interface}.

\setlength{\tabcolsep}{2.7pt}
\vspace*{-1.5mm}
\begin{table}
\vspace*{-2.5mm}
\centering
\begin{tabular}{ll@{\hskip 0.5in}ccccccccc}
 &  & Case & $N_x\times N_z$ & $Re_\tau$ & $\lambda_f$ & $s_x^+$ & $g_x^+$ & $s_z^+$ & $g_z^+$ & $h^+$ \\[0.15cm]
\multirow{8}{*}{Isotropic} & \multirow{8}{*}{\parbox[c]{1em}{\includegraphics[width=0.6in]{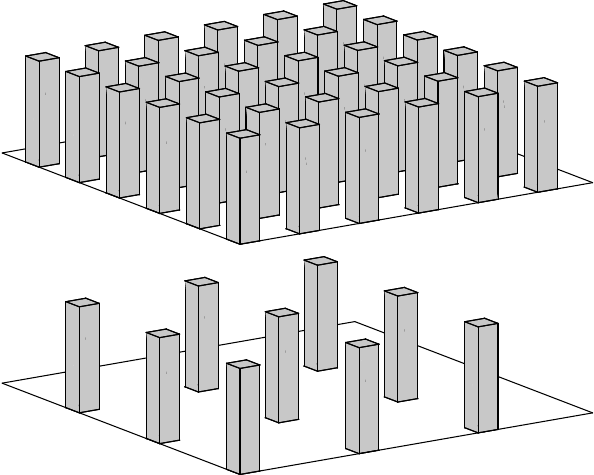}}} & $\mathrm{I_{S36\times36}}$ & 96$\times$48 & 550.2 & 2.04 & 36.0 & 12.0 & 36.0 & 12.0 & 110.0 \\
 &  & $\mathrm{I_{S54\times54}}^1$ & 64$\times$32 & 549.7 & 0.91 & 54.0 & 30.0 & 54.0 & 30.0 & 109.9 \\
 &  & $\mathrm{I_{S72\times72}}$ & 48$\times$24 & 548.2 & 0.51 & 71.8 & 47.8 & 71.8 & 47.8 & 109.6 \\
 &  & $\mathrm{I_{S108\times108}}^2$ & 32$\times$16 & 548.7 & 0.23 & 107.7 & 83.8 & 107.7 & 83.8 & 109.7 \\
 &  & $\mathrm{I_{S144\times144}}$ & 24$\times$12 & 546.9 & 0.13 & 143.2 & 119.3 & 143.2 & 119.3 & 109.4 \\
 &  & $\mathrm{I_{S216\times216}}^3$ & 16$\times$8 & 550.7 & 0.06 & 216.3 & 192.2 & 216.3 & 192.2 & 110.1 \\
 &  & $\mathrm{I_{S288\times288}}$ & 12$\times$6 & 551.1 & 0.03 & 288.5 & 264.5 & 288.5 & 264.5 & 110.2 \\
 &  & $\mathrm{I_{S432\times432}}$ & 8$\times$4 & 549.4 & 0.01 & 431.5 & 407.6 & 431.5 & 407.6 & 109.9 \\[0.2cm]
\multirow{5}{*}{Fence} & \multirow{5}{*}{\parbox[c]{1em}{\includegraphics[width=0.6in]{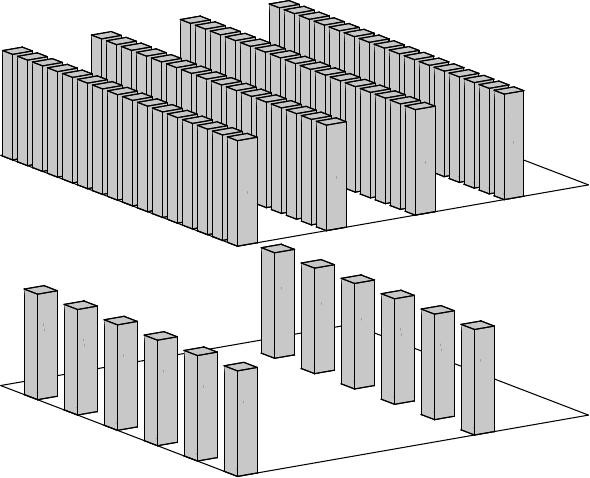}}} & $\mathrm{F_{S108\times27}}^4$ & 32$\times$64 & 548.9 & 0.91 & 107.8 & 83.8 & 26.9 & 3.0 & 109.8 \\
 &  & $\mathrm{F_{S144\times36}}$ & 24$\times$48 & 548.7 & 0.51 & 143.7 & 119.7 & 35.9 & 12.0 & 109.7 \\
 &  & $\mathrm{F_{S216\times54}}$ & 16$\times$32 & 553.4 & 0.23 & 217.3 & 193.2 & 54.3 & 30.2 & 110.7 \\
 &  & $\mathrm{F_{S288\times72}}$ & 12$\times$24 & 548.3 & 0.13 & 287.1 & 263.1 & 71.8 & 47.8 & 109.7 \\
 &  & $\mathrm{F_{S432\times108}}$ & 8$\times$16 & 548.9 & 0.06 & 431.1 & 407.1 & 107.8 & 83.8 & 109.8 \\[0.2cm]
\multirow{5}{*}{Canyon} & \multirow{5}{*}{\parbox[c]{1em}{\includegraphics[width=0.6in]{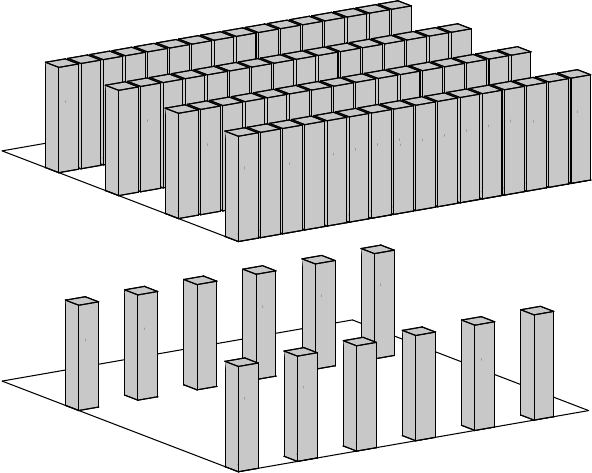}}} & $\mathrm{C_{S27\times108}}^5$ & 128$\times$16 & 552.9 & 0.91 & 27.1 & 3.0 & 108.6 & 84.4 & 110.6 \\
 &  & $\mathrm{C_{S36\times144}}$ & 96$\times$12 & 549.8 & 0.51 & 36.0 & 12.0 & 143.9 & 119.9 & 110.0 \\
 &  & $\mathrm{C_{S54\times216}}$ & 64$\times$8 & 553.6 & 0.23 & 54.3 & 30.2 & 217.4 & 193.2 & 110.7 \\
 &  & $\mathrm{C_{S72\times288}}$ & 48$\times$6 & 550.0 & 0.13 & 72.0 & 48.0 & 288.0 & 264.0 & 110.0 \\
 &  & $\mathrm{C_{S108\times432}}$ & 32$\times$4 & 548.5 & 0.06 & 107.7 & 83.8 & 430.8 & 406.9 & 109.7 \\[0.2cm]
\multirow{3}{*}{\begin{tabular}[c]{@{}l@{}}Fixed   $s_z$\\      ($s_z^+=108$)\end{tabular}} & \multirow{3}{*}{\parbox[c]{1em}{\includegraphics[width=0.6in]{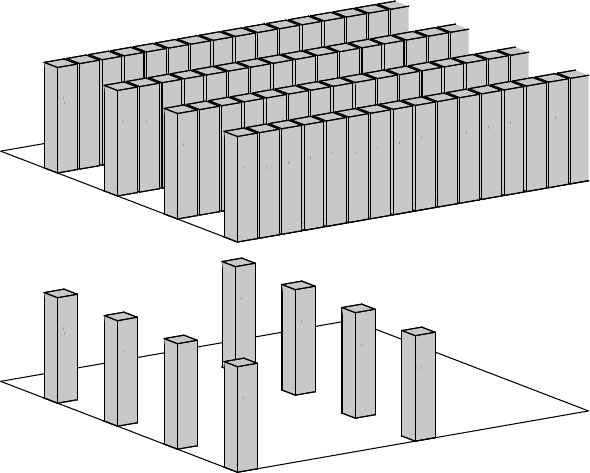}}} & $\mathrm{Z_{S27\times108}}^5$ & 128$\times$16 & 552.9 & 0.91 & 27.1 & 3.0 & 108.6 & 84.4 & 110.6 \\
 &  & $\mathrm{Z_{S108\times108}}^2$ & 32$\times$16 & 548.7 & 0.23 & 107.7 & 83.8 & 107.7 & 83.8 & 109.7 \\
 &  & $\mathrm{Z_{S216\times108}}$ & 16$\times$16 & 548.8 & 0.11 & 215.5 & 191.6 & 107.8 & 83.8 & 109.8 \\[0.2cm]
\multirow{3}{*}{\begin{tabular}[c]{@{}l@{}}Fixed   $s_x$\\      ($s_x^+=108$)\end{tabular}} & \multirow{3}{*}{\parbox[c]{1em}{\includegraphics[width=0.6in]{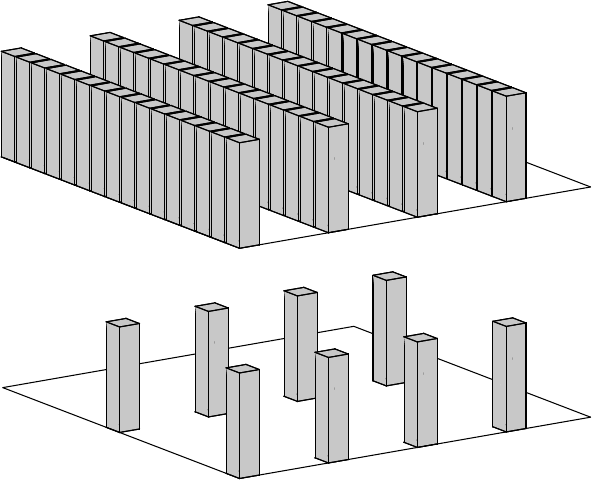}}} & $\mathrm{X_{S108\times27}}^4$ & 32$\times$64 & 548.9 & 0.91 & 107.8 & 83.8 & 26.9 & 3.0 & 109.8 \\
 &  & $\mathrm{X_{S108\times108}}^2$ & 32$\times$16 & 548.7 & 0.23 & 107.7 & 83.8 & 107.7 & 83.8 & 109.7 \\
 &  & $\mathrm{X_{S108\times216}}$ & 32$\times$8 & 548.0 & 0.11 & 107.6 & 83.7 & 215.2 & 191.3 & 109.6 \\[0.2cm]
\multirow{3}{*}{\begin{tabular}[c]{@{}l@{}}Fixed   gap\\      ($g^+\approx41$)\end{tabular}} & \multirow{3}{*}{\parbox[c]{1em}{\includegraphics[width=0.6in]{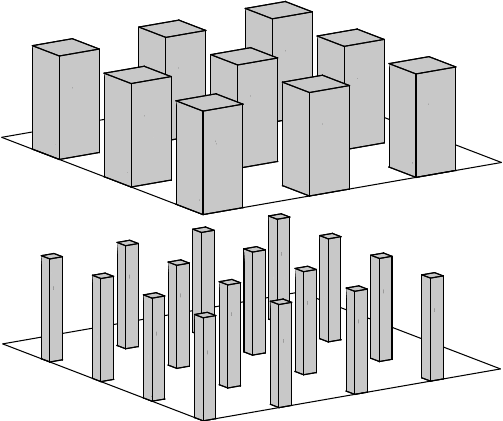}}} & $\mathrm{G41_{S66\times66}}$ & 48$\times$24 & 360.7 & 0.63 & 66.1 & 41.3 & 66.1 & 41.3 & 110.2 \\
 &  & $\mathrm{G41_{S53\times53}}$ & 48$\times$24 & 359.7 & 0.46 & 53.0 & 41.2 & 53.0 & 41.2 & 109.9 \\
 &  & $\mathrm{G41_{S47\times47}}^6$ & 48$\times$24 & 360.3 & 0.29 & 47.2 & 41.3 & 47.2 & 41.3 & 110.1 \\[0.2cm]
\multirow{4}{*}{\begin{tabular}[c]{@{}l@{}}Fixed   \\      pitch\\      ($s^+\approx47$)\end{tabular}} & \multirow{4}{*}{\parbox[c]{1em}{\includegraphics[width=0.6in]{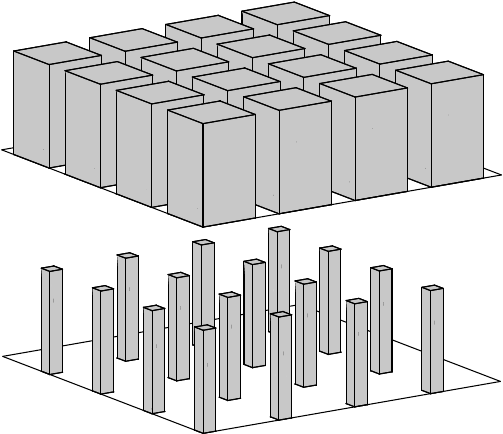}}} & $\mathrm{S47_{G14\times14}}$ & 48$\times$24 & 360.1 & 1.61 & 47.1 & 14.7 & 47.1 & 14.7 & 110.0 \\
 &  & $\mathrm{S47_{G27\times27}}$ & 48$\times$24 & 361.1 & 1.02 & 47.3 & 26.6 & 47.3 & 26.6 & 110.3 \\
 &  & $\mathrm{S47_{G35\times35}}$ & 48$\times$24 & 359.6 & 0.58 & 47.1 & 35.3 & 47.1 & 35.3 & 109.9 \\
 &  & $\mathrm{S47_{G41\times41}}^6$ & 48$\times$24 & 360.3 & 0.29 & 47.2 & 41.3 & 47.2 & 41.3 & 110.1 \\[0.2cm]
\multirow{4}{*}{\begin{tabular}[c]{@{}l@{}}Outer-\\      similar \\      geometry\end{tabular}} & \multirow{4}{*}{\parbox[c]{1em}{\includegraphics[width=0.6in]{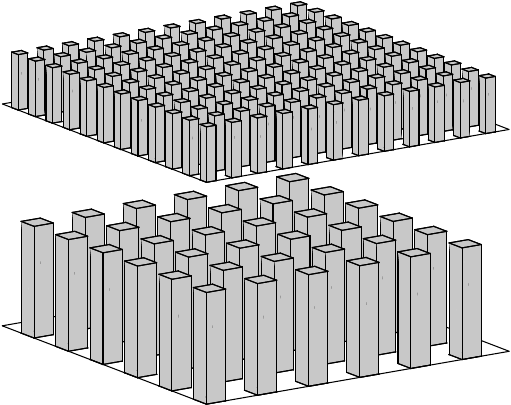}}} & $\mathrm{OS180_{G15\times15}}$ & 48$\times$24 & 179.8 & 0.70 & 23.5 & 14.7 & 23.5 & 14.7 & 44.1 \\
 &  & $\mathrm{OS360_{G30\times30}}$ & 48$\times$24 & 360.3 & 0.70 & 47.2 & 29.5 & 47.2 & 29.5 & 88.4 \\
 &  & $\mathrm{OS720_{G60\times60}}$ & 48$\times$24 & 725.5 & 0.70 & 95.0 & 59.4 & 95.0 & 59.4 & 178.0 \\
 &  & $\mathrm{OS1080_{G90\times90}}$ & 48$\times$24 & 1084.6 & 0.70 & 142.0 & 88.7 & 142.0 & 88.7 & 266.2 \\[0.2cm]
\multirow{9}{*}{\begin{tabular}[c]{@{}l@{}}Inner-\\      similar \\      geometry\end{tabular}} & \multirow{3}{*}{\parbox[c]{1em}{\includegraphics[width=0.6in]{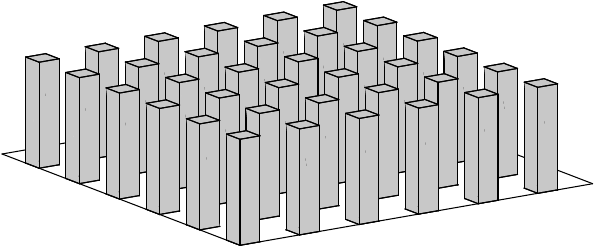}}} & $\mathrm{IS550_{G30\times30}}^1$ & 64$\times$32 & 549.7 & 0.91 & 54.0 & 30.0 & 54.0 & 30.0 & 109.9 \\
 &  & $\mathrm{IS900_{G30\times30}}$ & 96$\times$48 & 950.8 & 0.91 & 54.0 & 30.0 & 54.0 & 30.0 & 110.1 \\
 &  & $\mathrm{IS2000_{G30\times30}}$ & 192$\times$96 & 1991.9 & 0.91 & 53.8 & 29.9 & 53.8 & 29.9 & 109.6 \\
 & \multirow{3}{*}{\parbox[c]{1em}{\includegraphics[width=0.6in]{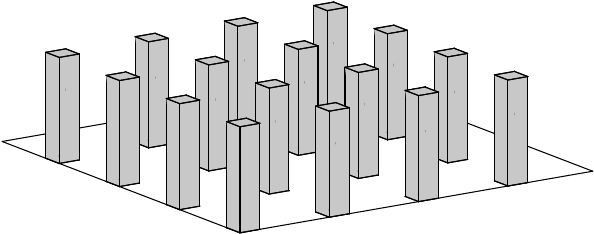}}} & $\mathrm{IS550_{G84\times84}}^2$ & 32$\times$16 & 548.7 & 0.23 & 107.7 & 83.8 & 107.7 & 83.8 & 109.7 \\
 &  & $\mathrm{IS900_{G84\times84}}$ & 48$\times$24 & 923.7 & 0.23 & 110.2 & 85.7 & 110.2 & 85.7 & 112.2 \\
 &  & $\mathrm{IS2000_{G84\times84}}$ & 96$\times$48 & 1974.0 & 0.23 & 111.8 & 87.0 & 111.8 & 87.0 & 113.9 \\
 & \multirow{3}{*}{\parbox[c]{1em}{\includegraphics[width=0.6in]{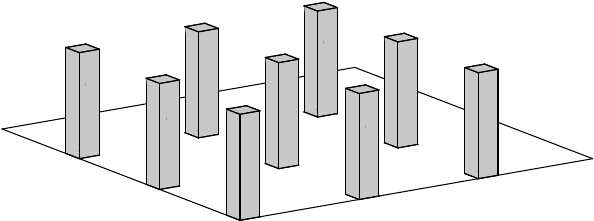}}} & $\mathrm{IS550_{G192\times192}}^3$ & 16$\times$8 & 550.7 & 0.06 & 216.3 & 192.2 & 216.3 & 192.2 & 110.1 \\
 &  & $\mathrm{IS900_{G192\times192}}$ & 32$\times$16 & 895.2 & 0.06 & 223.0 & 198.2 & 223.0 & 198.2 & 113.5 \\
 &  & $\mathrm{IS2000_{G192\times192}}$ & 48$\times$24 & 1934.2 & 0.06 & 228.8 & 203.4 & 228.8 & 203.4 & 116.4
\end{tabular}
\caption{Simulation parameters. $N_x$ and $N_z$ are the numbers of elements in the streamwise and spanwise directions, respectively; \rgm{$Re_\tau=\delta u_\tau / \nu$} is the friction Reynolds number; $\lambda_f$ is the frontal density;
$s_x$, $s_z$, $g_x$ and $g_z$ are the pitch and gap in the streamwise and spanwise directions, respectively\zsc{, and the element width is $w=s-g$}; $h$ is the canopy height.
\rgm{The~`+' superscript indicates viscous scaling, i.e. with the friction velocity $u_\tau$ and the kinematic viscosity $\nu$.} 
\rgm{Case names} with the same superscript are the same DNS.}
\label{tab:canopy_param}
\vspace*{-1mm}
\end{table}

\subsection{Simulation parameters}\label{sec:params}
We consider flows over canopies with rigid prismatic elements to thoroughly investigate turbulence penetration across various canopy layouts, element geometries and Reynolds numbers.
\rgm{We are mainly concerned with canopies fully immersed in boundary-layers, so our simulations fall in the range $\delta/h=5$–$20$. A few cases, in intermediate-to-dense
regimes with relatively small pitch, were run at $\delta/h \approx 3.3$, but for these closely packed canopies the roughness layer scales with element spacing,
rather than height \citep{chen2023examination}. Under the above conditions, an overlying turbulence, self-sustained independently of the presence of the canopy, can
properly develop, and whether it penetrates within the canopy or not can be discussed. We can then expect outer layer similarity to hold. In contrast, in canopies with $\delta/h\lesssim 2$, typical of aquatic
vegetation in shallow-submergence or emergent conditions, the whole flow is fully embedded within or directly affected by the canopy
\rtwo{\citep{Nepf2000,Zhao2024}}, and an overlying turbulence cannot be meaningfully defined. The latter are therefore out of the scope of the present work.}

Figure \ref{fig:lambda_parameter} illustrates the morphology of the canopies studied, and table \ref{tab:canopy_param} details the relevant simulation parameters.
\oldrev{We note that in the literature, the terms `spacing' and `gap' are sometimes used interchangeably, mainly in the context of canopies with thin filaments, where the element width $w$ is small compared to both $s$ and $g$, so that $s=g+w\approx g$ \citep{cui2003large, bailey2013turbulence}. Here, we consider both thin filaments ($w/s\approx0.06$) and thick and densely-packed elements ($w/s\approx0.67$). To avoid confusion, we refer to the `spacing' $s$, the sum of the gap and element width, as the `pitch' in the remainder of this work.}

All canopies investigated consist of collocated prismatic elements with a square wall-parallel cross-section, with $w_x=w_z$. To distinguish among different simulations, the prefix indicates the canopy characteristics, and the subscript denotes the approximate streamwise and spanwise spacing/pitch (`S') or gap (`G') in inner units. For example, case $\mathrm{C_{S27\times108}}$ refers to a canopy resembling a `Canyon' with streamwise-packed elements, where  $s_x^+\approx27$ and $s_z^+\approx108$.

In table \ref{tab:canopy_param}, the first five groups consist of canopies with identical element geometry, where $h^+ \approx 110$ and $w_x^+ = w_z^+ \approx 24$, but with different pitches and gaps. The first group, prefixed with `I', consists of isotropic canopies with equal streamwise and spanwise pitches $s_x^+=s_z^+$. These isotropic canopies are from \cite{chen2023examination}, which showed that turbulence effectively penetrates the sparse canopy $\mathrm{I_{S432\times432}}$ and perceives an origin at the floor, while turbulent eddies `skim' over the dense canopy $\mathrm{I_{S36\times36}}$ and perceives a smooth wall at the tips. The second and third groups consist of fence-like canopies prefixed with `F', with spanwise-packed elements, and canyon-like canopies prefixed with `C', with streamwise-packed elements, each matching the total number of posts of isotropic canopies $\mathrm{I_{S54\times54}}$ to $\mathrm{I_{S216\times216}}$ and therefore the value of $\lambda_f$. It will be demonstrated in \S\ref{sec:fencecanyon} that canopies with the same $\lambda_f$ can result in different turbulence penetration behaviours, depending on whether the elements are more closely distributed in the streamwise or spanwise direction. \zsc{\cite{cheer1987paddles} investigated flow through biological bristles, and showed that fence-like structures can behave as a `paddle' (a single solid obstacle) or a `rake' (a permeable one), depending on the gap size $g^+$.} In the fourth group, marked by `Z', we gradually reduce the number of posts in the streamwise direction $N_x$ from the canyon-like canopy $\mathrm{C_{S27\times108}}$, keeping all spanwise parameters ($N_z$, $s_z^+$ and $g_z^+$) fixed. Similarly, the fifth group, denoted by `X', consists of canopies with different $N_z$ but with the same streamwise parameters.
In \S\ref{sec:x_or_z}, with the `Z' and `X' groups, we investigate whether parameters in the streamwise or spanwise direction have more impact on flow penetration. To identify whether the gap or pitch matters, we study canopies with a fixed gap but different pitch (group `G') or with a fixed pitch but different gap (group `S') in \S\ref{sec:pitch_or_gap}. The number that follows prefixes `G' and `S' denotes the approximate value of the fixed gap and pitch in inner units, respectively. Finally, in \S\ref{sec:reynolds}, we investigate the effect of the Reynolds number using canopies with similar geometry in outer (same $s/\delta$, $g/\delta$ and $h/\delta$ in group `OS') or inner units (samex` $s^+$, $g^+$ and $h^+$ in group `IS').

\section{Measuring the penetration of turbulence into the canopy}\label{sec:penetration}

Before \oldrev{we embark in the analysis of} canopy density, \oldrev{we need to} establish how \oldrev{to measure} turbulence penetration from the overlying flow. In \S \ref{sec:structid}, we first discuss how we identify the penetrating structures that contribute to the momentum and turbulence transport in the vicinity of the canopy-tip plane. In \S \ref{sec:metric}, we introduce the metrics that we will use to quantify the penetration of these structures.

\subsection{Identification of background structures}\label{sec:structid}

\rgm{To investigate the penetration of the overlying background turbulence, we focus on structures of intense pointwise $u'v'$ \citep{lozano2012three,lozano2014time}, because of their importance in momentum transport.
The temporal evolution of these structures has been thoroughly studied by \cite{lozano2014time}; here we are only concerned with whether they penetrate into the canopy at any given instant, and the extent of their penetration, regardless of their dynamics, and as such we do not consider their preceding or subsequent evolution.}

\rgm{Just like the background turbulence can exhibit structures of intense $u'v'$, the other flow components in figure \ref{fig:poggi} can as well, including the
element-coherent flow. Its structures can merge with those from the background turbulence and, if the merger is identified as a single structure, misrepresent the 
background turbulence as penetrating more deeply into the canopy than it actually does.
}
To prevent this, 
we first filter out the element-coherent signal to isolate the background structures.

\begin{figure}
\vspace*{3mm}
    \centering
    \includegraphics[width=\textwidth]{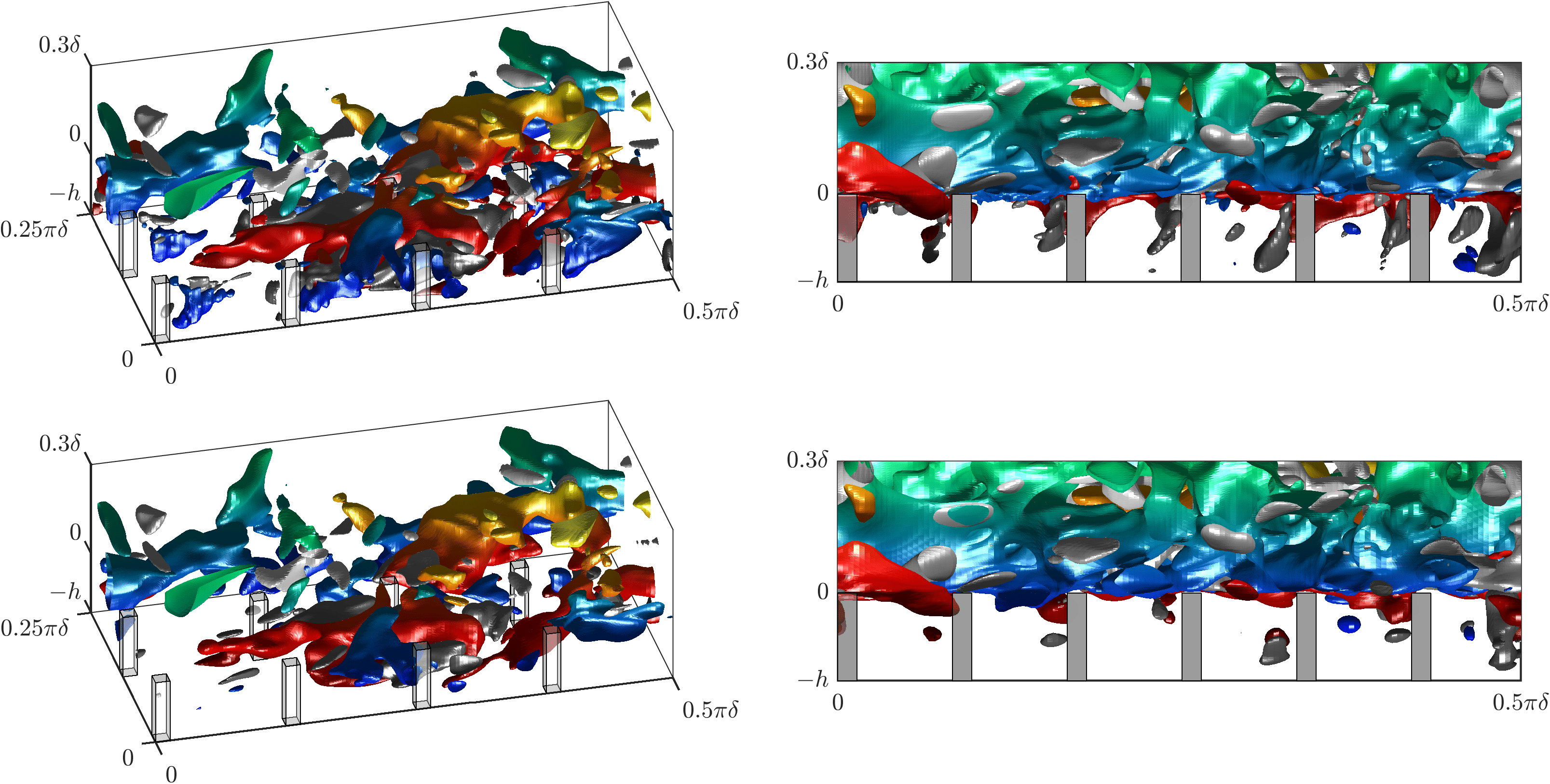}
    \put (-137mm,62mm) {(\aaa)}
    \put (-72mm ,62mm) {(\bbb)}
    \put (-137mm,27mm) {(\ccc)}
    \put (-72mm ,27mm) {(\ddd)}
    \caption{Instantaneous realisations of $u'v'$ structures for (\aaa, \ccc) case $\mathrm{I_{S216\times216}}$ and (\bbb, \ddd) case $\mathrm{F_{S144\times36}}$. Structures are coloured by distance to the floor, ejections in blue to green, sweeps in red to yellow, and outward and inward interactions in grey to white. (\aaa, \bbb), raw flow fields; (\ccc, \ddd), 
    flow fields spectrally filtered to remove the \rgm{element-coherent} flow.}
    \label{fig:uvster}
\end{figure}

As depicted in figure \ref{fig:uvster}(\aaa), \rgm{element-coherent} coherent flows typically are formed by stem-scale eddies in the wake of each obstacle. A similar pattern for stem-scale eddies is also observed in \cite{poggi2004effect} and \cite{sharma2020scaling_a} for sparse canopies. As shown in figure \ref{fig:uvster}(\bbb), \rgm{element-coherent} top-shedding eddies can also form. Note that the latter are different from the canopy-scale, mixing-layer-like eddies induced by the inflectional mean profile \citep{raupach1996coherent, nepf2012flow}.
The element-induced, element-coherent flows shown in figures \ref{fig:uvster}(\aaa, \bbb) are not part of the penetrating turbulence from the overlying background flow, and we therefore filter them out before measuring turbulence penetration.

A conventional method to obtain the background flow is through triple decomposition \citep{reynolds1972mechanics},
\begin{align}
    \mathbf{u}(x,y,z,t) &= \mathbf{U}(y)+\mathbf{u'}(x,y,z,t),\\
    \mathbf{u'}(x,y,z,t)&= \tilde{\mathbf{u}}(x,y,z)+\mathbf{u''}(x,y,z,t),
    \label{eq:tri_decomp}
\end{align}
where $\mathbf{U}$ is the mean velocity profile, and $\mathbf{u'}$ is the full temporal and spatial turbulent fluctuation.
The latter can be further decomposed into a time-averaged but spatially varying component, $\tilde{\mathbf{u}}$, and the remaining time-varying fluctuation, $\mathbf{u''}$. \rgm{The component $\tilde{\mathbf{u}}$ is often referred to as the dispersive flow \citep{raupach1982averaging} and assumed to embody the element-coherent flow.
\cite{abderrahaman2019modulation}, however, suggested that the element-coherent flow is modulated in amplitude by the overlying background turbulence for roughness elements up to $h^+\lesssim35$.}

\rgm{In any event,} as roughness size increases, wake unsteadiness intensifies, complicating the isolation of the background flow.
For the present canopies, with sizes up to $h^+\approx270$ and $w^+\approx30$, there is significant unsteady yet element-coherent flow, rendering (\ref{eq:tri_decomp}) ineffective as the time-dependent element-wake flow cannot be accurately represented by the temporally averaged term $\tilde{\mathbf{u}}$.
As a practical approximation, we thus use a spectral filter to isolate the background flow, aiming to retain most of the energy of the background-turbulent wavelengths while removing the \rgm{element-coherent} signal. The filtering procedure is detailed in appendix \ref{app:filter_sens}.

The instantaneous realisations of the filtered flow field in figures \ref{fig:uvster}(\ccc, \ddd) illustrate that both the stem-scale eddies of the sparse canopy $\mathrm{I_{S216\times216}}$ and the shedding eddies of the fence-like canopy $\mathrm{F_{S144\times36}}$ are removed \rgm{effectively by the filter}, while the background turbulent eddies remain essentially unmodified.
The applied filter slightly reduces the volume of the largest structure in figure \ref{fig:uvster}(\ccc) by approximately 5\% compared to its original size in figure \ref{fig:uvster}(\aaa), which we deem acceptable.
For case $\mathrm{F_{S144\times36}}$ in figures \ref{fig:uvster}(\bbb, \ddd), the large structures remain essentially unaffected by the spectral filter, as they do not significantly penetrate within the `fences' in any event.
\rgm{These flow visualisations suggest that the applied filtering can effectively isolate the background turbulence from the element-coherent flow.}

After filtering the element-induced, element-coherent flow, we define the background structures that contribute the most to the Reynolds shear stress as connected regions satisfying
\begin{equation}
|u'v'(x,y,z)|>H\tau_{tip},
\label{eq:qs}
\end{equation}
where $u'v'(x,y,z)$ is the pointwise Reynolds shear stress, \rgm{$H$ is a thresholding constant of order $\sim1$ and $\tau_{tip}$ is the total shear stress
evaluated at the canopy-tips plane.
The rigorous determination of a common value for $H$ across all our simulations
is discussed in appendix \ref{app:percolation}.
In the discretised space of the simulations, the connectivity is established between immediately neighbouring points along $x$, $y$, or $z$ that satisfy
(\ref{eq:qs}), to} \zsc{ensure that only physically contiguous regions of high $|u'v'|$ are identified as a single connected structure.}

\subsection{Metrics for the penetration of background-turbulence structures}\label{sec:metric}

\rgm{Once we have isolated the background turbulent structures from the full flow field, we can narrow our focus to the} structures that cross the canopy-tip plane, as they embody turbulence penetration from above.
\rgm{Of the whole background, overlying turbulence, these are the eddies that come into contact with the canopy.}
In this subsection, we introduce measures for the location, span and volume of these `tip-plane-intersecting' structures, which we will refer to as `interfacial eddies' \rgm{from here on}, aiming to characterise the extent and depth of their penetration into the canopy. 

The mean location of a structure can be characterised by the centroid position of its volume \zsc{$V$},
\begin{equation}
\mathbf{x}_{c, vol} = \frac{\int_\Omega \mathbf{x}(x,y,z) dV}{\int_\Omega dV},
\label{eq:vol_cen}
\end{equation}
and also by the centroid position of \oldrev{its $u'v'$ distribution,}
\begin{equation}
\mathbf{x}_{c, uv} = \frac{\int_\Omega u'v'(x,y,z)\,\mathbf{x}(x,y,z) dV}{\int_\Omega u'v'(x,y,z) dV},
\label{eq:uv_cen}
\end{equation}
where $\mathbf{x}$ is the pointwise position vector for all contiguous points within the volume of the structure, $\Omega$. The spatial extent of a structure can be characterised by its maximum ($\mathbf{x}_{max}$) and minimum ($\mathbf{x}_{min}$) position in $x, y$ and $z$, respectively. However, we typically observe that eddies are irregular, so the absolute \zsc{extrema} do not necessarily provide information about where the majority of the volume and $u'v'$ are located. To quantify the spatial distribution of eddies, we instead measure the position of the $n$th percentile for volume ($\mathbf{x}_{n\text{th}, vol}$),
\begin{equation}
\zsc{\frac{\int_{\mathbf{x}_{min}}^{\mathbf{x}_{n\text{th}, vol}} dV}{\int_\Omega dV}=\frac{n}{100},}
\label{eq:percentile_vol}
\end{equation}
and the position of the $n$th percentile based on $u'v'$ intensity ($\mathbf{x}_{n\text{th}, uv}$),
\begin{equation}
\frac{\int_{\mathbf{x}_{min}}^{\mathbf{x}_{n\text{th}, uv}} u'v'(x,y,z) dV}{\int_\Omega u'v'(x,y,z) dV}=\frac{n}{100}.
\label{eq:percentile_uv}
\end{equation}

\begin{figure}
\vspace*{1mm}
    \centering
    \includegraphics[width=\textwidth]{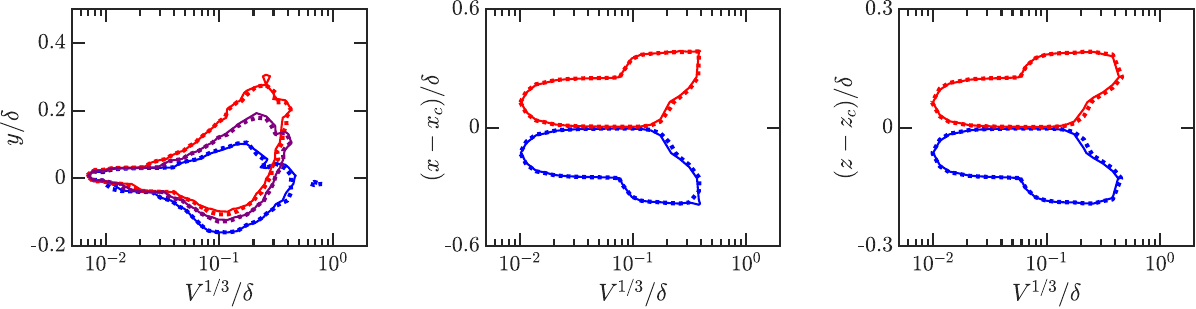}
    \put (-124mm ,29mm) {(\aaa)}
    \put (-78.5mm ,29mm) {(\bbb)}
    \put (-32mm ,29mm) {(\ccc)}
    \caption{\rgm{Probability density function of the (\aaa) $y$, (\bbb) $x$, and (\ccc)~$z$} distribution of interfacial eddies of intense $u'v'$    
    for canopy $\mathrm{I_{S216\times216}}$ \rgm{as a function of the eddy size. $V$ is the eddy volume and $\delta$ the half channel height}. The solid and dotted contour lines represent statistics for the volume and the density of $u'v'$, respectively, and enclose 95\% of the penetrating eddies;
    \rgm{The 25th, 50th (centroid) and 75th percentile are represented by the red, purple and blue lines, respectively.} In the wall-parallel directions, the percentile locations are measured relative to the position of the centroid, $x_c$-$z_c$.}
    \label{fig:den_eddy_stats}
\end{figure}

\begin{figure}
\vspace*{1mm}
    \centering
    \includegraphics[width=\textwidth]{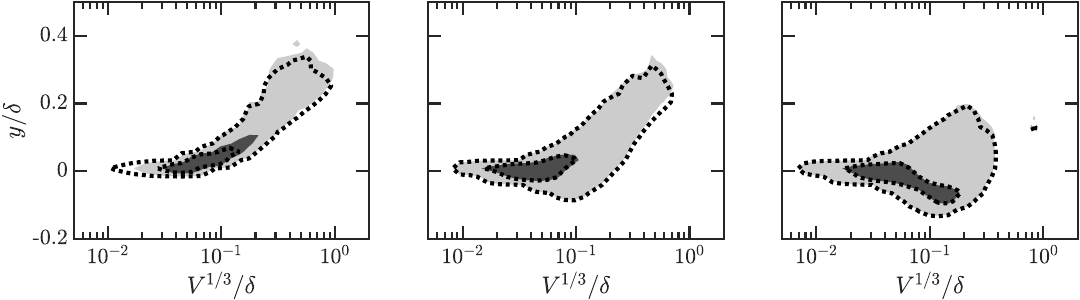}
    \put (-136mm ,36mm) {(\aaa)}
    \put (-87mm ,36mm) {(\bbb)}
    \put (-43mm ,36mm) {(\ccc)}
    \caption{Probability density function of the centroid \rgm{height $y$} for interfacial eddies \rgm{of intense $u'v'$ as a function of eddy size. $V$ is the eddy volume and $\delta$ the half channel height. Shaded contours are for the volume centroid, equation (\ref{eq:percentile_vol}), and dotted ones for the $u'v'$ centroid, equation (\ref{eq:percentile_uv}).} (\aaa), canopy $\mathrm{I_{S36\times36}}$ (dense); (\bbb), $\mathrm{I_{S72\times72}}$ (intermediate); (\ccc), $\mathrm{I_{S432\times432}}$ (sparse). The contours enclose 50\% and 95\% of the eddies.}
    \label{fig:ycentroid}
\end{figure}

An eddy with $y_{min}=0.1\delta$, $y_{5\text{th},uv}=0.3\delta$, $y_{95\text{th},uv}=0.5\delta$ and $y_{max}=0.6\delta$, for instance, would have an absolute extent in $y$ from $y=0.1\delta$ to $y=0.6\delta$, but $90\%$ of its total $u'v'$ stress \oldrev{would be} concentrated between $y=0.3\delta$ and $y=0.5\delta$, with only 5\% above and below. We denote the span of this $u'v'$-concentrated region as \oldrev{$y_{90, uv}^+=y_{95\text{th},uv}^+-y_{5\text{th},uv}^+=0.2\delta$.}
As an example, figure \ref{fig:den_eddy_stats} portrays the typical location of the centroid and the 25th and 75th percentiles for the penetrating eddies for a sparse canopy. \rgm{Let us define a generic charateristic lengthscale for eddies based on their volume as $V^{1/3}$.}
As shown in panel (\aaa), of the \oldrev{interfacial eddies, small ones} of size $V^{1/3}\lesssim0.03\,\delta$ tend to remain near the tips, as they cannot extend too much above or below this region due to their small \oldrev{size.
Meanwhile, the larger eddies,} of size $V^{1/3}\gtrsim0.2\,\delta$, can effectively penetrate into the canopy, as well as extend well above the tips.
\rgm{Along $x$, panel (\bbb) indicates that eddies extend relative to their centroid with upstream-downstream symmetry, with
$x_{50, vol}^+\approx x_{50, uv}^+\approx0.8\delta^+\approx440$; along $z$, panel (\ccc) indicates that they also extend with left-right symmetry, with
$z_{50, vol}^+\approx z_{50, uv}^+\approx0.4\delta^+\approx220$.
Figure \ref{fig:den_eddy_stats}, as well as figure \ref{fig:ycentroid},} also shows that the eddy-size statistics based on volume and $u'v'$ are very similar, indicating that either one can be used to characterise the location and topology of the \oldrev{interfacial eddies.} For simplicity, we present results based on volume only in the remainder of the paper, \rgm{and label e.g. $z_{50, vol}^+$ simply as $z_{50}^+$}.

The wall-normal distribution of the \oldrev{interfacial} eddies varies significantly with canopy density, providing useful metrics to characterise \oldrev{the density regime.} As shown in figure \ref{fig:ycentroid}(\aaa), a dense canopy like $\mathrm{I_{S36\times S36}}$ precludes \oldrev{the} overlying eddies from penetrating into it. In this case, small structures reside near the tips, while the larger ones reside essentially above the tips, with fewer than 1\% of the \oldrev{interfacial} eddies having centroids below the tips. In contrast, a sparse canopy like $\mathrm{I_{S432\times S432}}$ allows vigorous penetration of turbulent eddies into the canopy, with approximately 60\% of the \oldrev{interfacial} eddies having centroids below the tips, as depicted in figure \ref{fig:ycentroid}(\ccc). An intermediate canopy like $\mathrm{I_{S72\times S72}}$ in figure \ref{fig:ycentroid}(\bbb) lies between the two limits, with about 30\% of the \oldrev{interfacial} eddies having centroids below the tips.

\begin{figure}
\vspace*{1mm}
    \centering
    \includegraphics[width=\textwidth]{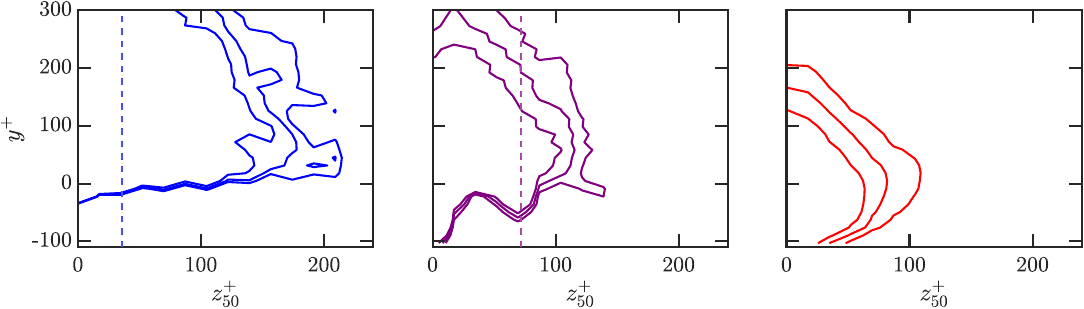}
    \put (-136mm ,36mm) {(\aaa)}
    \put (-87mm ,36mm) {(\bbb)}
    \put (-43mm ,36mm) {(\ccc)}
    \caption{Probability density function of the characteristic spanwise width \rgm{$z^+_{50}$ at each height $y^+$} of interfacial eddies for canopies (\aaa) $\mathrm{I_{S36\times36}}$ (dense); (\bbb) $\mathrm{I_{S72\times72}}$ (intermediate); and (\ccc) $\mathrm{I_{S432\times432}}$ (sparse). The contours enclose 70\%, 75\% and 80\% eddies. The dashed \rgm{vertical} lines mark the corresponding element pitch $s_z^+\approx36$, $72$, and $432$,
    with the latter is beyond the abscissa range displayed.}
    \label{fig:den_zy_example}
\end{figure}

\begin{figure}
\vspace*{1mm}
    \centering
    \includegraphics[width=0.4\textwidth]{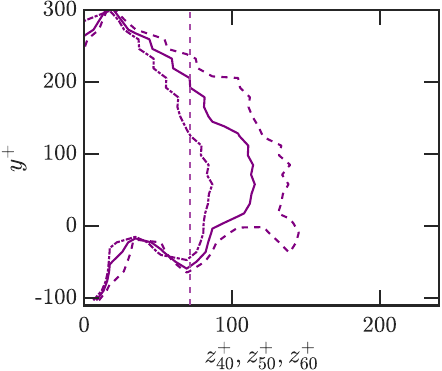}
    \caption{Probability density function of the spanwise width \rgm{at each height $y^+$} of interfacial eddies for canopy $\mathrm{I_{S72\times72}}$. The dash-dotted, solid and dashed contours represent \rgm{widths} $z_{40}^+$, $z_{50}^+$ and $z_{60}^+$, respectively; the vertical dashed line mark the element pitch $s_z^+\approx72$. The contours enclose 75\% of the eddies.}
     \label{fig:den_zy_example_validate}
\end{figure}

To further investigate the wall-normal distribution of eddy volume, we examine the streamwise and spanwise widths \oldrev{at each height of the interfacial eddies.} 
\rgm{For this, we use superficial averages at each height $y$ in (\ref{eq:percentile_vol}).}
Figure \ref{fig:den_zy_example} illustrates the typical spanwise width of eddies $z_{50}^+$ for dense to sparse canopies, $\mathrm{I_{S36\times S36}}$, $\mathrm{I_{S72\times S72}}$ and $\mathrm{I_{S432\times S432}}$.
\rgm{For the dense canopy, panel (\aaa) shows that the interfacial eddies generally have a large span compared to the element pitch, up to $z_{50}^+\approx180$, and few of these eddies can penetrate.
For the intermediate canopy, panel (\bbb) shows that penetration is partially possible for eddies with widths comparable to the canopy pitch, $z_{50}^+\approx s_z^+\approx72$, with narrower eddies with $z_{50}^+\ll s_z^+$ penetrating even deeper and reaching the floor --}
the influence of different canopy parameters, including $s_z^+$, on turbulence penetration will be further discussed in \S \ref{sec:results}.
\rgm{Finally, panel (\ccc) shows that, for the sparse canopy}, turbulence penetrates relatively unhindered, easily reaching the floor, and even deep within the canopy the characteristic eddy width remains comparable to the width at the tips. In figures \ref{fig:den_zy_example}(\aaa-\ccc), lower contour levels tend to capture the large-eddy tails within the probability distribution function, especially above the dense canopy and within the sparse canopy. In these regions with minimal obstruction, where the eddies are not constrained by the presence of canopy elements, turbulence spans a broader range of eddy sizes. Consequently, including a larger portion of eddies\oldrev{, by} using a lower contour \oldrev{level, increases} the contribution of the large but infrequent eddies.
In any event, as long as the $n$th percentile captures the majority of the volume for each eddy, the characteristic width within the canopy is essentially insensitive to the choice of percentile.
\zsc{For illustration, figure \ref{fig:den_zy_example_validate} shows that spanwise widths $z_{40}^+$, $z_{50}^+$ and $z_{60}^+$ are similar within the canopy, as eddy shapes in this region are constrained by the element geometry. Only farther above the canopy, where eddies broaden and are no longer constrained, these percentile-based widths begin to differ.} For simplicity, the discussion in \S \ref{sec:results} presents the characteristic width of eddies using $x_{50}^+$ and $z_{50}^+$, with a contour level that encloses 75\% of the eddies within the probability density function.

\begin{figure}
\vspace*{1mm}
    \centering
    \includegraphics[width=\textwidth]{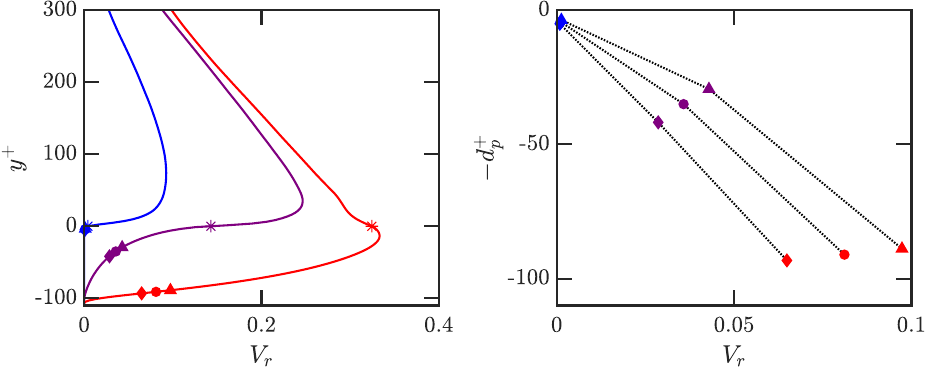}
    \put (-136mm,53mm) {(\aaa)}
    \put (-66mm ,53mm){(\bbb)}
    \caption{(\aaa) Relative volume $V_r$ occupied by interfacial eddies at each height \rgm{$y^+$}. \zsc{Blue to red lines are canopies $\mathrm{I_{S36\times36}}$ (dense), $\mathrm{I_{S72\times72}}$ (intermediate), and $\mathrm{I_{S432\times432}}$ (sparse), as in figure \ref{fig:den_zy_example}.} Symbols $*$, $\blacktriangle$, $\bullet$ and $\blacklozenge$ indicate the height where $V_r/V_{r,tip}=1.0$, $0.3$, $0.25$, and $0.2$, respectively. \rgm{The latter three are alternative choices for the definition of the penetration depth $d_p$, with $y=-d_p$, and are also portrayed in (\bbb), where} the dashed lines connect markers with the same value of $V_r/V_{r,tip}$ across different canopies.}
    \label{fig:den_vr_example}
\end{figure}

In addition to the typical location and size of structures, \oldrev{we are interested in quantifying penetration depth. We do so} based on the relative planar volume $V_r$, \oldrev{defined as the fraction of the domain at each height occupied by the interfacial eddies \citep{del2006self}.}
For a sparse canopy like $\mathrm{I_{S432\times S432}}$, where turbulence can freely penetrate within, figure \ref{fig:den_vr_example}(\aaa) shows that $V_r$ decays \oldrev{slowly} with depth into the canopy, remaining comparable to its footprint at the tips, $V_{r,tip}$, even deep within the canopy. For an intermediate canopy such as $\mathrm{I_{S72\times S72}}$, where turbulence penetration is partially obstructed, $V_r$ decays more rapidly.
\oldrev{For a dense canopy like $\mathrm{I_{S36\times S36}}$, eddies are essentially precluded from entering the canopy, and $V_r$ vanishes immediately below the tips}.
Based on these observations, we define the `penetration depth' \zsc{$d_p^+$} as the depth \zsc{below the tips} at which $V_r$ becomes small relative to its footprint at the tips.
The sensitivity analysis in figure \ref{fig:den_vr_example}(\bbb) demonstrates that for $V_r/V_{r,tip}=0.2$-$0.3$, \zsc{$d_p^+$} is essentially insensitive to the chosen threshold.
Dense canopies consistently result in minimal penetration, with \zsc{$d_p^+\approx0$,} while sparse canopies have penetration depths comparable to their height, \zsc{$d_p^+\approx h^+$.}
\oldrev{We thus use the threshold $V_r/V_{r,tip}=0.25$ to determine \zsc{$d_p^+$} from here on.} This single-point indicator quantifies the depth of turbulence penetration and enables \oldrev{a simple} comparative analysis across different canopies.

\section{Results and discussion}\label{sec:results}

With the metrics to quantify the extent and depth of turbulence penetration proposed above, we now apply them to our DNSs, examining the influence of one canopy parameter at a time.
In \S \ref{sec:fencecanyon}, we first explore the effect of canopy layout by making the elements closely packed in either the streamwise or spanwise direction while keeping frontal density $\lambda_f$ constant. This allows us to demonstrate that the same value of $\lambda_f$ does not necessarily correspond to the same density regime, as advanced in \S \ref{sec:intro}.
In \S \ref{sec:x_or_z}, we vary the pitch between elements in either the spanwise or streamwise direction to identify in which direction variations have a greater influence on turbulence penetration.
In \S \ref{sec:pitch_or_gap}, we vary either the element pitch or gap for isotropic canopies \oldrev{to determine which of the two has the most influence.} In \S \ref{sec:reynolds}, we investigate the same canopy at different Reynolds numbers to determine the length scale governing turbulence penetration. Finally, in \S 
\ref{sec:definedense}, we put together all the preceding results to propose a single parameter to encapsulate canopy density.

\subsection{Different density regimes for the same frontal density $\lambda_f$}\label{sec:fencecanyon}
In this subsection, we demonstrate that frontal density $\lambda_f$ alone cannot fully characterise the density for canopies with an anisotropic layout.
Figure \ref{fig:uvster_can_fen} illustrated that for canopies with the same number of elements per unit area and identical $\lambda_f$, the arrangement of elements can significantly influence how much blockage they produce on the eddies from the overlying turbulence, affecting the degree of their penetration. This difference in penetration can now be quantified with the metrics proposed in \S \ref{sec:penetration}.

\begin{figure}
\vspace*{3mm}
    \centering
    \includegraphics[width=\textwidth]{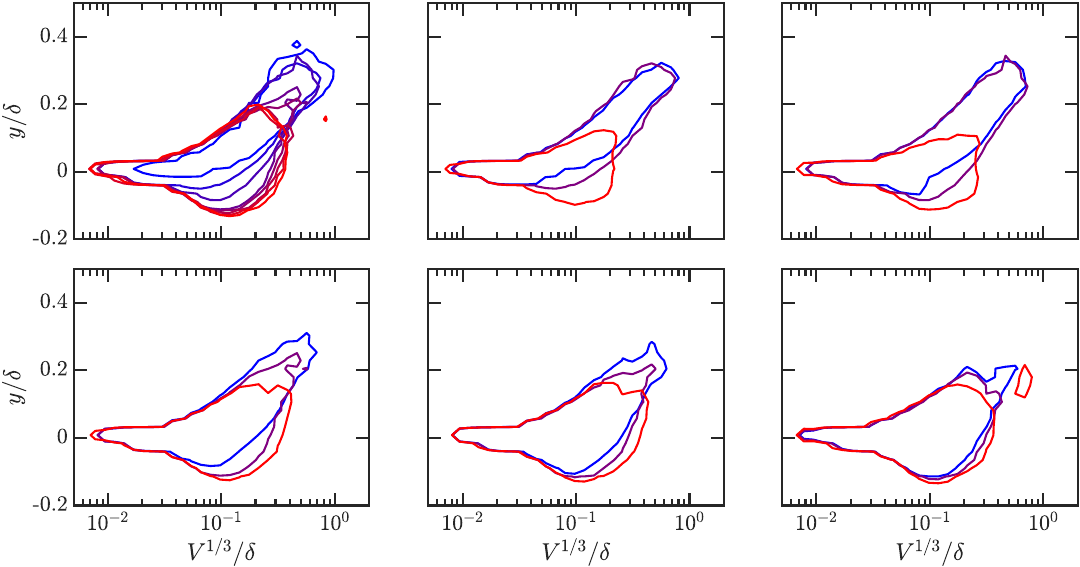}
    \put (-136mm ,69mm) {(\aaa)}
    \put (-87mm ,69mm) {(\bbb)}
    \put (-43mm ,69mm) {(\ccc)}
    \put (-136mm ,36mm) {(\ddd)}
    \put (-87mm ,36mm) {(\eee)}
    \put (-43mm ,36mm) {(\fff)}
    \put(-99mm,44mm){\vector(-1,1){8mm}}
    \put(-97mm,44mm){$\lambda_f$}
    \put(-79mm,65mm){$\lambda_f=0.91$}
    \put(-35mm,65mm){$\lambda_f=0.51$}
    \put(-124mm,32mm){$\lambda_f=0.23$}
    \put(-78mm,32mm){$\lambda_f=0.13$}
    \put(-35mm,32mm){$\lambda_f=0.06$}
    \caption{Probability density function of the \rgm{centroid height $y$} of interfacial eddies \rgm{as a function of eddy size. $V$ is the eddy volume and $\delta$ the half channel height}. (\aaa),~isotropic-layout canopies with increasing frontal density $\lambda_f\approx0.01$-$2.04$ from red to blue\zsc{, increasing in the direction of the arrow}.
    (\bbb)~through (\fff), decreasing frontal density $\lambda_f\approx0.91$-$0.06$;
     blue, magenta and red indicate fence-like, isotropic and canyon-like canopies, respectively, with the same $\lambda_f$. The contours enclose 75\% of the eddies.}
    \label{fig:den_eddy_stats_lamf}
\end{figure}

Figure \ref{fig:den_eddy_stats_lamf} depicts the typical location and size of the penetrating eddies for both isotropic and anisotropic canopies. As shown in figure \ref{fig:den_eddy_stats_lamf}(\aaa), canopy $\mathrm{I_{S36\times36}}$, with the highest density value of $\lambda_f\approx2.04$, behaves as dense, as less than $1\%$ of the eddies have centroids below the canopy-tip plane.
As the elements are spaced further apart, canopies $\mathrm{I_{S54\times54}}$ and $\mathrm{I_{S72\times72}}$ allow eddies to penetrate moderately into the canopy. For these two intermediate canopies, eddies of size $V^{1/3}\gtrsim0.05\,\delta$ can extend well above and below the tips, and more than 20\% eddies have centroids below the tips. 
For canopies $\mathrm{I_{S108\times108}}$ to $\mathrm{I_{S432\times432}}$, with pitch $s^+>100$ and $\lambda_f\approx0.23$-$0.01$, turbulence penetrates with little obstruction, with more than 60\% of the eddies having centroids below the tips.
\cite{sharma2020scaling_a} performed DNSs for isotropic layouts of filaments, and argued that canopies were sparse if the pitch between elements was $s^+>100$, which allowed sufficient space for near-wall streaks to fit within.
However, while $\lambda_f$ captures the overall trends of turbulence penetration in dense and intermediate canopies, it fails to distinguish the sparser canopies $\mathrm{I_{S108\times108}}$ to $\mathrm{I_{S432\times432}}$ in figure \ref{fig:den_eddy_stats_lamf}(\aaa), which essentially have similar distributions of centroids.
More importantly, figures \ref{fig:den_eddy_stats_lamf}(\bbb-\fff) illustrate that the same value of $\lambda_f$ does not necessarily result in the same density regime. For instance, canopies with $\lambda_f\approx0.91$ would typically be classified as dense based on $\lambda_f$ alone \citep{nepf2012flow, brunet2020turbulent}.
However, as shown in figure \ref{fig:den_eddy_stats_lamf}(\bbb), although the isotropic and the fence-like canopies $\mathrm{I_{S54\times54}}$ and $\mathrm{F_{S108\times27}}$ behave as dense, the canyon-like canopy $\mathrm{C_{S27\times108}}$, with the same $\lambda_f$, behaves as sparse, \oldrev{with} approximately 65\% of the \oldrev{interfacial eddies having} centroids below the tips. At lower densities, the isotropic and fence-like canopies allow eddies to penetrate more effectively, see figures \ref{fig:den_eddy_stats_lamf}(\ccc-\eee) , while canyon-like canopies consistently result in the most significant turbulent penetration for a given $\lambda_f$. Eventually, for low enough frontal densities $\lambda_f\lesssim0.06$, turbulence can enter freely into the canopies regardless of the \oldrev{canyon, fence or isotropic} layout.

\begin{figure}
\vspace*{1mm}
    \centering
    \includegraphics[width=\textwidth]{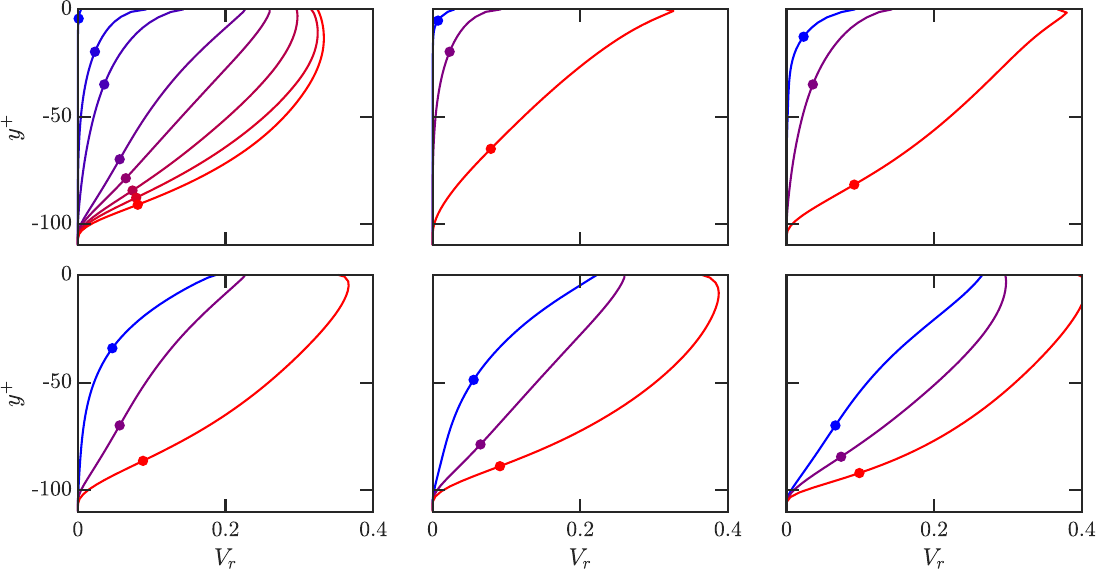}
    \put (-136mm ,69mm) {(\aaa)}
    \put (-87mm ,69mm) {(\bbb)}
    \put (-43mm ,69mm) {(\ccc)}
    \put (-136mm ,36mm) {(\ddd)}
    \put (-87mm ,36mm) {(\eee)}
    \put (-43mm ,36mm) {(\fff)}
    \put(-105mm,49mm){\vector(-1,1){12mm}}
    \put(-103mm,48mm){$\lambda_f$}
    \put(-60mm,42mm){$\lambda_f=0.91$}
    \put(-16mm,42mm){$\lambda_f=0.51$}
    \put(-104mm,10mm){$\lambda_f=0.23$}
    \put(-60mm,10mm){$\lambda_f=0.13$}
    \put(-16mm,10mm){$\lambda_f=0.06$}
    \caption{Relative volume $V_r$ occupied by interfacial eddies at each height \rgm{$y^+$}. Markers $\bullet$ indicate the penetration depth \rgm{$y^+=-d_p^+$} where $V_r/V_{r,tip}=0.25$. The canopies portrayed in each panel and the colour scheme are the same as in the respective panels in figure \ref{fig:den_eddy_stats_lamf}:
    \rgm{(\aaa),~isotropic-layout canopies with increasing frontal density $\lambda_f\approx0.01$-$2.04$ from red to blue, increasing in the direction of the arrow.
    (\bbb)~through (\fff), decreasing frontal density $\lambda_f\approx0.91$-$0.06$;
     blue, magenta and red indicate fence-like, isotropic and canyon-like canopies, respectively, with the same $\lambda_f$.}}
    \label{fig:den_vr_lamf}
\end{figure}

In addition to examining the distribution of centroids, we also analyse the penetration depth \zsc{$d_p^+$} of eddies \oldrev{below the canopy-tip plane}, as portrayed in \oldrev{figure} \ref{fig:den_vr_lamf}.
In agreement with figure \ref{fig:den_eddy_stats_lamf}(\aaa), figure \ref{fig:den_vr_lamf}(\aaa) illustrates that turbulence only penetrates moderately into the dense to intermediate canopies $\mathrm{I_{S36\times36}}$ to $\mathrm{I_{S72\times72}}$, with \zsc{$d_p^+\gtrsim35$.} In contrast, turbulence penetrates significantly deeper into the sparser canopies $\mathrm{I_{S108\times108}}$ to $\mathrm{I_{S432\times432}}$, where \zsc{$d_p^+\approx70$-$91$.} 
Figures \ref{fig:den_vr_lamf}(\bbb-\fff) show that the influence of element layout is more noticeable for canopies with a higher value of frontal density $\lambda_f\gtrsim0.51$, where the isotropic and the fence-like layouts induce more blockage on the overlying eddies compared to the canyon-like layout. 
However, once the frontal density is small, $\lambda_f\lesssim0.06$, the canopy behaves as sparse regardless of its layout anisotropy, suggesting that $\lambda_f$ provides 
only a notional measure of density for denser canopies.
As depicted in figures \ref{fig:den_vr_lamf}(\bbb, \ccc), the isotropic and the fence-like canopies essentially preclude the eddies from entering the canopy, whereas the canyon-like canopy allows effective turbulence penetration, with \zsc{$d_p^+\approx65$} even at $\lambda_f\approx0.91$.
As $\lambda_f$ decreases, the isotropic and the fence-like canopies in figures \ref{fig:den_vr_lamf}(\ddd, \eee) allow turbulence to penetrate within, but they consistently behave as denser compared to the corresponding canyon-like canopies. In the sparse limit, $\lambda_f\lesssim0.06$, turbulence penetrates freely into the canopy, regardless of the layout. We also note that the canyon-like canopies in figures \ref{fig:den_vr_lamf}(\bbb-\fff) appear as sparse for all $\lambda_f$ considered, with \zsc{$d_p^+\gtrsim65$-$92$.}
The observations in figures \ref{fig:den_eddy_stats_lamf} and \ref{fig:den_vr_lamf} suggest that the depth of turbulence penetration has a dependence on whether the canopy elements are closely packed in the streamwise or spanwise direction, with the former behaving as sparser.
In \S \ref{sec:x_or_z} and \S \ref{sec:pitch_or_gap}, we will further examine the effects of element pitch and gap in different directions.

\subsection{Influence of streamwise and spanwise pitch}\label{sec:x_or_z}

We have shown that, for a fixed $\lambda_f$, the location of interfacial eddies and their penetration depth depend on whether the canopy elements are more closely packed in the streamwise or spanwise direction.
To further investigate the effect of the streamwise and spanwise arrangement, we now vary the element pitch \rgm{i.e. spacing} separately in the streamwise or spanwise direction while keeping the element geometry fixed. This allows us to examine the effect of $s_x^+$ and $s_z^+$ independently, and to better understand how canopy anisotropy affects turbulence penetration. \rgm{We thus focus on the results for canopies
$\mathrm{Z_{S27\times108}}$, $\mathrm{Z_{S108\times108}}$ and $\mathrm{Z_{S216\times108}}$ on one hand, and $\mathrm{X_{S108\times27}}$, $\mathrm{X_{S108\times108}}$ and $\mathrm{X_{S108\times216}}$ on the other.}

\begin{figure}
\vspace*{2mm}
    \centering
    \includegraphics[width=\textwidth]{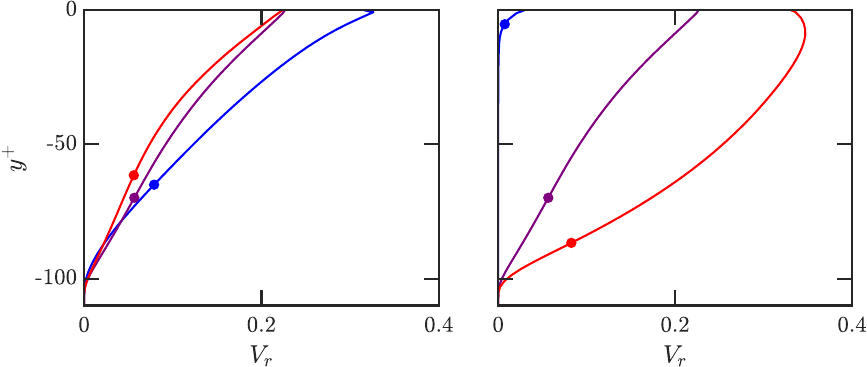}
    \put (-135mm,55mm) {(\aaa)}
    \put (-64mm ,55mm){(\bbb)}
    \caption{Relative volume $V_r$ occupied at each height \rgm{$y^+$} by interfacial eddies: canopies with (\aaa) fixed $s_z^+\approx108$ and (\bbb) fixed $s_x^+\approx108$.
    \rgm{Blue, purple and red lines are for canopies with frontal densities $\lambda_f\approx0.91$, $0.23$ and $0.11$, respectively.}
    Markers $\bullet$ indicate the penetration depth \rgm{$y^+=-d_p^+$} where $V_r/V_{r,tip}=0.25$.
    }
    \label{fig:den_vr_xz}
\end{figure}

\rgm{Detailed, quadrant-wise results for the penetration of \rtwo{$u'v'$} eddies for these canopies are presented and discussed in appendix \ref{app:quadrants}.
The canopies with a fixed spanwise pitch, $s_z^+\approx100$, allow turbulence to penetrate effectively within, with no significant influence of the value of $s_x^+$. 
In contrast, canopies with a fixed streamwise pitch, $s_x^+\approx100$, exhibit different penetration behaviours depending on their spanwise pitch.
The canopy with $s_z^+\approx25$ significantly restricts turbulence penetration, while the canopies with $s_z^+\approx100$ and $s_z^+\approx200$ allow turbulence
to penetrate increasingly more. This is also shown in
figure \ref{fig:den_vr_xz}(\aaa), which portrays eddy volume reduction and penetration depth into the canopy for the above cases.} All the canopies with \rgm{common} 
spanwise pitch $s_z^+\approx100$ behave as sparse, regardless of their frontal density $\lambda_f\approx0.11$-$0.91$ \rgm{and streamwise pitch $s_x^+\approx25$-$200$},
allowing eddies to effectively penetrate into the canopy with a penetration depth \zsc{$d_p^+\approx65$, as portrayed in figure \ref{fig:den_vr_xz}(\aaa).}
In contrast, figure \ref{fig:den_vr_xz}(\bbb) highlights that when the pitch in $x$ is fixed to $s_x^+\approx108$, the canopies behave \oldrev{as} increasingly sparser as the spanwise pitch increases from $s_z^+\approx25$ to $s_z^+\approx\rgm{200}$.
This transition in density regime is marked by a substantial increase in penetration depth, from \zsc{$d_p^+\approx5$} for the \oldrev{dense} canopy $\mathrm{X_{S108\times27}}$ to \zsc{$d_p^+\approx87$} for the \oldrev{sparse} canopy $\mathrm{X_{S108\times216}}$.

The correlation between turbulence penetration and spanwise pitch is \oldrev{not surprising; as} turbulence is advected downstream, eddies \oldrev{will} be obstructed by the \oldrev{elements they encounter. In turn,} when the elements are spaced further apart in the spanwise direction, \oldrev{once eddies have penetrated within the canopy they will travel through the canyons without encountering any significant obstruction.} \rgm{We note that this argument is based on turbulent eddies being advected downstream, essentially travelling with the local mean flow, rather than on their length or length-to-span aspect ratio. Eddies penetrating significantly within the canopy need pathways through which they can travel relatively unhindered, rather than gaps where they can fit, as if motionlessly, in between canopy elements.}

\subsection{Influence of element pitch versus gap width}\label{sec:pitch_or_gap}

We have shown that as the canopy elements are spaced further apart in the spanwise direction, the canopy transitions to a sparser regime, whereas variations in the streamwise direction have little impact on the density regime. However, it remains to be determined whether this difference in turbulence penetration is due to the increase in gap\rgm{, $g_z^+$,} or in pitch \rgm{i.e. spacing, $s_z^+$}. We address this question by examining canopies with either a fixed inter-element gap and varying element pitch or a fixed pitch and varying gap. 

\begin{figure}
\vspace*{2mm}
    \centering
    \includegraphics[width=\textwidth]{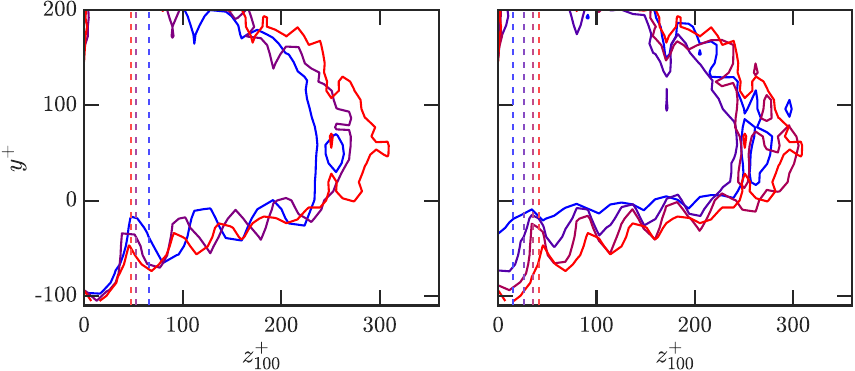}
    \put (-135mm,55mm) {(\aaa)}
    \put (-64mm ,55mm){(\bbb)}
    \caption{Probability density function of the total spanwise width of interfacial eddies, $z_{100}^+=z_{max}^+-z_{min}^+$, \rgm{at each height $y^+$} for canopies with (\aaa) fixed gap $g_z^+\approx41$ and (\bbb) fixed pitch $s_z^+\approx47$. The contours from blue to red represent canopies with decreasing frontal densities, $\lambda_f\approx0.63$, $0.46$, $0.29$ in (\aaa), and $\lambda_f\approx1.61$, $1.02$, $0.58$, $0.29$ in (\bbb); the dashed lines mark the corresponding pitch $s_z^+\approx66$, $53$, $47$ in (\aaa) and gap $g_z^+\approx15$, $27$, $35$, $41$ in~(\bbb). The contours enclose 75\% of the interfacial eddies.}
    \label{fig:den_eddy_zy_gs_all}
\end{figure}

\rgm{To illustrate how $s_z^+$ and $g_z^+$ influence the `accessible space' that penetrating eddies must fit within,
figure \ref{fig:den_eddy_zy_gs_all} depicts the total spanwise width of interfacial eddies $z_{100}^+$ at each height;
further, quadrant-wise results are presented and discussed in appendix \ref{app:quadrants}.}
As shown in panel (\aaa), within the canopies with a fixed gap $g_z^+\approx41$, the distribution of $z_{100}^+$ remains largely similar, regardless of the variations in $s_z^+$.
In contrast, panel (\bbb) shows that increasing $g_z^+$ while keeping $s_z^+$ constant enables eddies with larger widths to penetrate more easily into the canopy, suggesting that gap width, rather than pitch, determines the `accessibility' of the canopy for turbulence penetration.
For all canopies considered in figure \ref{fig:den_eddy_zy_gs_all}, only eddies with widths smaller than the gap size, $z_{100}^+<g_z^+$, can penetrate all the way into the canopy.
When this gap is very small ($g_z^+\lesssim15$), \oldrev{not even the smallest turbulent structures, of size $\lambda^+\sim\mathcal{O}(20)$ \citep{jimenez2012cascades, jimenez2013near}, can fit in between the elements,
and thus turbulence penetration is essentially fully impeded,} as evidenced by the results for canopy $\mathrm{S47_{G14\times14}}$ in figure \ref{fig:den_eddy_zy_gs_all}(\bbb).
\rgm{We note that eddies with a span between $g_z$ and $s_z$, which would directly conflict with the canopy elements of width $w_z=s_z-g_z$, are particularly less likely to penetrate as deep as those slightly smaller and, to a lesser extent, those slight larger, which would still need to circumvent obstructing elements. This gives
the \rtwo{contours of the probability density function} their characteristic serrated shape.}

\begin{figure}
\vspace*{2mm}
    \centering
    \includegraphics[width=\textwidth]{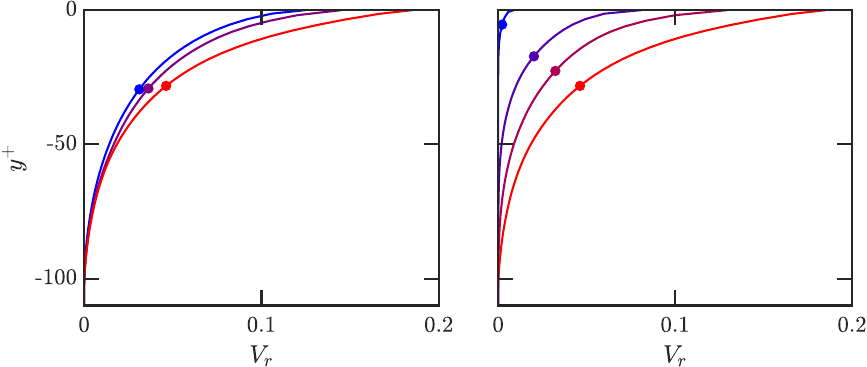}
    \put (-135mm,55mm) {(\aaa)}
    \put (-64mm ,55mm){(\bbb)}
    \caption{Relative volume $V_r$ occupied at each height \rgm{$y^+$} by interfacial eddies. Canopies with (\aaa) fixed gap $g_z^+\approx41$ and (\bbb) fixed pitch $s_z^+\approx47$. Markers $\bullet$ indicate the penetration depth \rgm{$y^+=-d_p^+$} where $V_r/V_{r,tip}=0.25$.
    The colour scheme is as in figure \ref{fig:den_eddy_zy_gs_all}\rgm{: from blue to red, canopies with decreasing frontal densities, (\aaa) $\lambda_f\approx0.63$, $0.46$, $0.29$ and (\bbb) $\lambda_f\approx1.61$, $1.02$, $0.58$, $0.29$.}}
    \label{fig:den_vr_zy_gs}
\end{figure}

In agreement with \zsc{figure \ref{fig:den_eddy_zy_gs_all},} figure \ref{fig:den_vr_zy_gs} portrays the dependence of the penetration depth on the spanwise gap between elements.
As shown in figure \ref{fig:den_vr_zy_gs}(\aaa), the penetration depth remains similar for cases with fixed gap size, \zsc{$d_p^+\approx30$,} independently of variations in pitch.
However, figure \ref{fig:den_vr_zy_gs}(\bbb) demonstrates that canopies behave as sparser with increasing spanwise gap $g_z^+$.
As the spanwise gap widens from $g_z^+\approx15$ to \oldrev{$g_z^+\approx40$,} the penetration depth increases from \zsc{$d_p^+\approx5$ to $d_p^+\approx30$.}
The main contribution to the net eddy volume within the canopies is from \rgm{fourth-quadrant, sweep structures},
which typically have a larger width within the canopy compared to those from the other quadrants,
as demonstrated in figures \ref{fig:den_eddy_zy_gs}(\ddd, \hhh) \rgm{in appendix \ref{app:quadrants}}.

\subsection{Influence of the Reynolds number}\label{sec:reynolds}

We have so far argued that the penetration of the overlying turbulence depends on whether
the span of canopy `canyons' is sufficient for the interfacial turbulent eddies to fit within.
This inherently involves not only the canopy dimensions, but also the characteristic dimensions of those eddies.
Assuming that their dimensions are set by the overlying turbulent dynamics, classical scaling would
imply that they scale in inner units if they are sufficiently embedded in the near-wall region,
and with the height above a notional wall if they lie farther above and into the log layer.
The governing parameter would be the zero-plane-displacement \rgm{depth below the plane of tips, $d^+_0$,} the depth perceived as the
origin by the overlying-turbulence eddies, which we have recently reported in detail for a wide range of
canopy densities \citep{chen2023examination}.
For most of the present cases, with low to moderate Reynolds numbers and $d^+_0\lesssim100$, we would expect interfacial
eddies to scale in inner units. 
To investigate this, 
we examine canopies with either the same dimensions in inner
units (same $s^+$, $g^+$ and $h^+$) or in outer units (same $s/\delta$, $g/\delta$ and $h/\delta$), for different frontal densities $\lambda_f$ and across different Reynolds numbers $Re_\tau$.

\begin{figure}
\vspace*{2mm}
    \centering
    \includegraphics[width=\textwidth]{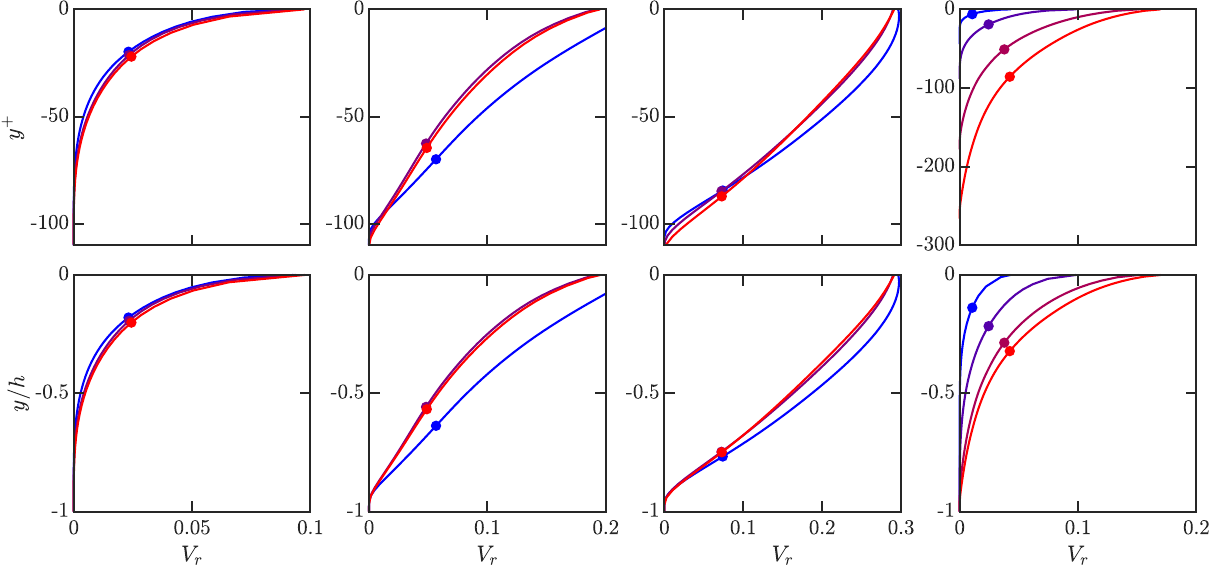}
    \put (-135mm ,62mm) {(\aaa)}
    \put (-100mm ,62mm) {(\bbb)}
    \put (-67mm ,62mm) {(\ccc)}
    \put (-34.5mm ,62mm) {(\ddd)}
    \put (-135mm ,32mm) {(\eee)}
    \put (-100mm ,32mm) {(\fff)}
    \put (-67mm ,32mm) {(\ggg)}
    \put (-34.5mm ,32mm) {(\hhh)}
\vspace*{1mm}
    \caption{Relative volume $V_r$ occupied at each height \rgm{$y^+$} by interfacial eddies. (\aaa, \eee), (\bbb, \fff) and (\ccc, \ggg) are, respectively for $\lambda_f\approx0.91$, $0.23$, and $0.06$, `inner-similar' canopies with the same $s^+$, $g^+$ and $h^+$. (\ddd, \hhh) are `outer-similar' canopies with $\lambda_f\approx0.70$ and the same $s/\delta$, $g/\delta$ and $h/\delta$.
    Lines from blue to red indicate increasing Reynolds number $Re_\tau\approx550$-$2000$.
    Markers $\bullet$ indicate the penetration depth \rgm{$y=-d_p$} where $V_r/V_{r,tip}=0.25$.}
    \label{fig:den_vr_re}
\end{figure}

Figure \ref{fig:den_vr_re} depicts the decay rate of the eddy volume and the penetration depth for
both the `inner-similar' and the `outer-similar' canopies from table \ref{tab:canopy_param}.
As shown in panels (\aaa-\ccc) and (\eee-\ggg), when a canopy remains
the same in inner units for varying $Re_\tau$, be it dense as in panels (\aaa, \eee), intermediate as in (\bbb, \fff), or sparse as in (\ccc, \ggg), it exhibits similar
turbulent penetration. This is consistent with the scaling of interfacial eddies in viscous units mentioned above.
Scaling of the flow in inner units over canopies and rough surfaces has also been reported by \cite{chan2015systematic} and \cite{sharma2020scaling_a, sharma2020turbulent_b} \oldrev{at $Re_\tau$ and $s^+$-$h^+$ values comparable to the present ones.}

In contrast, a canopy that remains the same in outer units, i.e. with fixed $s/h$ and $g/h$ but varying $g^+$ and all other dimensions in viscous units, can 
exhibit different behaviours as $Re_\tau$ varies. This is portrayed
for the canopy of $\mathrm{OS180_{G15\times15}}$ in figures \ref{fig:den_vr_re}(\ddd, \hhh). This canopy behaves as dense at $Re_\tau\approx180$, for which \zsc{$d_p^+\approx5$} and the turbulence essentially skims above, with limited penetration.
However, for increasing Reynolds numbers, the canopy behaves as increasingly sparser,
as the penetration depth increases gradually to \zsc{$d_p^+\approx85$} at $Re_\tau\approx1080$, for which \zsc{$d_p^+$} is of the order of the zero-plane displacement \zsc{$d_0^+$.}
The increase in \zsc{$d_p^+$} is not linear with the canopy dimensions in inner units, e.g. with $g_z^+$, which would
result in a collapse of the results in panel \ref{fig:den_vr_re}(\hhh). For
$g_z^+ \approx 15$, eddy penetration is negligible, and only for $g_z^+ \gtrsim 30$ it begins to approach \zsc{$d_p\approx g_z$} and an eventual collapse in panel (\hhh).
This is what one would expect if the impedance exerted by the canopy elements on the interfacial eddies was essentially viscous
for $g_z^+ \approx 15$, but became increasingly inertial for \zsc{$d_0^+\approx50$-$100$;} interfacial eddies would then scale with \zsc{$d_0$,}
and the vanishing effect of viscosity would result in \zsc{$d_0 \sim d_p \sim g_z$.} Finally, we note that the increase in penetration 
depth has been observed for a canopy with \zsc{$d_p\approx0.15\,h\,$-$\,0.35\,h$,} 
and would not occur for a sparse canopy with
\zsc{$d_p\approx h$,} for which no further increase of \zsc{$d_p$} would be possible due to the presence of the flow. The conclusion is that canopies that are sparse at low
$Re_\tau$ remain sparse as $Re_\tau$ increases, while dense ones can behave as increasingly sparse as the interfacial eddy size
decreases relative to the canopy gap. We expect, however, that this effect eventually saturates, as the penetration depth asymptotes
to \zsc{$d_p\approx g_z$;} verifying this will require measurements at Reynolds numbers higher than our current computational capability.

\subsection{What makes a canopy dense or sparse?}\label{sec:definedense}

\begin{figure}
\vspace*{3mm}
    \centering
    \includegraphics[width=\textwidth]{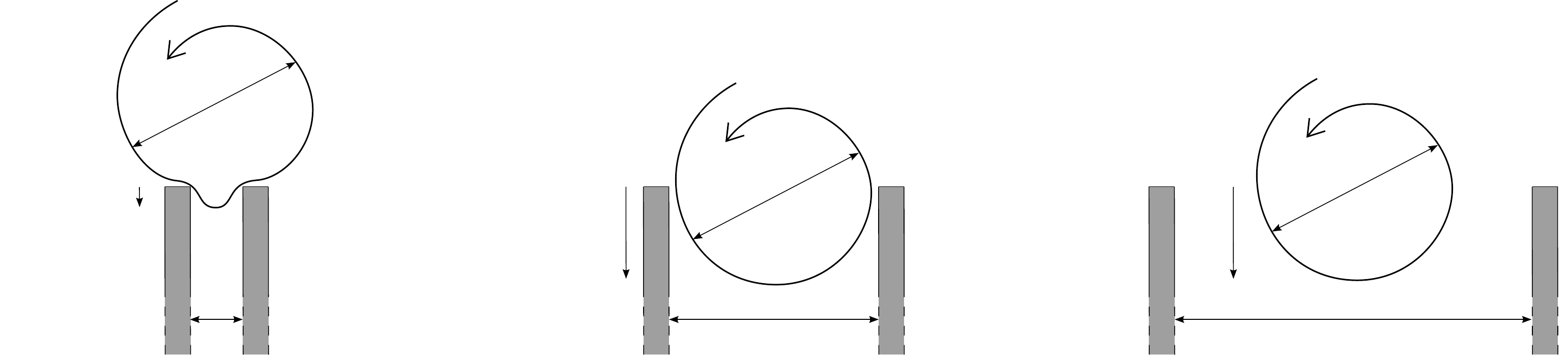}
    \put (-128mm,13mm) {$d_p$}
    \put (-119mm,23mm) {$\ell_e$}
    \put (-118mm,00mm) {$g_z$}
    \put (-85mm ,10mm) {$d_p$}
    \put (-71mm ,15mm) {$\ell_e$}
    \put (-70mm ,00mm) {$g_z$}
    \put (-33mm ,09mm) {$d_p$}
    \put (-20mm ,16mm) {$\ell_e$}
    \put (-19mm ,00mm) {$g_z$}
    \put (-135mm ,25mm) {(\aaa)}
    \put (-89mm  ,25mm) {(\bbb)}
    \put (-41mm  ,25mm) {(\ccc)}
    \vspace*{2mm}
    \caption{Sketch of the penetration depth \zsc{$d_p$} for interfacial eddies of size $\ell_e$, depending on the value of $\ell_e$ relative to the spanwise gap between canopy elements, $g_z$. (\aaa), $\ell_e>g_z$; (\bbb), $\ell_e\sim g_z$; (\ccc), $\ell_e<g_z$.}
    \label{fig:eddy_sketch}
\end{figure}

The preceding discussion suggests that, depending on the ratio between canopy gap $g_z$ and the width of interfacial eddies $\ell_e$, the
eddies will be in one of the situations sketched in figure \ref{fig:eddy_sketch}. For closely packed elements,
$g_z<\ell_e$, the eddies would not fit in the gaps, and their penetration would be limited as sketched in panel (\aaa) -- to a few viscous units if $g_z^+\lesssim15$ or to order \zsc{$d_p \sim g_z$} otherwise. For a wider separation between elements, such that $g_z\sim\ell_e$ as in panel (\bbb),
eddies could penetrate into the canopy, but the depth of the penetration would be limited by their own size, as the
span-to-height ratio of wall-turbulent eddies is of order unity \citep{Flores2010}. We would then have \zsc{$d_p \sim g_z$} once more.
For widely separated elements, $g_z>\ell_e$ as in panel (\ccc), interfacial eddies could traverse the plane of the tips freely and penetrate into
the canopy undisturbed, but the penetration depth would again be limited by the eddy size itself, yielding \zsc{$d_p \sim \ell_e < g_z$.}
The latter case is only possible if the zero-plane displacement \zsc{$d_0$,} which sets the scale for the eddy size $\ell_e$, is
small enough to result in $\ell_e<g_z$. This in turn is only possible if \zsc{$d_0$} is restricted by the proximity of the floor,
rather than by $g_z$, that is, for sparse canopies with \zsc{$d_p \approx h$.}

\begin{figure}
\vspace*{3mm}
    \centering
    \includegraphics[width=\textwidth]{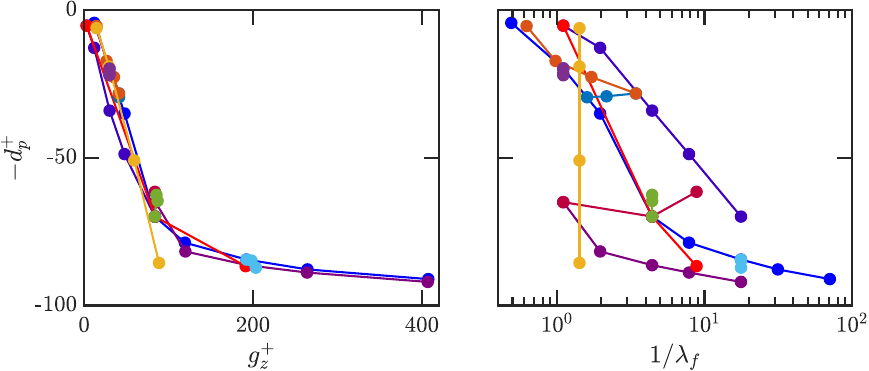}
    \put (-135mm,55mm) {(\aaa)}
    \put (-64mm ,55mm){(\bbb)}
    \caption{Penetration depth \zsc{$d_p^+$} versus (\aaa) spanwise gap $g_z^+$ and (\bbb) frontal density $\lambda_f$. \textcolor{col1}{$\bullet$},~isotropic-layout canopies; \textcolor{col2}{$\bullet$}, `fence-like' canopies; \textcolor{col3}{$\bullet$}, `canyon-like' canopies; \textcolor{col4}{$\bullet$}, fixed-$s_z$ canopies; \textcolor{col5}{$\bullet$}, fixed-$s_x$ canopies; \textcolor{col6}{$\bullet$}, fixed-gap canopies; \textcolor{col7}{$\bullet$}, fixed-pitch canopies; \textcolor{col8}{$\bullet$}, `outer-similar' canopies; \textcolor{col9}{$\bullet$}, `inner-similar' canopies with $\lambda_f\approx0.91$; \textcolor{col10}{$\bullet$} `inner-similar' canopies with $\lambda_f\approx0.23$; \textcolor{col11}{$\bullet$} `inner-similar' canopies with $\lambda_f\approx0.11$.
    \rgm{The coloured lines connect cases in the corresponding set.}}
    \label{fig:den_penetration_depth_lam_gap}
\vspace*{2mm}
\end{figure}

Given the limited $Re_\tau$ in our simulations, \rgm{let us} first assume that across our canopies the interfacial-eddy size broadly scales in viscous units. If that is the case, $g_z^+$ would be a reasonable surrogate for the ratio $g_z/\ell_e$, and can therefore be used to
characterise eddy penetration. This is illustrated in figure \ref{fig:den_penetration_depth_lam_gap}, which shows that
$g_z^+$ is a good predictor for \zsc{$d_p^+$} and significantly outperforms the classic frontal density $\lambda_f$.
Starting at small $g_z^+$, the penetration depth \zsc{$d_p^+$} increases monotonically with $g_z^+$ until it eventually becomes comparable
to the canopy height, \rgm{which is} $h^+\approx100$ for most of our canopies. The only exception are the `outer-similar' canopies discussed in
\S \ref{sec:reynolds}, which have $h^+\approx50$-$270$ and for which the last case has \zsc{$d_p^+\approx85$,} still far from the
full canopy height $h^+\approx270$. This case appears as a slight outlier in figure \ref{fig:den_penetration_depth_lam_gap}(\aaa).

\setlength{\tabcolsep}{3.7pt}
\begin{table}
\centering
\begin{tabular}{ll@{\hskip 0.5in}ccccccccc}
 & \multicolumn{1}{c}{} & Case & $N_x\times N_z$ & $Re_\tau$ & $\lambda_f$ & $g_z^+$ & $g_x^+$ & $s_z^+$ & $s_x^+$ & $h^+$ \\[2.5mm]
\multirow{6}{*}{\begin{tabular}[c]{@{}l@{}}Different \\ height\end{tabular}} & \multirow{6}{*}{\parbox[c]{1em}{\vspace{2.5mm}\includegraphics[width=0.6in]{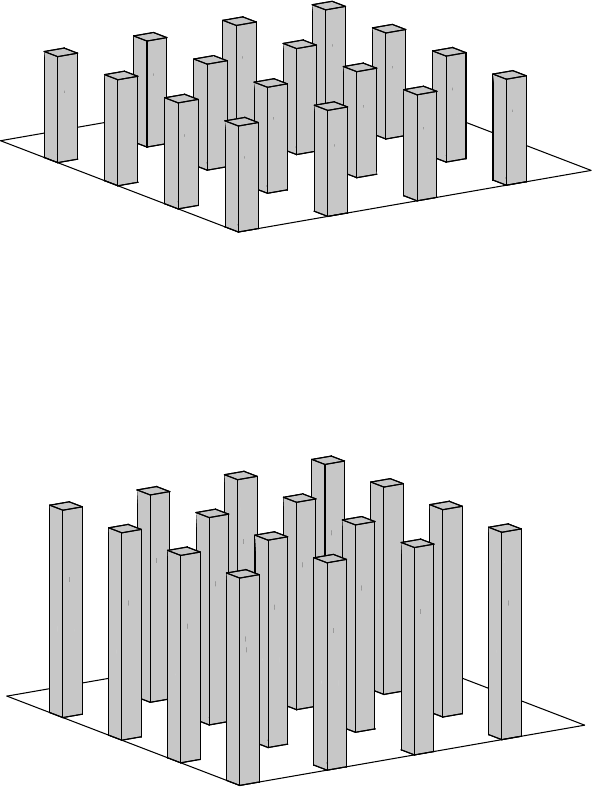}}} & $\mathrm{I_{S54\times54}}$ & 64$\times$32 & 549.7 & 0.91 & 30.0 & 30.0 & 54.0 & 54.0 & 109.9 \\
 &  & $\mathrm{I_{S108\times108}}$ & 32$\times$16 & 548.7 & 0.23 & 83.8 & 83.8 & 107.7 & 107.7 & 109.7 \\
 &  & $\mathrm{I_{S216\times216}}$ & 16$\times$8 & 550.7 & 0.06 & 192.2 & 192.2 & 216.3 & 216.3 & 110.1 \\[0.3cm]
 &  & $\mathrm{H220_{S54\times54}}$ & 64$\times$32 & 549.2 & 1.81 & 30.0 & 30.0 & 53.9 & 53.9 & 219.7 \\
 &  & $\mathrm{H220_{S108\times108}}$ & 32$\times$16 & 550.9 & 0.45 & 84.1 & 84.1 & 108.2 & 108.2 & 220.4 \\
 &  & $\mathrm{H220_{S216\times216}}$ & 16$\times$8 & 551.6 & 0.11 & 192.6 & 192.6 & 216.6 & 216.6 & 220.7
\end{tabular}
\caption{Simulation parameters for canopies with height doubled to $h^+\approx220$. The corresponding simulations with $h^+\approx110$ from table \ref{tab:canopy_param} are also included for reference.}
\label{tab:canopy_h}
\vspace*{1mm}
\end{table}

\begin{figure}
\vspace*{1mm}
    \centering
    \includegraphics[width=\textwidth]{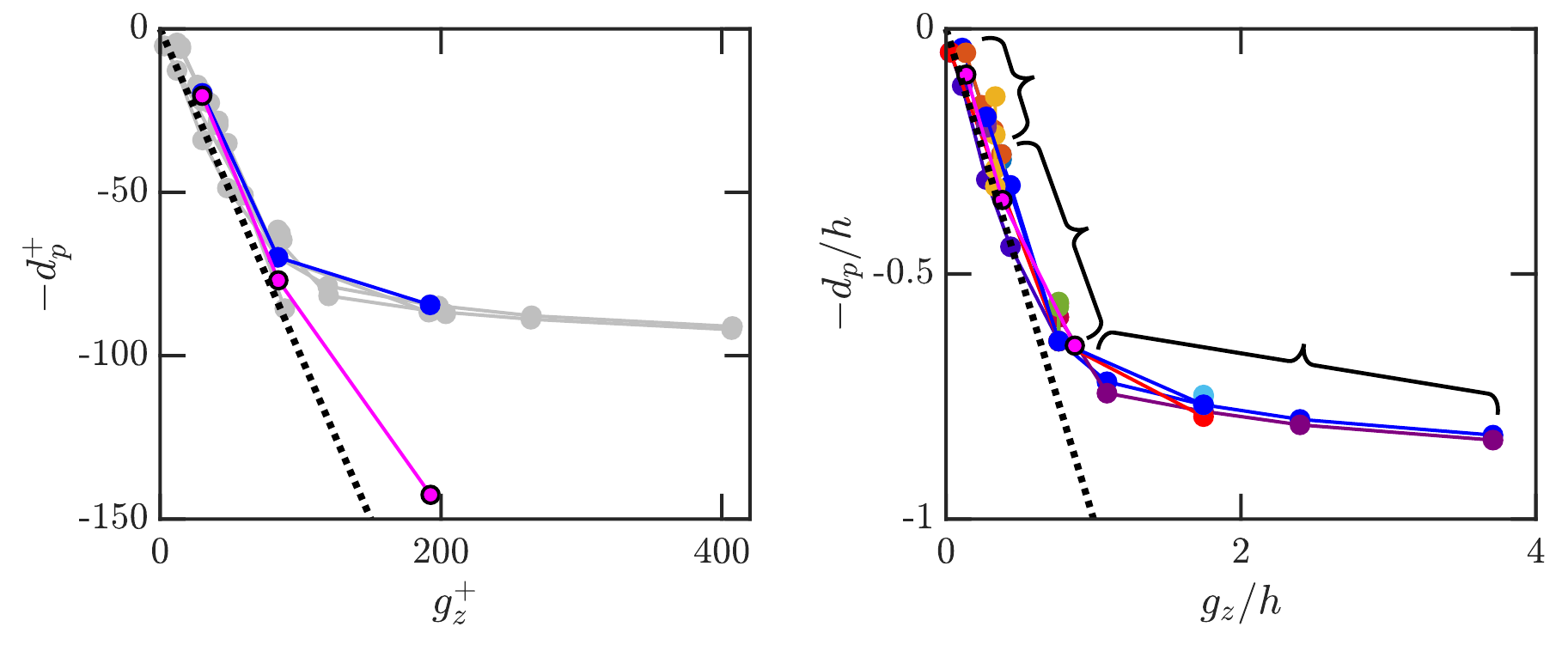}
    \put (-132mm,53mm){(\aaa)}
    \put (-63mm ,53mm){(\bbb)}
    \put (-44.5mm,48mm) {dense}
    \put (-40mm,35.5mm) {intermediate}
    \put (-27mm,28mm) {sparse}
\vspace*{-2mm}
    \caption{Penetration depth versus spanwise gap, scaled (\aaa) in viscous units as in figure \ref{fig:den_penetration_depth_lam_gap}(\aaa), and (\bbb) with the canopy height $h$. 
    \textcolor{colblue}{$\bullet$}~and~\protect\mdot, cases `I' and `H220' from table \ref{tab:canopy_h}, respectively.    
    In (\aaa), \textcolor{colgrey}{$\bullet$} is used for cases already portrayed in figure \ref{fig:den_penetration_depth_lam_gap}. In (\bbb), colours are as in figure \ref{fig:den_penetration_depth_lam_gap}.
    \rgm{The coloured lines connect cases in the corresponding set.}
    The dotted lines are \zsc{$d_p=g_z$.} 
    }
    \label{fig:yp_dense_depth}
\end{figure}

The question that arises then is if, for our sparse canopies with eddy penetration essentially all the way to the floor,
that penetration is restricted by the presence of the floor, or if is it is already as much as it would be if the floor was
further below. To address this, we have run three additional simulations for a dense, an intermediate and a sparse canopy
in which we doubled the canopy height. The parameters of the simulations are summarised in table \ref{tab:canopy_h},
which also includes the original simulations for reference.
The results for \zsc{$d_p^+$} are shown in figure \ref{fig:yp_dense_depth}(\aaa).
For dense and intermediate canopies, doubling the canopy height from $h^+ \approx 110$ to $h^+ \approx 220$ results in
no significant change in the penetration depth \zsc{$d_p^+$,} which remains essentially \zsc{$d_p\approx g_z$.} This is consistent with
the representations for eddy penetration in figures \ref{fig:eddy_sketch}(\aaa-\bbb). For the sparse canopy, however,
doubling the canopy height results in an increase in \zsc{$d_p^+$,}  and an approach to the \zsc{$d_p\approx g_z$} trend of denser
canopies. This would correspond to a transition from the representation of figure \ref{fig:eddy_sketch}(\ccc) to the one in figure \ref{fig:eddy_sketch}(\bbb), and
illustrates a secondary role of canopy height in the penetration of the overlying turbulence. We have so far considered the penetration of interfacial eddies as driven
essentially by the dynamics and canopy properties at the interface with the overlying flow. For sparse canopies with \zsc{$d_p\approx h$,} however, increasing their height can
have a direct result of increasing the depth of the zero-plane displacement. This, in turn, results in an increase of the typical size of interfacial eddies, and thus of their penetration depth,
\zsc{$d_p\sim \ell_e$,} which can continue until they reach $\ell_e\approx g_z$, when the penetration depth begins to be limited by the canopy canyon width.

Ultimately, canopy density can be characterised by what proportion of the canopy the overlying turbulence can penetrate, that is, by \zsc{$d_p/h$.}
From the above discussion, we expect \zsc{$d_p/h$} to be roughly proportional to $g_z/h$ for dense and intermediate canopies, and 
to approach \zsc{$d_p/h\approx 1$} for sparse ones. This suggests that density can be characterised in a graph of \zsc{$d_p/h$} vs. $g_z/h$,
as portrayed in figure \ref{fig:yp_dense_depth}(\bbb). The figure is similar to figure \ref{fig:den_penetration_depth_lam_gap}(\aaa), but
the largest `outer-similar' case, as well as the simulations with doubled height of figure \ref{fig:yp_dense_depth}(\aaa), no longer appear as outliers, and collapse with the rest.
The figure shows that canopies lie along the line \zsc{$d_p/h=g_z/h$} for $g_z/h\lesssim0.5$, and tend to a constant \zsc{$d_p/h\approx0.9$} for
$g_z/h\gtrsim1$.
\rgm{In any event, if we characterise density in terms of what proportion of the canopy the overlying turbulence can penetrate,}
dense canopies would be those with small penetration depth relative to the canopy height, e.g. \zsc{$d_p/h\lesssim0.2$,} sparse ones those with
substantial penetration, e.g. \zsc{$d_p/h\gtrsim0.7$,} and those in-between would have intermediate density and connect the two aforementioned
limit trends. \rgm{From the inspection of figure \ref{fig:yp_dense_depth}(\bbb), intermediate canopies mostly lie along the $d_p\approx g_z$ trend of dense ones; the difference between the two is not in their trend in the figure, but merely in percentage of the canopy accessible to the overlying turbulence.
Expressed} in terms of gap-to-height ratio, dense canopies would then have $g_z/h\lesssim0.2$, intermediate ones $0.2\lesssim g_z/h\lesssim 1$,
and sparse ones $g_z/h\gtrsim 1$.

\subsection{\rgm{Beyond collocated layouts}}

\rgm{The present work has focused solely on canopies of collocated, i.e. streamwise- and spanwise-aligned, elements. In this case, the concept of a spanwise gap as a `canyon width' through which turbulent eddies can travel unobstructed is straightforward to define.
While the concept of a canyon width determining the extent of eddy penetration should apply to other element layouts, quantifying this may be less straightforward and is left for future work.
As a first, preliminary step to how this would play out, let us consider staggered layouts. For this, we have run a few additional simulations, listed in table \ref{tab:canopy_stg},
where we have taken some of our original isotropic canopies, also listed for reference, with spanwise gaps $g_z^+\approx30$, 50, 85 and 120, and shifted consecutive rows of elements by half a period 
either in $x$ (canopies SX) or in $z$ (canopies SZ), as indicated in the sketches in the table.}

\setlength{\tabcolsep}{4pt}
\begin{table}
\centering
\begin{tabular}{ll@{\hskip 0.9in}llllll}
 &  & Case & $\lambda_f$ & $s_z^+$ & $g_z^+$ & $g_z^{\prime +}$ & $\widehat{g}_z^+$ \\[0.2cm]
\multirow{4}{*}{Isotropic} & \multirow{4}{*}{\parbox[c]{1em}{\includegraphics[width=0.85in]{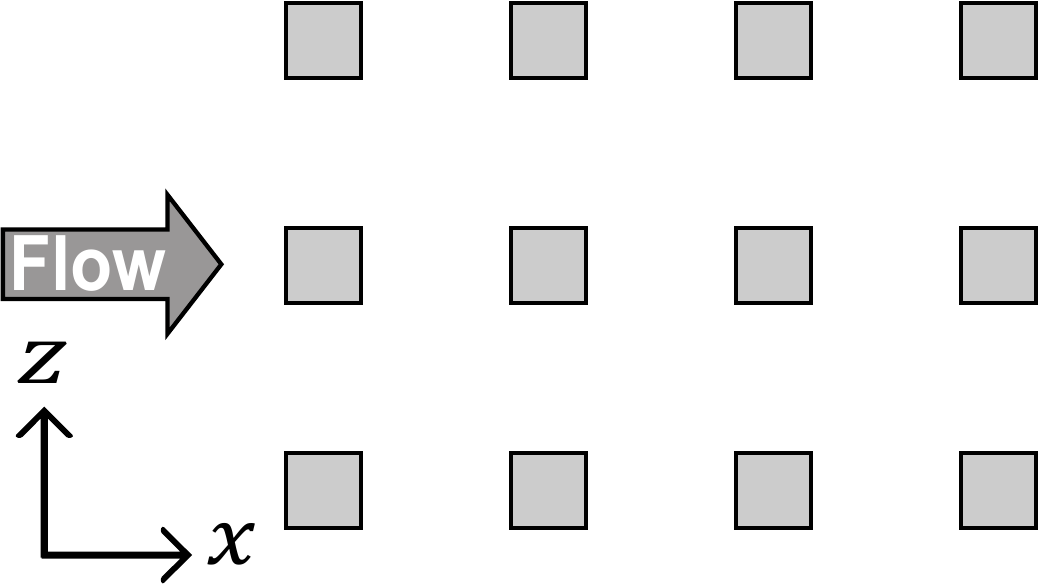}}}
    & $\mathrm{I_{S54\times54}}$ & 0.91 & 54.0 & 30.0 & 54.0 & 30.0\\
 &  & $\mathrm{I_{S72\times72}}$ & 0.51 & 71.8 & 47.8 & 71.8 & 47.8\\
 &  & $\mathrm{I_{S108\times108}}$ & 0.23 & 107.7 & 83.8 & 107.7 & 83.8\\
 &  & $\mathrm{I_{S144\times144}}$ & 0.13 & 143.2 & 119.3 & 143.2 & 119.3\\[0.3cm]
\multirow{3}{*}{\begin{tabular}[c]{@{}l@{}}Staggered \\      in X\end{tabular}} & \multirow{3}{*}{\parbox[c]{1em}{\includegraphics[width=0.85in]{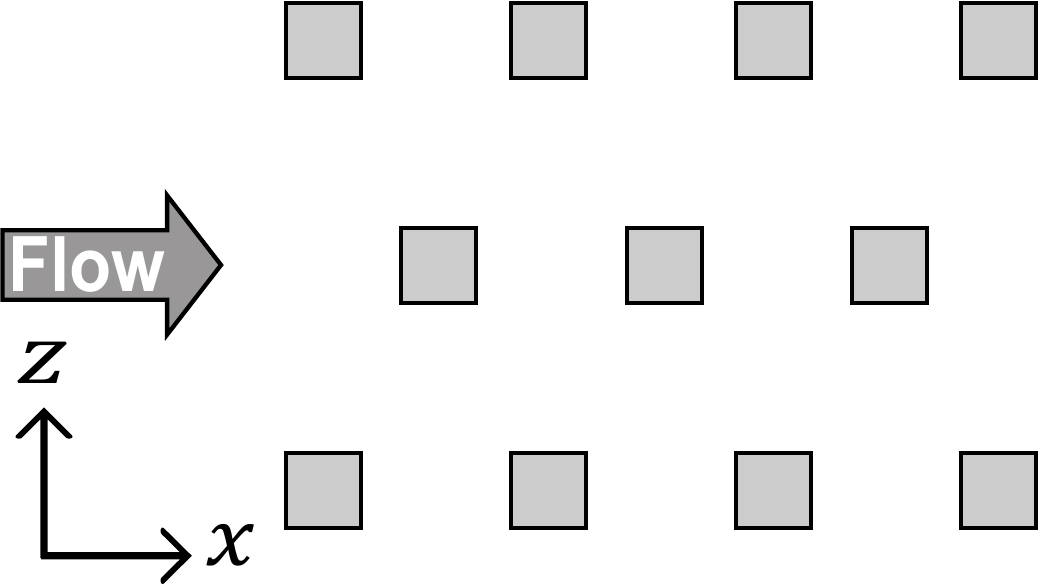}}}
    & $\mathrm{SX_{S27\times54}}$ & 0.91 & 54.1  & 30.1 & 30.1 & 16.1 \\
 &  & $\mathrm{SX_{S36\times72}}$ & 0.51 & 72.1  & 48.0 & 48.0 & 32.2\\
 &  & $\mathrm{SX_{S54\times108}}$ & 0.23 & 108.1 & 84.1 & 84.1 & 64.4\\[0.3cm]
\multirow{4}{*}{\begin{tabular}[c]{@{}l@{}}Staggered \\      in Z\end{tabular}} & \multirow{4}{*}{\parbox[c]{1em}{\includegraphics[width=0.85in]{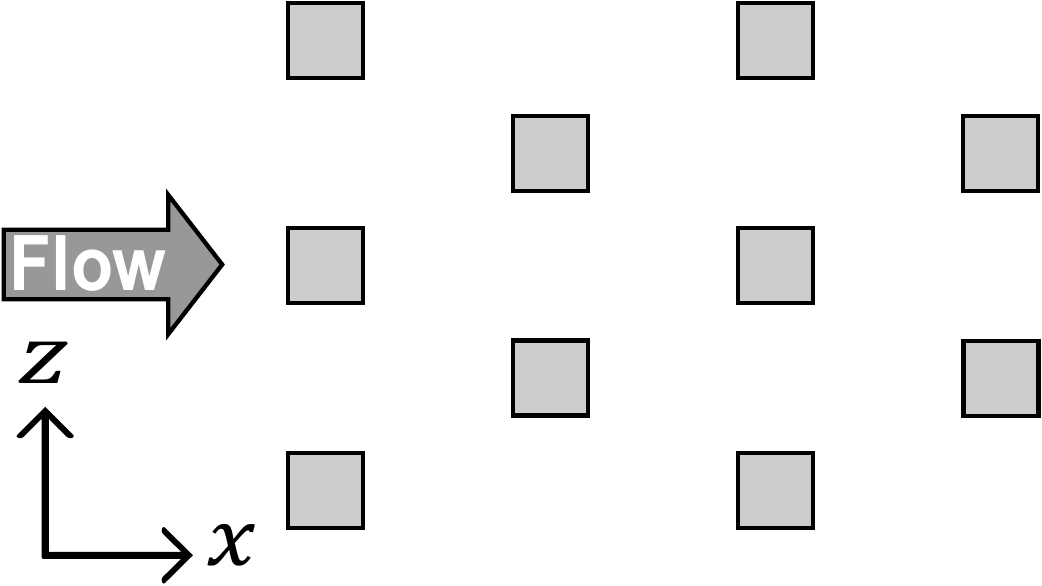}}}
    & $\mathrm{SZ_{S54\times27}}$ & 0.91 & 53.9 & 29.9 & 3.0 & 16.1\\
 &  & $\mathrm{SZ_{S72\times36}}$ & 0.51 & 72.0 & 48.0 & 12.0 & 32.2\\
 &  & $\mathrm{SZ_{S108\times54}}$ & 0.23 & 108.2 & 84.2 & 30.1 & 64.4\\
 &  & $\mathrm{SZ_{S144\times72}}$ & 0.13 & 143.5 & 119.6 & 47.8 & 96.3\\[0.2cm]
\end{tabular}
\caption{\rgm{Simulation parameters for canopies with staggered layouts. The corresponding isotropic canopies from table \ref{tab:canopy_param} are also included for reference. The staggered arrangements are offset by half a canopy pitch in either the streamwise ($x$) or spanwise ($z$) direction, as illustrated in the top-view sketches. The pitch $s^+$ and the gap $g^+$ are defined based on the values for the corresponding isotropic configuration, and the frontal and effective gaps, $g_z^{\prime +}$ and $\widehat{g}_z^+$, are defined according to figure \ref{fig:flow_alley}.}}
\label{tab:canopy_stg}
\vspace*{1mm}
\end{table}

\begin{figure}
    \centering
    \includegraphics[width=.75\textwidth]{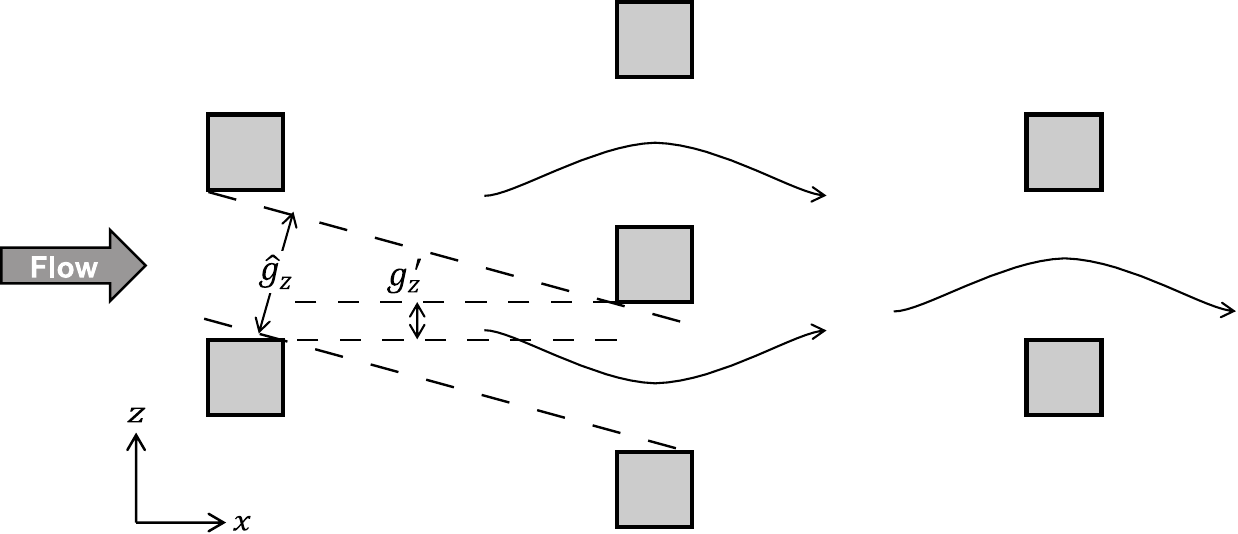}
\vspace*{3mm}
    \caption{\rtwo{Sketch of a staggered-layout canopy comparing the frontal gap, $g_z^\prime$, and an effective gap reflecting the flow meandering, $\widehat{g}_z^+$.}}
    \label{fig:flow_alley}
\end{figure}

\rgm{In terms of `canyon width', we would expect the staggering in $x$ not to have an effect on the effective $g_z$.
In turn, for elements staggered in $z$, we would in principle assume
the effective gap to be the one when viewing the `canyons' from the front, $g_z^{\prime +}$. This frontal gap, however, can quickly become vanishingly small,
even when there is still significant room for the flow to travel in between elements, as sketched in figure \ref{fig:flow_alley}. Intuitively, we would then expect that the effective gap was larger than the frontal one.}

\begin{figure}
\vspace*{1mm}
    \centering
    \includegraphics[width=\textwidth]{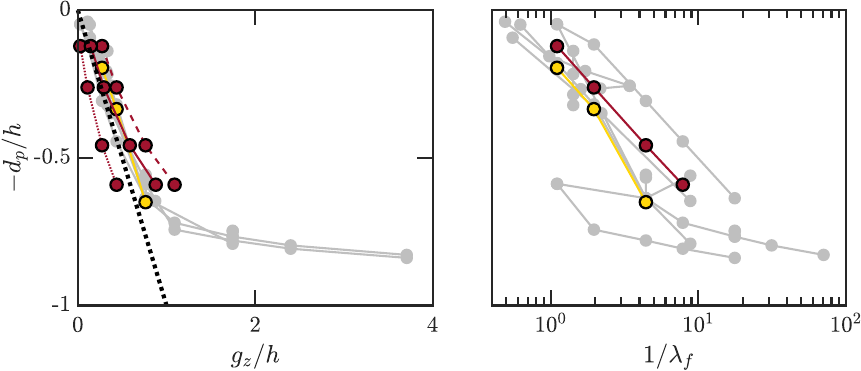}
    \put(-135mm,55mm){(\aaa)}
    \put(-64mm ,55mm){(\bbb)}
    \put(-338,60){\vector(1,4){5}}
    \put(-300,55){\vector(-1,2){13}}
    \put(-277,74){\vector(-3,1){25}}
    \put(-120mm,18.5mm){$g_z$}
    \put(-106mm,16.5mm){$\widehat{g}_z$}
    \put(-97mm,25mm){$g_z^\prime$}
    \caption{\rgm{Penetration depth $d_p$ versus spanwise gap $g_z$ and frontal density $\lambda_f$, with lengths scaled with the canopy height $h$.
    \protect\ydot, x-staggered canopies `SX';
    \protect\rdot, z-staggered canopies `SZ';
    \textcolor{colgrey}{$\bullet$},~cases already portrayed in figure \ref{fig:yp_dense_depth}.
    The coloured lines connect cases in the corresponding set;
    \rtwo{dotted lines connect markers for the original gap from the isotropic layout, $g_z$;
    dashed lines markers for the frontal gap, $g_z^\prime$;
    solid lines markers for the effective gap, $\widehat{g}_z$, defined in figure \ref{fig:flow_alley}.}
    The black dotted lines are $d_p=g_z$.}}
    \label{fig:yp_dense_depth_staggered}
\end{figure}

\rgm{Figure \ref{fig:yp_dense_depth_staggered} portrays eddy penetration for $x$- and $z$-staggered canopies vs. $g_z$ and $g_z^\prime$.
For $x$-staggered canopies, the results suggest that there is indeed no effective change in gap with respect to collocated ones.
For $z$-staggered canopies, the results suggest that the effective gap width is smaller than $g_z$, but larger than $g_z^{\prime +}$.
There is always some degree of meandering in wall-turbulent eddies and, in addition, from Taylor's frozen turbulence hypothesis \citep{Taylor1938} we would expect
eddies to loosely follow the path of mean-flow streamlines, which within the canopy meander themselves around the canopy elements.
A rigorous calculation of the width for such `streamline pathways' is in principle possible using the ensemble average flow of
(\ref{eq:tri_decomp}), but in the present first approximation we use the simple, rough estimate of an effective, oblique gap $\widehat{g}_z$, calculated from the canopy geometry alone, as indicated in figure \ref{fig:flow_alley}.
We note that the same oblique gap can also be calculated for $x$-staggered canopies, but in that case it would be narrower than the $g_z$ of their streamwise-aligned canyons.
In general, we would expect the larger of $g_z$, $g_z^\prime$ and $\widehat{g}_z$ to be the effective gap in terms of eddy penetration, which implicitly entails that,
for $\widehat{g}_z$ to be the effective gap, the associated oblique paths must have limited yaw.
Figure \ref{fig:yp_dense_depth_staggered} suggests that $\widehat{g}_z$ is a reasonable approximation for the effective gap for the present $z$-staggered canopies,
which prefigures how to approach the problem for more complex canopy layouts.}

\section{Conclusions}\label{sec:conclusions}

In the present work, we have identified and characterised canopy density regimes based on the penetration of \oldrev{eddies} from the
overlying turbulence. In the sparse regime, turbulence penetrates with little obstruction into the canopy \oldrev{all the way to the canopy floor},
whereas in the dense regime, this penetration is limited. We have examined the conventional measure of canopy density using frontal
density $\lambda_f$, and \oldrev{found} that $\lambda_f$ alone cannot \oldrev{fully} characterise canopy density, as it cannot capture
\oldrev{e.g.} the effect of \oldrev{different streamwise and spanwise spacing of the canopy elements} or of the Reynolds number.

To measure turbulence penetration, we focus on \oldrev{flow} structures of intense \oldrev{local} Reynolds shear stress $u'v'$, as these
structures are responsible for momentum transport and turbulence diffusion.
\oldrev{We separate overlying-turbulence structures from element-coherent ones by spectral filtering, and choose a suitable $u'v'$ threshold
to delimit them through a percolation analysis. We then focus on the structures that are, at least partially, in the immediate vicinity
of the canopy-tip plane, referring to them as `interfacial eddies', and propose metrics to quantify their location, span and volume.}
We apply these metrics to analyse a series of canopies across a range of frontal densities $\lambda_f\approx0.01$-$2.04$, heights
$h^+\approx44-266$, and Reynolds numbers $Re_\tau\approx180-2000$.

Our results show that\oldrev{, for the same $\lambda_f$,} streamwise-packed canopies with large spanwise gaps allow eddies to penetrate
effectively, and thus behave as sparser compared to the corresponding isotropic and spanwise-packed canopies\oldrev{, which limit eddy
penetration to a significantly greater degree. Streamwise-packed canopies with frontal densities as high as $\lambda_f\approx0.9$
still behave as sparse, while only in the limit of very small frontal density, $\lambda_f\lesssim0.06$, do canopies behave as sparse regardless of their anisotropy,} 
\oldrev{Our results} further show that turbulence penetration remains similar when the spanwise gap \oldrev{between canopy elements} is fixed,
regardless of their pitch \zsc{(spacing)} and \zsc{width}. For closely packed canopies with small spanwise gaps, $g_z^+\lesssim15$, turbulence
penetration is \oldrev{restricted to a few visocus units}. As the spanwise gaps widen, \oldrev{canopies behave} as
\oldrev{increasingly} sparse, allowing eddies to penetrate deeper and more vigorously.
\rgm{This occurs as the eddies are increasingly unhindered as they are advected downstream  through the canyons that appear between canopy elements.}
\oldrev{Turbulence penetration does not only depend on the canopy topology, but also on the Reynolds number.}
A canopy with fixed geometry can \oldrev{inhibit turbulence penetration} at low Reynolds numbers, $Re_\tau\approx180$,
\oldrev{when its dimensions in viscous units, mainly $g_z^+$, are small,} but
\oldrev{permit increasing penetration} as $Re_\tau$ \oldrev{and $g_z^+$ increase. In turn,}
canopies with the same dimensions in inner units exhibit similar density regimes and penetration depth across different $Re_\tau$.

Overall, our results suggest that turbulence penetration 
depends on the \rtwo{effective} spanwise gap \rgm{of the canyons} between canopy elements, and how it compares to the typical spanwise size of interfacial eddies.
The latter scales generally with the zero-plane displacement depth, resulting in a penetration depth \zsc{$d_p \sim g_z$} unless further limited
by the immediate vicinity of the canopy bed -- something that occurs for sparse canopies. This leads to a characterisation of canopy
density in terms of \zsc{$d_p/h$,} the proportion of the canopy that the overlying turbulence can penetrate, as a function of the
canyon-width-to-element-height ratio $g_z/h$. Dense canopies have 
$g_z/h\lesssim0.2$, their turbulence penetration is small and follows roughly \zsc{$d_p\approx g_z$.}
Intermediate canopies have $0.2\lesssim g_z/h\lesssim 1$, and experience an eventual departure from \zsc{$d_p\approx g_z$} for large
$g_z/h$. Sparse canopies have $g_z/h\gtrsim 1$, and their turbulence penetration approaches the full height of the canopy, \zsc{$d_p/h\approx0.9$.}

\rgm{The present study has focused on canopies of collocated elements.
Nevertheless, preliminary results for staggered layouts suggest that the concept of canyons through which eddies travel can be extended to more complex layouts, which would allow estimates for their effective gap, the degree of eddy penetration, and the corresponding density regime.}

\vspace*{4mm}



\backsection[Funding]{This work was supported in part by the UK Engineering and Physical Sciences Research Council (EPSRC) under grant EP/S013083/1.
Computational resources were provided by the University of Cambridge Research Computing Service under EPSRC Tier-2 grant EP/P020259/1
(project cs155), and by the UK 'ARCHER2' system under PRACE project pr1u1702 and EPSRC Access to HPC projects e776 and e800. 
For the purpose of open access, the authors have applied a Creative Commons Attribution
(CC BY) licence to any Author Accepted Manuscript version arising from this submission.
}

\backsection[Declaration of interests]{The authors report no conflict of interest.}


\appendix

\section{\rgm{Assessment of other density characteristic lengthscales as predictors for eddy penetration depth}}\label{app:other_densities}

\begin{figure}
    \centering
    \includegraphics[width=.94\textwidth]{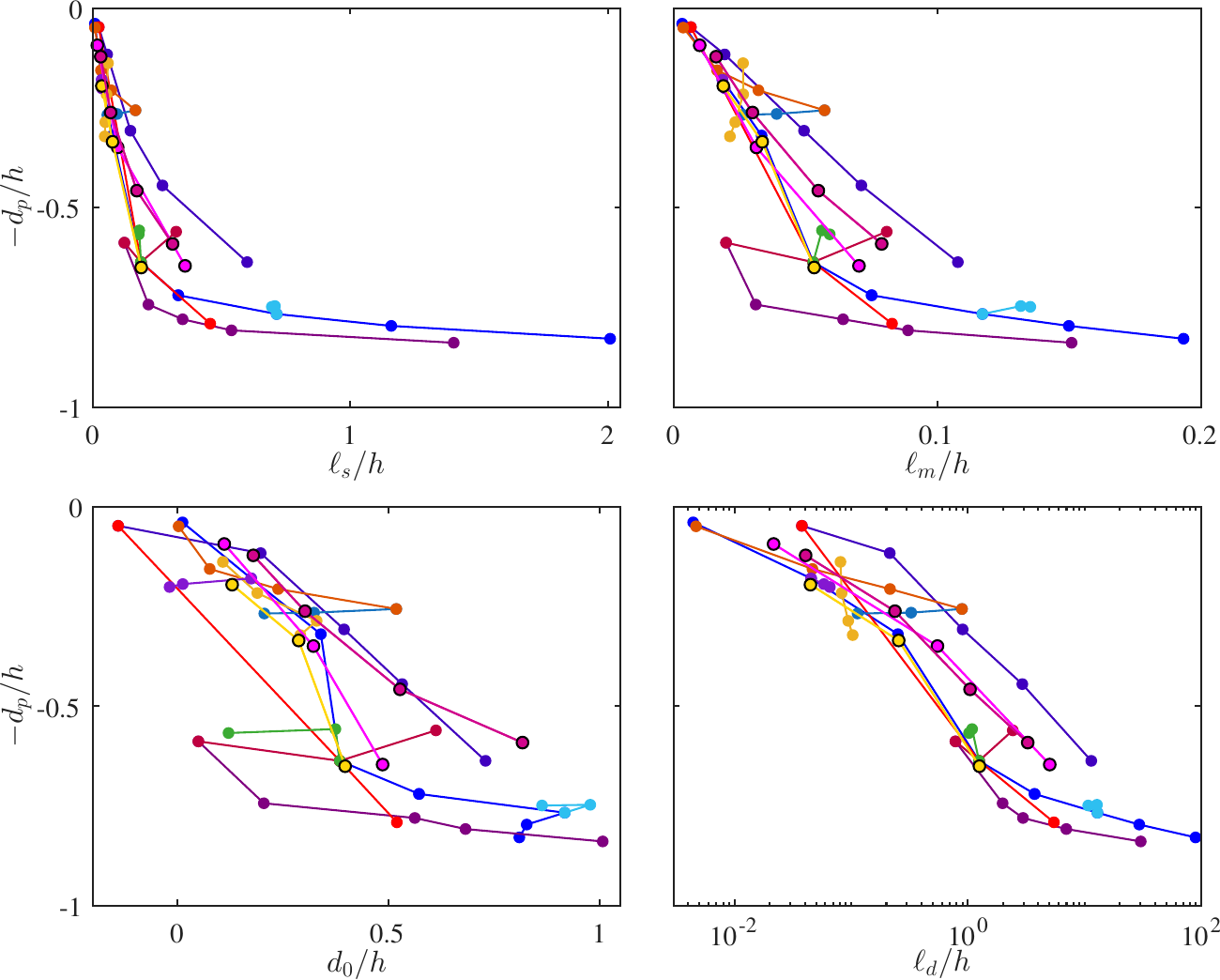}
    \put(-69mm,97mm){(\aaa)}
    \put(-08mm,97mm){(\bbb)}
    \put(-69mm,45mm){(\ccc)}
    \put(-08mm,45mm){(\ddd)}
    \vspace*{-1mm}
    \caption{\rgm{Penetration depth $d_p$ versus (\aaa) shear length $\ell_s$, (\bbb) mixing length $\ell_m$, (\ccc) zero-plane-displacement depth $d_0$, and (\ddd)  drag length $\ell_d$, relative to the canopy height $h$. Canopies are \textcolor{col1}{$\bullet$},~isotropic; \textcolor{col2}{$\bullet$}, `fence-like'; \textcolor{col3}{$\bullet$}, `canyon-like'; \textcolor{col4}{$\bullet$}, fixed $s_z$; \textcolor{col5}{$\bullet$}, fixed $s_x$; \textcolor{col6}{$\bullet$}, fixed gap; \textcolor{col7}{$\bullet$}, fixed pitch; \textcolor{col8}{$\bullet$}, `outer-similar'; \textcolor{col9}{$\bullet$}, \textcolor{col10}{$\bullet$}, and \textcolor{col11}{$\bullet$}, `inner-similar' with $\lambda_f\approx0.91$, 0.23 and 0.11, respectively; \protect\mdot, `doubled-height', $h^+\approx220$; \protect\ydot, streamwise-staggered; \protect\rdot, spanwise-staggered.
    \rgm{The coloured lines connect cases in the corresponding set.}}}
    \label{fig:other_densities}
\vspace*{3mm}
    \centering
    \includegraphics[width=.95\textwidth]{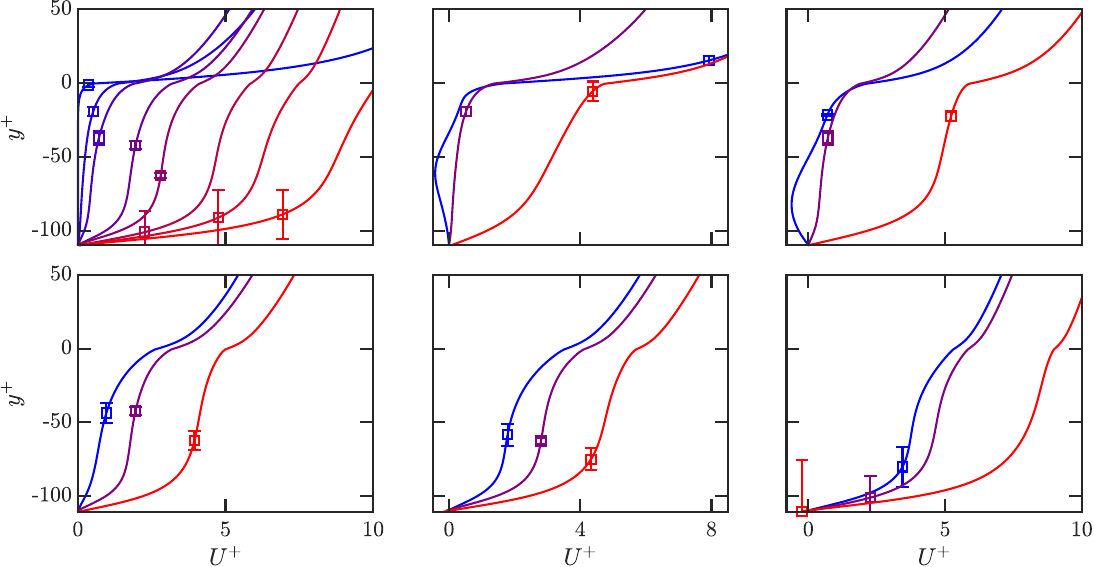}
    \put (-118mm ,61mm) {(\aaa)}
    \put (-76mm ,61mm) {(\bbb)}
    \put (-35mm ,61mm) {(\ccc)}
    \put (-118mm ,30mm) {(\ddd)}
    \put (-76mm ,30mm) {(\eee)}
    \put (-35mm ,30mm) {(\fff)}
    \put(-119mm ,55mm){\vector(3,-1){24mm}}
    \put(-94mm,47mm){$\lambda_f$}
    \put(-58mm,40mm){$\lambda_f=0.91$}
    \put(-16mm,40mm){$\lambda_f=0.51$}
    \put(-100mm,9mm){$\lambda_f=0.23$}
    \put(-58mm,9mm){$\lambda_f=0.13$}
    \put(-16mm,9mm){$\lambda_f=0.06$}
    \caption{\rgm{Mean-velocity profiles $U^+$ versus height $y^+$ for the same canopies of figure \ref{fig:den_eddy_stats_lamf}: (\aaa),~isotropic layouts with increasing frontal
    density, from red to blue $\lambda_f\approx0.01$-$2.04$; (\bbb)~through (\fff), fence-like (blue), isotropic (magenta)
    and canyon-like (red) layouts with $\lambda_f\approx0.06$-$0.91$. The symbols and errorbars are for the corresponding values of the penetration depth $d_p$.}}
    \label{fig:mean_profile}
\end{figure}

\rgm{In \S \ref{sec:intro} we have discussed different lengthscales that can be used to characterise canopy density,
such as the drag length, $\ell_d=(U)^2/D$, the shear length, $\ell_s=U/(dU/dy)$, the turbulent mixing length,
$\ell_m=[\overline{u'v'}]^{1/2}/(dU/dy)$, all measured at the plane of the canopy tips, or the zero-plane-displacement depth
below the that plane, $d_0$. All of these lengthscales are based on properties of the mean flow, and thus all give a measure of
its degree of penetration within the canopy. Throughout the paper, in contrast, we've advocated for characterising the
penetration of turbulence in terms of the penetration of turbulent eddies, rather than of the mean velocity profile.
Form this, we have proposed the use of the eddy canyon gap $g_z$ as a predictor for eddy penetration and thus canopy
density. Figures \ref{fig:yp_dense_depth}(\bbb) and \ref{fig:yp_dense_depth_staggered}(\aaa) portray the 
collapse of $d_p$ with $g_z$, summarising the paper. For comparison, figure \ref{fig:other_densities} portrays results
for the above four lengthscales vs $d_p$. The scatter observed is significantly larger than for $g_z$, and comparable to that
for the forntal density $\lambda_f$, shown for instance in figure
\ref{fig:yp_dense_depth_staggered}(\bbb). 
}

\rgm{Figure \ref{fig:mean_profile} portrays results for the mean velocity profile $U(y)$ for a wide variety of canopy layouts, and includes values for $d_p$ for reference. Again, although the shape of the mean velocity profile exhibits a generally monotonic evolution with increasing canopy density, this evolution is not always consistent with that of the penetration depth $d_p$. This highlights that the coupling between the penetration of the mean velocity and of the turbulence are not completely coupled.}

\section{Filtering out the element-coherent flow}\label{app:filter_sens}

\begin{figure}
\vspace*{3mm}
    \centering
    \includegraphics[width=\textwidth]{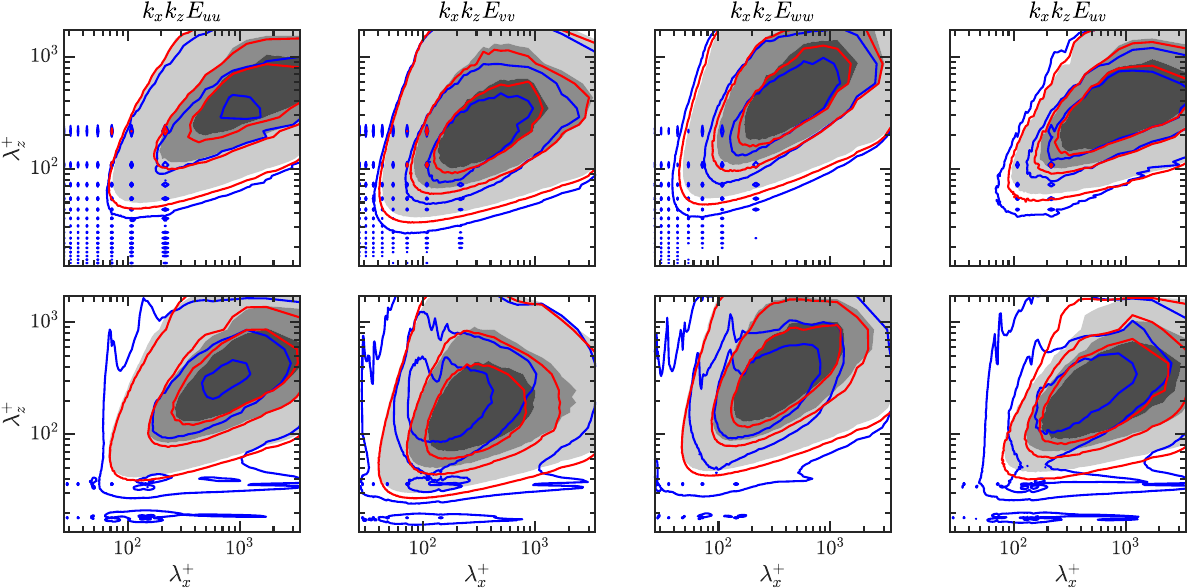}
    \put (-137mm ,62mm) {(\aaa)}
    \put (-100mm ,62mm) {(\bbb)}
    \put (-66mm ,62mm) {(\ccc)}
    \put (-33mm ,62mm) {(\ddd)}
    \put (-137mm ,32mm) {(\eee)}
    \put (-100mm ,32mm) {(\fff)}
    \put (-66mm ,32mm) {(\ggg)}
    \put (-33mm ,32mm) {(\hhh)}
\vspace*{1mm}
    \caption{Spectral energy densities \rgm{versus streamwise and spanwise wavelengths $\lambda_x^+$ and $\lambda_z^+$;}
    (\aaa-\ddd), case $\mathrm{I_{S216\times216}}$; (\eee-\hhh), case $\mathrm{F_{S144\times36}}$. The blue, red and shaded contours are at heights \zsc{$y=(0,1,2)\,\ell_d$} above the tips, respectively, where $l_d^+=w^+\approx24$.
    The contours are at 0.01, 0.05 and 0.1 times the r.m.s. level.}
    \label{fig:sp_abv_tip}
\vspace*{6mm}
    \centering
    \includegraphics[width=\textwidth]{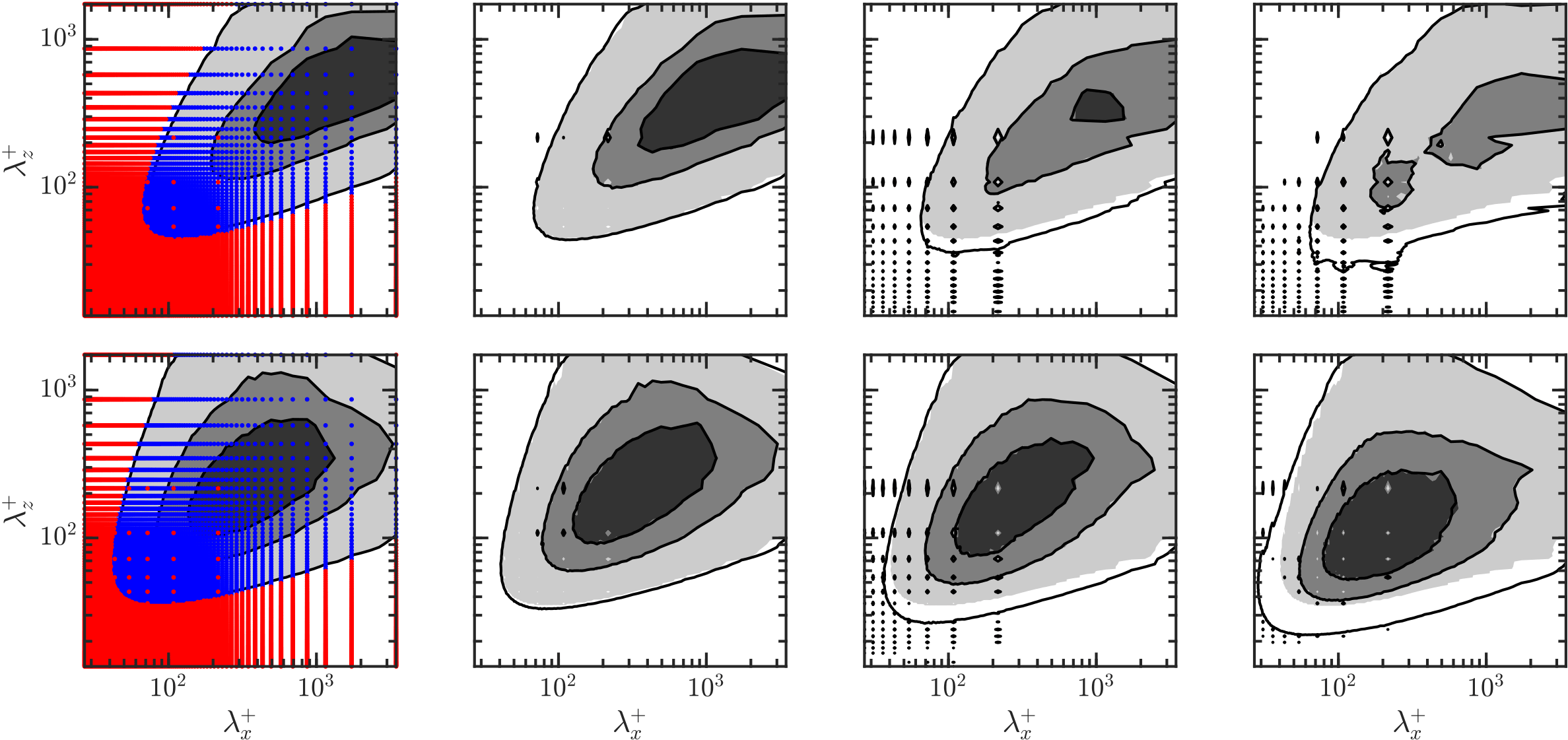}
    \put (-137mm ,62mm) {(\aaa)}
    \put (-100mm ,62mm) {(\bbb)}
    \put (-66mm ,62mm) {(\ccc)}
    \put (-33mm ,62mm) {(\ddd)}
    \put (-137mm ,32mm) {(\eee)}
    \put (-100mm ,32mm) {(\fff)}
    \put (-66mm ,32mm) {(\ggg)}
    \put (-33mm ,32mm) {(\hhh)}
\vspace*{1mm}
    \caption{Spectral filters and spectral energy densities \rgm{versus streamwise and spanwise wavelengths $\lambda_x^+$ and $\lambda_z^+$}
    for case $\mathrm{I_{S216\times216}}$. (\aaa-\ddd), $k_xk_zE_{uu}$; (\eee-\hhh), $k_xk_zE_{vv}$. Contours from left to right represent energy densities at $y^+\approx 48$, $24$, $0$ and $-55$. Line contours represent the energy density of the raw velocity, and  shaded contours that of the filtered flow fields. The filters are shown in $(a,e)$, where red dots indicate the wavelengths filtered-out and blue dots those retained, containing up to 97\% of the total $(u'^+)^2$ and $(v'^+)^2$, respectively. The contour levels are 0.01, 0.05 and 0.1 times the r.m.s. level.}
    \label{fig:sp_filter}
\end{figure}

\rgm{This appendix details the spectral filter applied in \S \ref{sec:structid} to remove the element-coherent flow and isolate the background-turbulence eddies.
Figure \ref{fig:sp_abv_tip} portrays premultiplied spectra, e.g. $k_x k_z E_{uu}$, for the three velocity components and the shear Reynolds stress at various heights above the tip plane for two sample canopies.
These maps portray the spectral energy density across different wavelengths (or lengthscales) in $x$ and $z$.} As shown in panels \ref{fig:sp_abv_tip}(\aaa-\ddd), the \rgm{element-coherent} footprint appears as concentrated energy at the canopy pitch \zsc{$s^+$} and its harmonics.
This footprint decays naturally with height above the canopy-tip plane, leaving only the \oldrev{background-turbulence wavelengths.}
Thus, we can use these background wavelengths to construct a spectral filter that selectively removes the \rgm{element-coherent} signal while retaining the background flow field.

To establish this filter, we first identify the height at which the \rgm{element-coherent} signal \oldrev{has decayed} naturally.
Near the tips, the scale of the \rgm{element-coherent} eddies is generally determined by the element width or gap, which provides the size of the `active' region for canopy-turbulence interaction \citep{sharma2020turbulent_b}. When the element width is larger than the gap ($w^+>g^+$), the \rgm{element-coherent} eddies are approximately of the size of the element width \citep{poggi2004effect, sadique2017aerodynamic, sharma2020scaling_a}. Conversely, when the gap is larger than the element width ($w^+<g^+$),  the \rgm{element-coherent} eddies are approximately of the size of the gap \citep{macdonald2018direct}.
Thus, the reference height for the spectral filter depends on a `decay length', \rgm{which characterises the height over which the element coherent flow vanishes,} defined as $\ell_d^+ = min(w^+,g^+)$.
For \oldrev{our anisotropic canopies with $s_x\neq s_z$,} we use $\ell_d^+ = w^+$ for simplicity.
As shown in figure \ref{fig:sp_abv_tip}, the \rgm{element-coherent} footprint, \rgm{in the form of spectral spikes at the streamwise and/or spanwise wavelengths of the canopy \citep{abderrahaman2019modulation}}, essentially \oldrev{vanishes} at a height of two `decay lengths', $y^+\approx2\ell_d^+\approx48$, above both the isotropic canopy $\mathrm{I_{S216\times216}}$ and the fence-like canopy $\mathrm{F_{S144\times36}}$, suggesting that \oldrev{this as a suitable reference height to capture the background signal alone.}

As depicted in figures \ref{fig:sp_filter}(\aaa, \eee), the filter removes the energy at the canopy pitch and its harmonics, while retaining the main spectral `lobe' of background-turbulence wavelengths.
The energy within this `lobe' accounts for $97\%$ of $(u'^+)^2$ and $(v'^+)^2$, respectively.
Care must be taken when choosing this filter percentage because if too many wavelengths remain unfiltered, too little \rgm{element-coherent} flow is removed.
However, if too many wavelengths are removed, the background flow within the substrate can be distorted or significantly reduced, artificially removing the signature of the eddies that are actually penetrating.

The spectral filter is applied within the canopy region ($y^+\leq0$) to remove the \rgm{element-coherent} flow, while it is switched off at height $y^+\geq 2\ell_d^+$, where the \rgm{element-coherent} flow has already decayed.
To avoid an abrupt transition, we apply a half-Hanning window for the filter intensity at height $0\leq y^+\leq 2\ell_d^+$, gradually reducing the filtered energy from $100\%$ at $y^+\leq0$ to $0\%$ at $y^+\geq2\ell_d^+$ for each filtered wavelength. As illustrated in figure \ref{fig:sp_filter}, the proposed filter removes the energy in the canopy harmonics and the quiescent wavelengths.
\oldrev{We note that this also removes any background-turbulence signal that may exist at the canopy harmonics, but this contribution is generally small compared to the remainder.}

\begin{figure}
\vspace*{4mm}
    \centering
    \includegraphics[width=\textwidth]{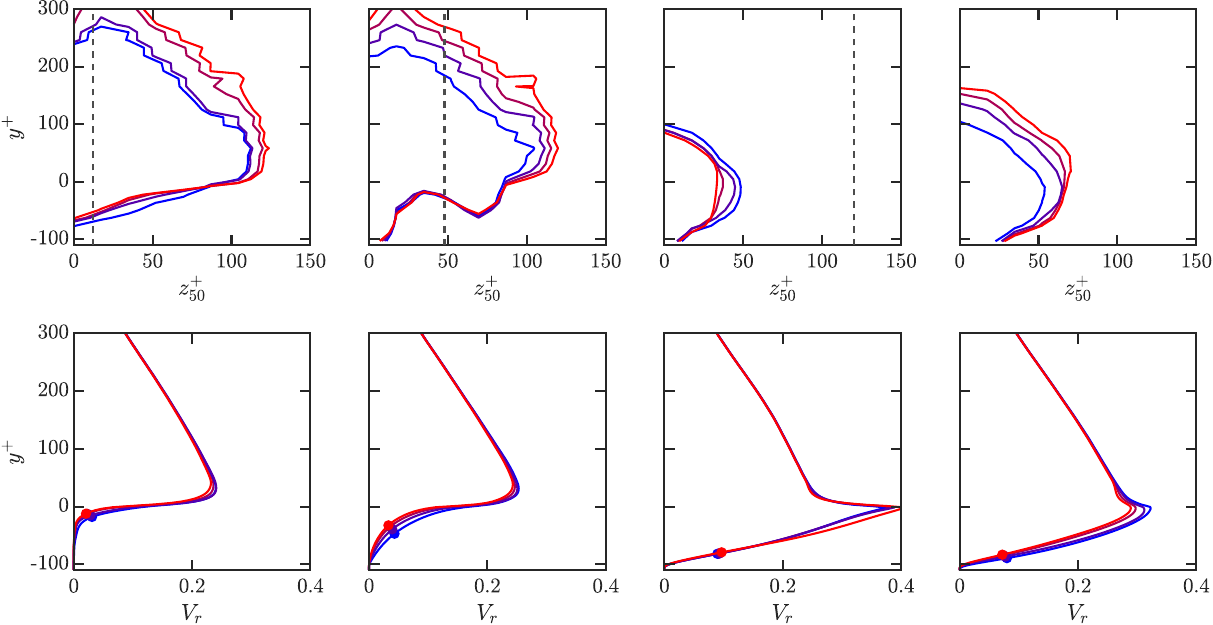}
    \put (-136mm ,68mm) {(\aaa)}
    \put (-99mm ,68mm) {(\bbb)}
    \put (-66mm ,68mm) {(\ccc)}
    \put (-33mm ,68mm) {(\ddd)}
    \put (-136mm ,31mm) {(\eee)}
    \put (-99mm ,31mm) {(\fff)}
    \put (-66mm ,31mm) {(\ggg)}
    \put (-33mm ,31mm) {(\hhh)}
\vspace*{2mm}
    \caption{Characteristic spanwise width $z_{50}^+$ and relative volume $V_r$ 
    \rgm{at each height $y^+$ for different levels of filtering.
     From blue to red, the contours are for filter levels $100\%$ (no filter), $99\%$, $97\%$ and $95\%$.    Cases portrayed are (\aaa, \eee), $\mathrm{F_{S144\times36}}$; (\bbb, \fff), $\mathrm{I_{S72\times72}}$; (\ccc, \ggg), $\mathrm{C_{S36\times144}}$; (\ddd, \hhh),~$\mathrm{I_{S216\times216}}$.
     In (\aaa-\ddd), the p.d.f. contours enclose 75\% of the penetrating eddies, and the dashed lines mark the corresponding inter-element gap \rgm{$g_z^+\approx12$, $48$, $120$, and $192$, when within the range displayed.
     In (\eee-\hhh),} markers $\bullet$ indicate the penetration depth where $V_r/V_{r,tip}=0.25$.}
     }
     \label{fig:app_sens_f}
\end{figure}

\begin{figure}
\vspace*{1mm}
    \centering
    \includegraphics[width=.96\textwidth]{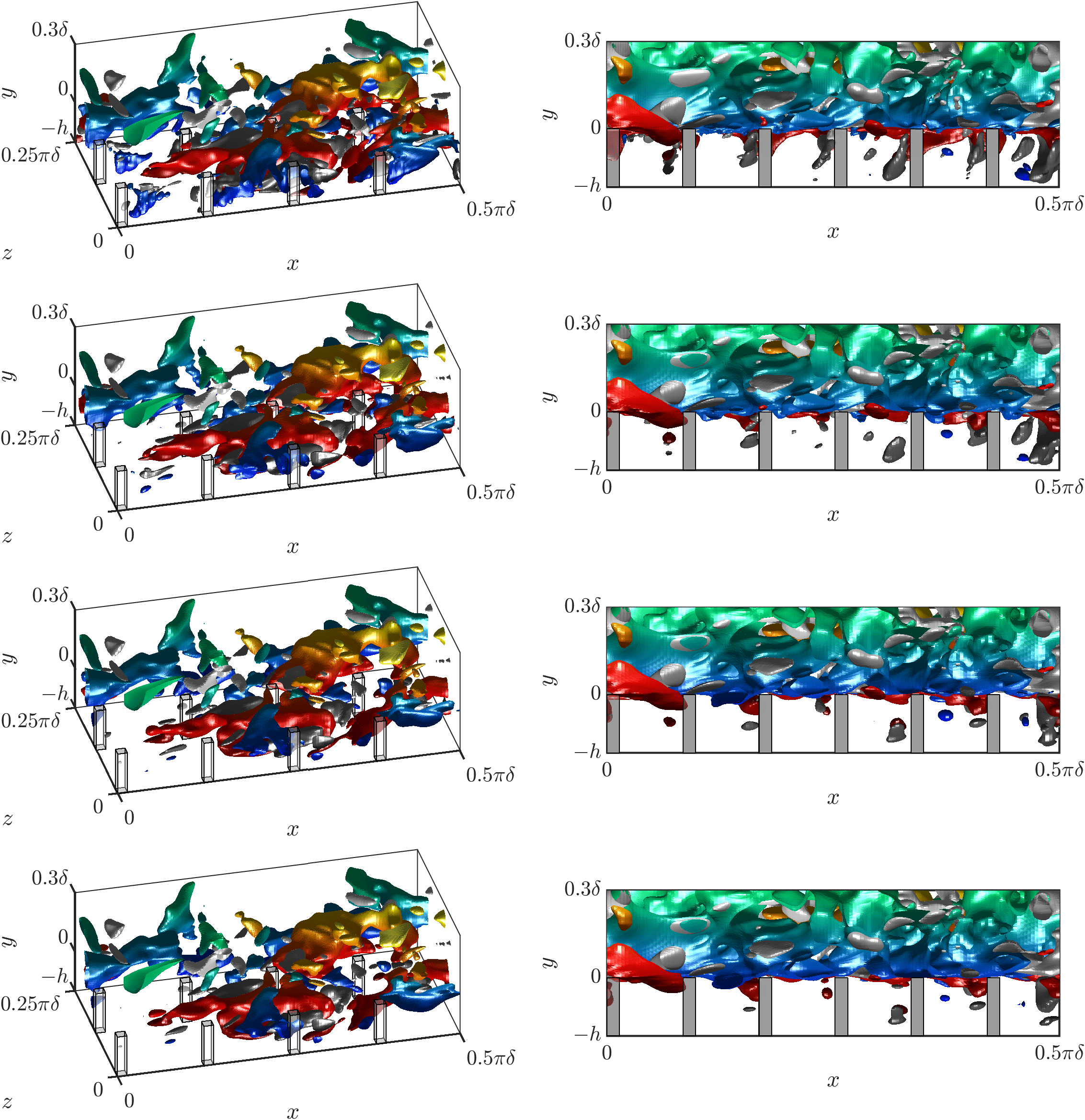}
    \put (-131mm,131mm) {(\aaa)}
    \put (-67mm ,131mm) {(\bbb)}
    \put (-131mm,98mm) {(\ccc)}
    \put (-67mm ,98mm) {(\ddd)}
    \put (-131mm,64mm) {(\eee)}
    \put (-67mm ,64mm) {(\fff)}
    \put (-131mm,30mm) {(\ggg)}
    \put (-67mm ,30mm) {(\hhh)}
    \caption{Instantaneous realisations of $u'v'$ structures for canopies $\mathrm{I_{S216\times216}}$ (\aaa, \ccc, \eee, \ggg)  and $\mathrm{F_{S144\times36}}$ (\bbb, \ddd, \fff, \hhh). The filter levels are $100\%$ (no filter) in (\aaa, \bbb), $99\%$ in (\ccc, \ddd), $97\%$ in (\eee, \fff) and $95\%$ in (\ggg, \hhh). The structures are coloured by distance to the floor, ejections blue to green, sweeps red to yellow, and outward and inward interactions grey to white.}
    \label{fig:app_uvster}
\end{figure}

The filter intensity must be chosen carefully so that the background eddies are not eliminated or significantly distorted. Figures \ref{fig:sp_filter}(\aaa, \eee) showed that, for the filter applied throughout the paper, the filtered flow retained 97\% of the total energy of the turbulent fluctuations $(u'^+)^2$ and $(v'^+)^2$. Here, we assess the sensitivity of the filtered $u'v'$ structures by comparing results based on filter levels that retain $100\%$ (no filter), $99\%$, $97\%$ and $95\%$ of the fluctuating energy.
The goal is to ensure that the location and topology of the background-turbulence structures remain consistent \oldrev{regardless of the filter intensity,} while the element-coherent signal is effectively removed. 
Results are shown in figures \ref{fig:app_sens_f}(\aaa-\ddd), where the characteristic width $z_{50}^+$ \rgm{at each height} of eddies within the canopy remains similar across all filter levels. The collapse of the relative volume $V_r$ and the penetration depth \zsc{$d_p^+$,} shown in figures \ref{fig:app_sens_f}(\eee-\hhh), further confirms that the applied filter essentially isolates the background turbulence without altering it significantly.
As illustrated in figure \ref{fig:app_uvster}, both the stem-scale wakes behind the elements in panel (\aaa) and the shedding eddies between the fences in panel (\bbb) are effectively removed for filter levels $\leq99\%$, while the background-turbulence structures remain mostly unaffected.

\section{Determining the extent of $u'v'$ eddies}\label{app:percolation}

\rgm{This appendix details the methodology used to set the threshold value used in equation (\ref{eq:qs}) to determine the extent in space of the structures
of intense point-wise $u'v'$ from the background turbulence that come into contact with the canopy.}
The choice of the right-hand side in (\ref{eq:qs}) can significantly influence the resulting topology of the eddies. A low threshold value results in large, interconnected eddies spanning the entire flow field, while an excessively high threshold may filter out too many actual eddies, leaving a thin skeleton for only the most intense ones \citep{nagaosa2003statistical, del2006self}. In smooth-wall studies, three-dimensional flow structures are typically identified using height-dependent normalisation factors \rgm{to reflect the intense variation of mean shear Reynolds stress with height} \citep{jimenez2018coherent, marusic2019attached}. The early work of quadrant analysis by \cite{willmarth1972structure} used the mean Reynolds shear stress at each height \rgm{$|\overline{u'v'(y)}|$},
\begin{equation}
|u'v'(x,y,z)|>H_{u'v'(y)}|\overline{u'v'(y)}|.
\label{eq:mean uv}
\end{equation}

However, this approach encounters a limitation when $|\overline{u'v'(y)}|$ tends toward zero near the channel centre, leading to an exaggerated representation of the eddies in this region. This limitation could be circumvented by using the r.m.s. of velocity fluctuations $u'(y)$ and $v'(y)$, as in \cite{wallace1972wall}, \cite{lu1973measurements} and \cite{bogard1986burst},
or the r.m.s. of Reynolds stress fluctuation \citep{narasimha2007turbulent},
In the present case, the height-dependent normalisation factors $|\overline{u'v'(y)}|$, $u'(y)v'(y)$ and $(uv)'(y)$ 
decrease rapidly with \rgm{depth into} the canopy, which would result in an over-representation of \oldrev{the eddies within.} 
\cite{sharma2020turbulent_b} reported that turbulent fluctuations decayed exponentially within the canopy, with the slowest decay occurring for the wall-normal component $v'$, even when persistent footprints of Kelvin-Helmholtz \oldrev{rollers} were still observed at half-height below the tips for dense canopies.
Consequently, as the \rgm{above height-dependent} normalisation factors become vanishingly small within a dense canopy, weak fluctuations would be identified as strong eddies, with their significance overemphasised.
To avoid this, we normalise the pointwise Reynolds shear stress with a \oldrev{$y$-constant value,} as indicated in (\ref{eq:qs}), using the total shear stress at the canopy-tip plane, $\tau_{tip}$.
This normalisation factor allows a consistent statistical analysis of canopies across various density regimes using a constant $H$, circumventing the need to recalibrate the value of $H$ for each canopy, as $\tau_{tip}$ provides the scale for the flow in the vicinity of the tips for both dense and sparse canopies.
\oldrev{In contrast,} alternative normalisation factors, such as $u'v'(y_{tip})$ and $u'(y_{tip})v'(y_{tip})$, can yield large values for $H$ for dense canopies and small values for sparse canopies, complicating the comparative analysis across different canopies.

\begin{figure}
\vspace*{1mm}
    \centering
    \includegraphics[width=\textwidth]{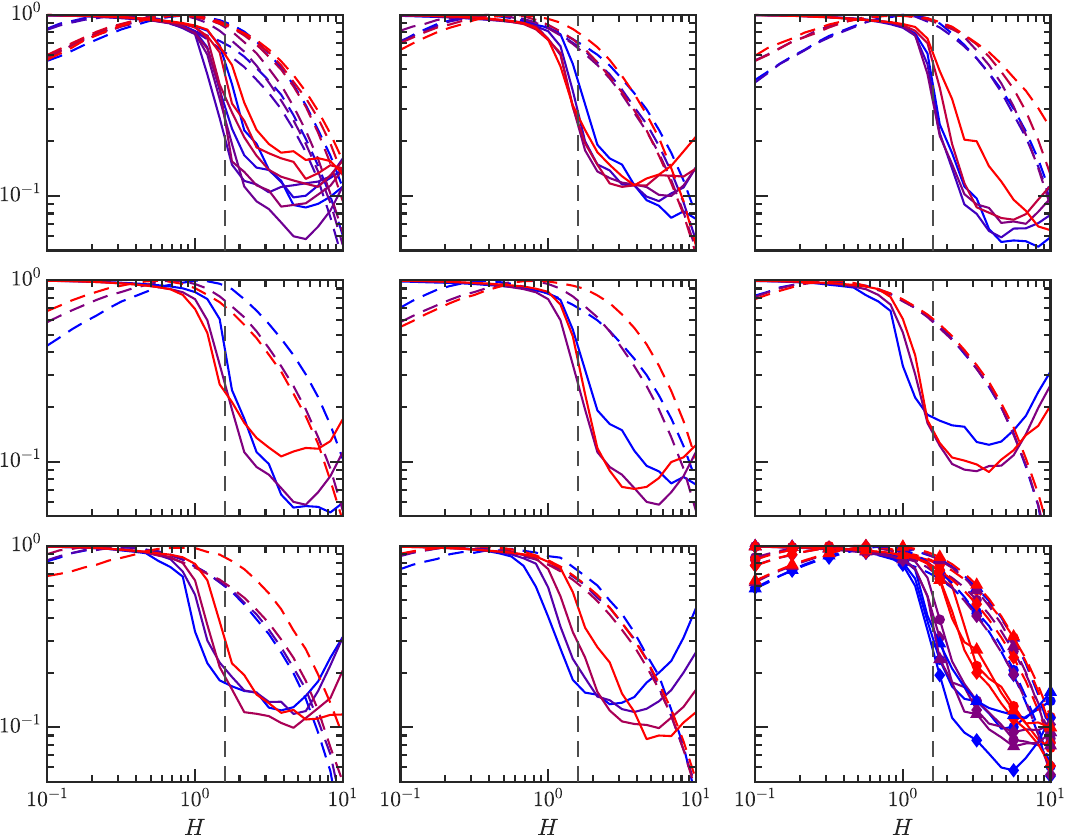}
    \put (-139mm,103mm) {(\aaa)}
    \put (-89mm ,103mm) {(\bbb)}
    \put (-44mm ,103mm) {(\ccc)}
    \put (-139mm,69mm) {(\ddd)}
    \put (-89mm ,69mm) {(\eee)}
    \put (-44mm ,69mm) {(\fff)}
    \put (-139mm,35mm) {(\ggg)}
    \put (-89mm ,35mm) {(\hhh)}
    \put (-44mm ,35mm) {(\iii)}
    \caption{\rgm{Percolation diagrams for the identification of $u'v'$ structures for a broad set of canopies as a function of the threshold value $H$ in (\ref{eq:qs}).
    Solid lines represent the volume ratio between the largest eddy and the total volume of all eddies, $V_{largest}/V_{total}$; dashed lines represent the ratio of the number of identified eddies to the maximum number of eddies, $N/N_{max}$.}
    (\aaa)~canopies with isotropic layout; (\bbb) fence-like layout; (\ccc) canyon-like layout; (\ddd) fixed $s_z$; (\eee)~fixed $s_x$; (\fff) fixed gap; (\ggg) fixed pitch; (\hhh) similar geometry in outer units; and (\iii) similar geometry in inner units. From blue to red, the lines represent canopies with decreasing $\lambda_f$ in (\aaa-\ggg) and increasing $Re_\tau$ in (\hhh, \iii).     
    In (\iii), the symbols $\bullet$, $\blacklozenge$, and $\blacktriangle$, mark the cases with the same geometry as $\mathrm{IS550_{G30\times30}}$, $\mathrm{IS550_{G84\times84}}$, and $\mathrm{IS550_{G192\times192}}$, respectively.
    The chosen threshold $H=1.6$ is indicated by a vertical dashed line.}
    \label{fig:percolation}
\end{figure}

The value of the threshold $H$ in (\ref{eq:qs}) is determined based on a percolation analysis \citep{del2006self}, as depicted in figure \ref{fig:percolation}.
\zsc{This percolation analysis examines how the size and connectivity between identified eddies vary as a function of $H$.}
At high values of $H$ ($H\gtrsim3$), only a few eddies, corresponding to the core of the most intense structures, are identified. As $H$ decreases, new structures are identified while the existing ones expand in size.
For $H\lesssim3$, the largest eddy expands proportionally with the total volume, and the volume ratio between the largest eddy and all eddies, $V_{largest}/V_{total}$, remains roughly constant.
As $H$ continues to decrease, the identified objects begin to merge, significantly increasing the ratio $V_{largest}/V_{total}$. For $H\lesssim1$, the total number of eddies, normalised with its maximum, $N/N_{max}$ decreases, and $V_{largest}/V_{total}$ approaches unity as most \rgm{eddies merge into a single one}. \zsc{This sudden transition is the percolation crisis, at which separate turbulent structures coalesce into one large connected cluster, and the distinction between individual eddies is lost \rtwo{\citep{del2006self}}.}
Figure \ref{fig:percolation} illustrates that a percolation crisis occurs when $0.5\lesssim H\lesssim3$ for all cases. We therefore use $H=1.6$ throughout to ensure that individual structures remain separate while avoiding excessive volume shrinkage.

\begin{figure}
\vspace*{3mm}
    \centering
    \includegraphics[width=\textwidth]{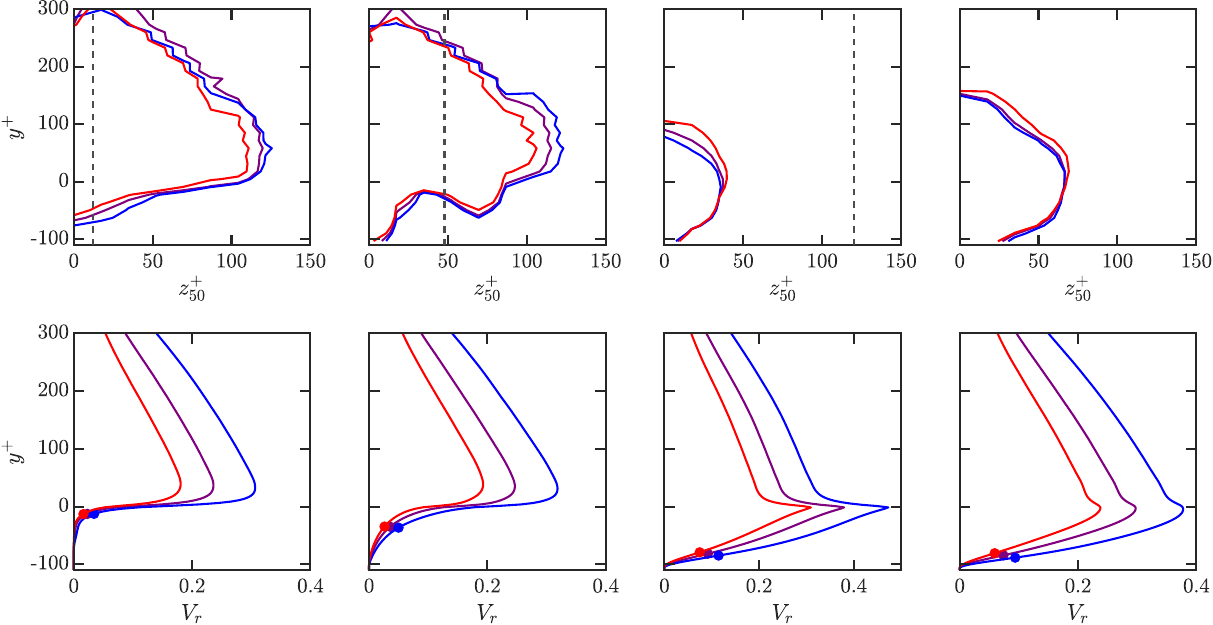}
    \put (-136mm ,68mm) {(\aaa)}
    \put (-99mm ,68mm) {(\bbb)}
    \put (-66mm ,68mm) {(\ccc)}
    \put (-33mm ,68mm) {(\ddd)}
    \put (-136mm ,31mm) {(\eee)}
    \put (-99mm ,31mm) {(\fff)}
    \put (-66mm ,31mm) {(\ggg)}
    \put (-33mm ,31mm) {(\hhh)}
    \caption{Characteristic spanwise width $z_{50}^+$ and relative volume $V_r$ 
    \rgm{at each height $y^+$ for different values of the threshold $H$ in (\ref{eq:qs}). From blue to red, thresholds $H=1.2$, $1.6$ and $2.0$. Cases portrayed are (\aaa, \eee), $\mathrm{F_{S144\times36}}$; (\bbb, \fff), $\mathrm{I_{S72\times72}}$; (\ccc, \ggg), $\mathrm{C_{S36\times144}}$; (\ddd, \hhh), $\mathrm{I_{S216\times216}}$.
    In (\aaa-\ddd), the \rgm{p.d.f.} contours enclose 75\% of the penetrating eddies, and the dashed lines mark the corresponding inter-element gap, $g_z^+\approx12$, $48$, $120$, and $192$, when within the range displayed. In (\eee-\hhh), markers $\bullet$ indicate the penetration depth where $V_r/V_{r,tip}=0.25$.}
    }
    \label{fig:app_sens_h}
\end{figure}

For $H$ within the range of the percolation crisis, the size and location of the penetrating eddies are largely insensitive to the exact choice of $H$.
Although the eddy volume at each height \oldrev{increases slightly} for larger $H$, the penetration depth remains unaffected.
\rgm{We assess this by comparing results with} thresholds $H = 1.2,1.6,2.0$. As $H$ decreases, new structures emerge while the existing ones expand in size, leading to a slight increase in the relative volume $V_r$ at each height, as shown in figures \ref{fig:app_sens_h}(\eee-\hhh). Nevertheless, the penetration depth \zsc{$d_p^+$} remains consistent across all examined values of $H$.
Similarly, figures \ref{fig:app_sens_h}(\aaa-\ddd) show that the characteristic spanwise width $z_{50}^+$ at each height within and immediately above the canopies are similar for different $H$,
\oldrev{and only exhibit differences far above the canopy-tip plane. We can conclude that
the influence of the choice of $H$ in the range $H = 1.2$-$2$ has no significant influence on our results and the ensuing conclusions.}

\section{\rgm{Quadrant-wise analysis of eddy penetration}}\label{app:quadrants}

\rgm{The analysis of structures of intense pointwise Reynolds stress $u'v'$ can also distinguish between structures in different quadrants,} 
depending on the signs of the fluctuating velocity components $u'$ and $v'$ \citep{wallace1972wall, willmarth1972structure, lozano2012three, lozano2014time}.
The quadrants Q1, Q2, Q3, and Q4 represent outward interactions $(u'>0, v'>0)$, ejections $(u'<0, v'>0)$, inward interactions \rgm{$(u'<0, v'<0)$}, and sweeps $(u'>0, v'<0)$, respectively. Depending on the density regime, turbulent structures from different quadrants can have different intensities and locations \citep{brunet2020turbulent}.

\begin{figure}
\vspace*{1mm}
    \centering
    \includegraphics[width=\textwidth]{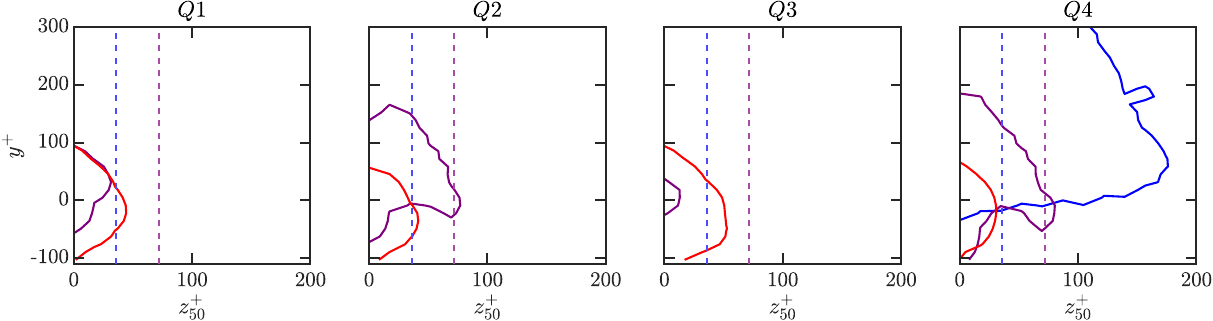}
    \put (-136mm,32mm) {(\aaa)}
    \put (-99mm ,32mm){(\bbb)}
    \put (-66mm ,32mm) {(\ccc)}
    \put (-33mm ,32mm) {(\ddd)}
    \caption{\rgm{Probability density function of the spanwise width $z^+_{50}$  
    vat each height $y^+$ of interfacial eddies from different $u'v'$ quadrants.
    Blue contours are for dense canopy $\mathrm{I_{S36\times36}}$;
    magenta contours for intermediate canopy $\mathrm{I_{S72\times72}}$;
    and red contours for sparse canopy $\mathrm{I_{S432\times432}}$.
    The dashed vertical lines mark the corresponding element pitch $s^+\approx36$, $72$, and $432$, with the latter beyond the abscissa range displayed.
    The contour level for each canopy encloses 75\% of the eddies from the corresponding quadrant. \zsc{In (\aaa-\ccc), the occurrence of interfacial eddies of canopy $\mathrm{I_{S36\times36}}$ is too low to be captured.}}}
     \label{fig:den_zy_q_example}
\end{figure}

Figure \ref{fig:den_zy_q_example} depicts the typical \rgm{spanwise size $z_{50}^+$ at each height} of eddies across all four quadrants for isotropic-layout canopies 
with $s^+\approx35$, $70$, and $430$, using the same contour level used in figure \ref{fig:den_zy_example} \rgm{for all $u'v'$ structures together}.
This analysis provides insights into how structures \oldrev{from} different quadrants, e.g. sweeps and ejections, penetrate differently.
For the dense canopy $\mathrm{I_{S36\times S36}}$, sweeps dominate the turbulent structures.
This is expected because the stress balance in a turbulent channel requires $\overline{u'v'}<0$, which enhances the significance of sweeps and ejections with $u'v'<0$.
Furthermore, near the canopy-tip plane, ejections and outward interactions with $v'>0$ are significantly \oldrev{hindered} by the canopy drag \citep{huang2009effects, sharma2020turbulent_b}.
As a result, only large sweeping structures are observed to penetrate slightly into the dense canopy \oldrev{with very few ejections springing out of it.} For the intermediate canopy, sweeps and ejections exhibit larger spanwise widths and occur more frequently than inward and outward interactions.
\zsc{The dominance of sweeps and ejections within and above vegetation canopies has long been recognised, from early studies such as \cite{nakagawa1977prediction}, \cite{finnigan1979turbulence}, and \cite{raupach1981turbulence}, to later work including \cite{gardiner1994wind}, \cite{novak2000wind}, and \cite{dupont2010modelling}, consistent with our observations.} 
These studies noted that sweeps generally contribute more significantly to Reynolds shear stress compared to ejections.
Finally, for a sparse canopy like $\mathrm{I_{S432\times432}}$, structures from different quadrants exhibit similar spanwise widths and occur with comparable frequency, suggesting a more uniform contribution from all four quadrants.

\begin{figure}
\vspace*{1mm}
    \centering
    \includegraphics[width=\textwidth]{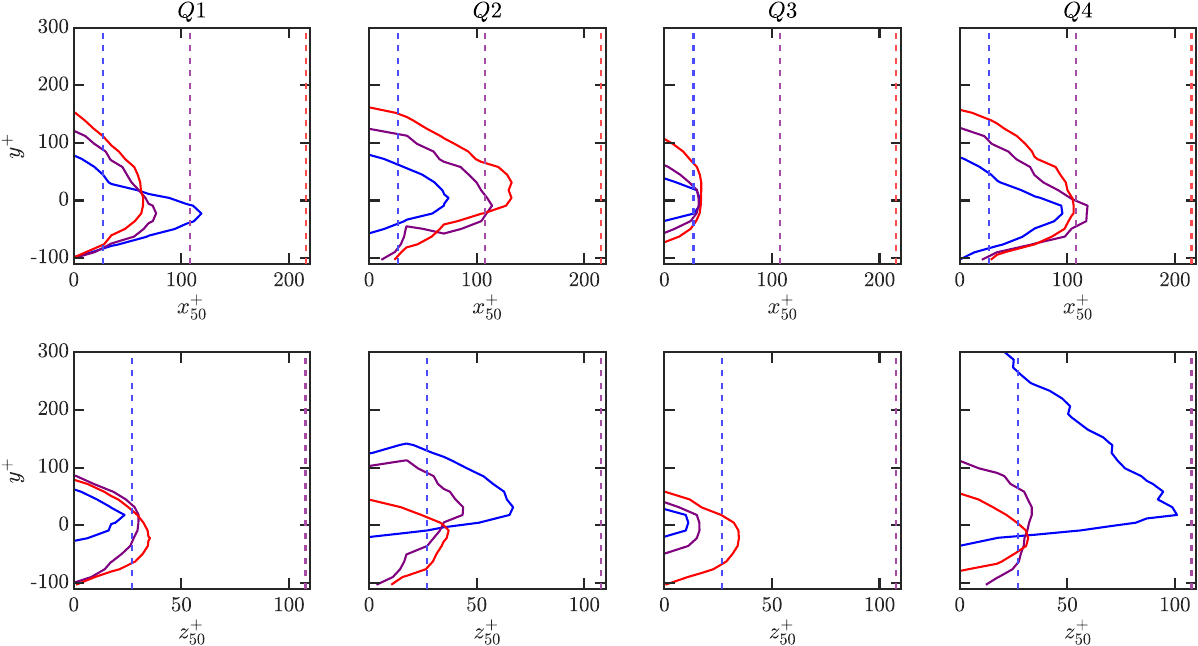}
    \put (-137mm ,69mm) {(\aaa)}
    \put (-100mm ,69mm) {(\bbb)}
    \put (-66mm ,69mm) {(\ccc)}
    \put (-33mm ,69mm) {(\ddd)}
    \put (-137mm ,32mm) {(\eee)}
    \put (-100mm ,32mm) {(\fff)}
    \put (-66mm ,32mm) {(\ggg)}
    \put (-33mm ,32mm) {(\hhh)}
    \caption{Probability density function of the characteristic wall-parallel sizes 
    \rgm{$x^+_{50}$ and $z^+_{50}$ at each height $y^+$}    
    of interfacial eddies from different quadrants. (\aaa-\ddd), eddy streamwise length for canopies with a fixed $s_z^+\approx108$; (\eee-\hhh), eddy spanwise width for canopies with a fixed $s_x^+\approx108$.
    The \rgm{blue, purple and red contours are for canopies with frontal densities $\lambda_f\approx0.91$, $0.23$ and $0.11$, respectively}; the dashed lines mark the canopy streamwise and spanwise pitch, $s_x^+,s_z^+\approx27$, $108$, and $216$. The contours enclose 75\% overall eddies.}
    \label{fig:den_eddy_xzy}
\end{figure}

\rgm{Figure \ref{fig:den_eddy_xzy} portrays the characteristic streamwise $x_{50}^+$ and spanwise $z_{50}^+$ size at ecah height of structures in different quadrants for
the canopies discussed in \S \ref{sec:x_or_z}, with fixed element geometry and pitch in one direction and varying pitch in the other.}
Panels (\aaa-\ddd) show that canopies with a fixed spanwise pitch $s_z^+\approx108$ allow turbulence to penetrate effectively into the canopy, and the eddies within can elongate up to $x_{50}^+\approx120$, with no significant influence of the value of $s_x^+$. However, in contrast to these fixed-$s_z$ canopies, where turbulence penetration is largely unimpeded, canopies with a constant streamwise pitch $s_x^+\approx108$ exhibit different penetration behaviours depending on their spanwise pitch, as illustrated in figures \ref{fig:den_eddy_xzy}(\eee-\hhh).
A fence-like canopy such as $\mathrm{X_{S108\times27}}$ significantly restricts turbulence penetration, allowing limited Q2 and Q4 structures to penetrate, with Q4 structures having the highest occurrence and the largest width, up to $z_{50}^+\approx100$.
This suggests that sweeps dominate turbulence penetration within fence-like canopies, which is consistent with previous observations on obstructing vegetation canopies \citep{poggi2004effect, yue2007comparative, bailey2013turbulence}. \cite{bailey2013turbulence} reported that the dominance of sweeps persists even when the streamwise pitch between the fences increases to $s_x/h\approx3$. 
In contrast, as $s_z^+$ increases, canopies such as $\mathrm{X_{S108\times108}}$ and $\mathrm{X_{S108\times216}}$ allow turbulence to penetrate more effectively, and structures in different quadrants have a more uniform contribution to turbulence penetration.

\begin{figure}
    \centering
    \includegraphics[width=\textwidth]{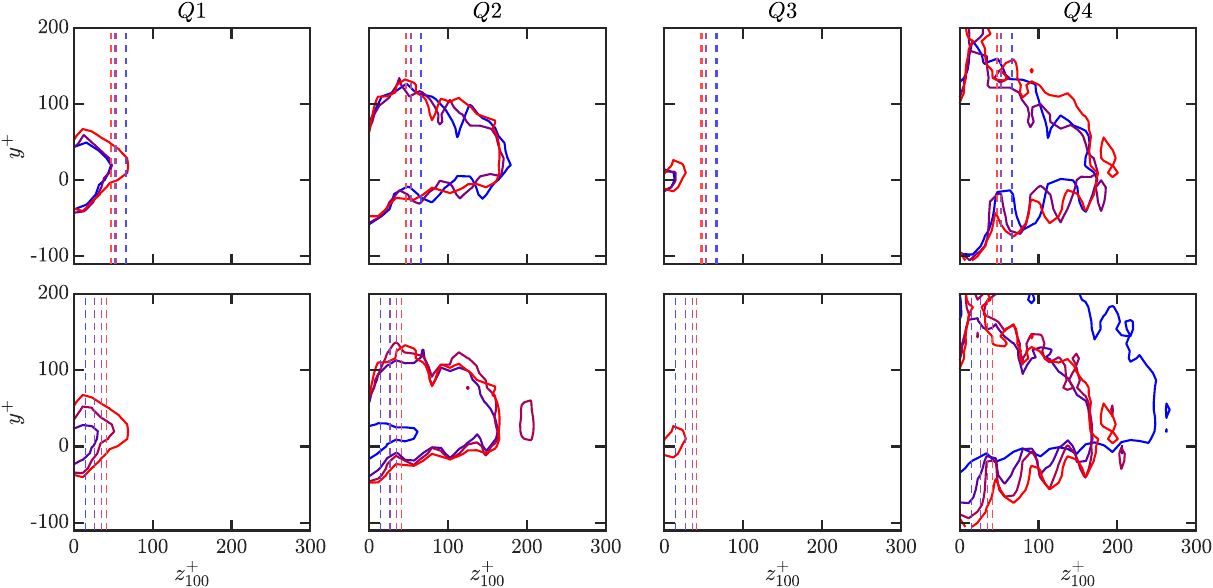}
    \put (-137mm ,62mm) {(\aaa)}
    \put (-100mm ,62mm) {(\bbb)}
    \put (-66mm ,62mm) {(\ccc)}
    \put (-33mm ,62mm) {(\ddd)}
    \put (-137mm ,32mm) {(\eee)}
    \put (-100mm ,32mm) {(\fff)}
    \put (-66mm ,32mm) {(\ggg)}
    \put (-33mm ,32mm) {(\hhh)}
    \caption{Probability density function of the total spanwise width 
    \rgm{$z^+_{100}$ at each height $y^+$}
    of eddies from different $u'v'$ quadrants, for canopies with (\aaa-\ddd) fixed gap $g_z^+\approx41$ and (\eee-\hhh) fixed pitch $s_z^+\approx47$. Colours and contour levels are as in figure \ref{fig:den_eddy_zy_gs_all}.}
    \label{fig:den_eddy_zy_gs}
\end{figure}

\rgm{In turn, figure \ref{fig:den_eddy_zy_gs} illustrates the relative contributions from different quadrants to \oldrev{the} overall turbulence penetration 
for
the canopies discussed in \S \ref{sec:pitch_or_gap}, with either a fixed element gap and varying pitch or fixed pitch and varying gap.
For all the canopies portrayed,}
sweeps and ejections dominate the penetrating structures, and sweeps are generally more intense and penetrate deeper within the canopy compared to ejections. This observation aligns with the findings by \cite{collineau1993detection}, who investigated flows over an obstructing pine forest and found that the ratio of the total stress fraction transferred by sweeps to that transferred by ejections is up to \oldrev{2-3} at height $y/h\approx0.8$. The dominance of sweeps has also been reported on other obstructing vegetation canopies \citep{novak2000wind, dupont2011long, dupont2012influence}, as summarised in \cite{brunet2020turbulent}. 
Figures \ref{fig:den_eddy_zy_gs}(\ddd, \hhh) show that large sweep-type eddies with width $z_{100}^+>g_z^+$ cannot fully enter the canopy, but their `feet' may partially penetrate if they fit within the gap between elements.
As an example, for canopy $\mathrm{S47_{G41\times41}}$ in figure \ref{fig:den_eddy_zy_gs}(\hhh), for large eddies that span multiple canopy elements, their width $z_{100}^+$ exhibits harmonics of the element pitch $s_z^+\approx47$ within the canopy.
Eddies with increasing widths $z_{100}^+\approx70,115,160$ have decreasing penetration depths $y^+\approx75,60,40$, respectively, demonstrating a reduced penetration depth for larger footprints.

\bibliographystyle{jfm}
\bibliography{canopies_zishen}

\end{document}